\documentclass[a4paper,11pt]{article}
\usepackage{jheppub}
\usepackage{amsfonts,amsmath,amssymb}
\usepackage{graphicx}
\usepackage{physics}
\usepackage{float}
\usepackage{orcidlink}

\title{Multiplicity distributions in QCD jets and jet topics}

\author[a,b,c]{Xiang-Pan Duan\orcidlink{0009-0001-0471-8832}, }
\author[c]    {Lin Chen\orcidlink{0000-0003-2082-3533}, }
\author[a,b]  {Guo-Liang Ma\orcidlink{0000-0002-7002-8442}, }
\author[c,d]  {Carlos A. Salgado\orcidlink{0000-0003-4586-2758}, }
\author[c]    {and Bin Wu\orcidlink{0000-0002-3320-0442}}
\affiliation[a]{Key Laboratory of Nuclear Physics and Ion-beam Application (MOE), Institute of Modern Physics, Fudan University, Shanghai 200433, China}
\affiliation[b]{Shanghai Research Center for Theoretical Nuclear Physics, NSFC and Fudan University, Shanghai 200438, China}
\affiliation[c]{Instituto Galego de F\'isica de Altas Enerx\'ias IGFAE, Universidade de Santiago de Compostela, E-15782 Galicia, Spain}
\affiliation[d]{Axencia Galega de Innovación (GAIN), Xunta de Galicia, Galicia, Spain}
\emailAdd{xpduan20@fudan.edu.cn}
\emailAdd{lin.chen@usc.es}
\emailAdd{glma@fudan.edu.cn}
\emailAdd{carlos.salgado@usc.es}
\emailAdd{b.wu@cern.ch}

\abstract{
We evaluate the Koba–Nielsen–Olesen (KNO) scaling functions for quark- and gluon-initiated jets by incorporating energy conservation into the Double Logarithmic Approximation (DLA). The resulting modified DLA (MDLA) expressions differ substantially from the DLA predictions and qualitatively align with the recently proposed QCD-inspired expressions, albeit with some quantitative differences. By fixing the two parameters in the MDLA expressions, we show that the inclusive charged-particle multiplicity distributions of the two leading jets in $pp$ collisions at $\sqrt{s} = 13$ TeV, measured by ATLAS over a wide jet $p_T$ range of $0.1$–$2.5$ TeV, are well described within experimental uncertainties and consistent with {\tt PYTHIA} simulations. This conclusion is further supported by direct comparisons with quark- and gluon-initiated jet distributions extracted via jet topics, though the propagated uncertainties from experimental data remain sizable.
}

\begin{document}

\maketitle
\flushbottom

\section{Introduction}
\label{sec:intro}

QCD jets are collimated sprays of hadrons with fluctuating particle numbers, and their multiplicity distributions, that is, the probability distribution $P(n)$ for producing $n$ particles, constitute an essential feature. Studying these distributions provides valuable insights into the dynamics of Quantum Chromodynamics (QCD), particularly the mechanisms of parton branching and hadronization in the high-virtuality regime. An expected simplification of multiplicity distributions in high-energy QCD jets arises from the concept of \emph{Koba–Nielsen–Olesen} (KNO) scaling~\cite{Koba:1972ng}. This scaling hypothesis states that at asymptotically high energy, the probability distribution of final-state multiplicities $P(n)$ becomes self-similar when rescaled by the mean multiplicity $\Bar{n}$, such that $\Bar{n}P(n)$ depends only on the scaling variable $x=n/\Bar{n}$ and approaches a universal function $\Psi(x)$, that is,
\begin{align}\label{eq:KNO_old}
    \bar{n} P(n) = \Psi\left({n}/{\bar{n}}\right)
\end{align}
with $\int dx~\Psi(x)=1$. It was originally proposed on general grounds of scale invariance in quantum field theory~\cite{Polyakov:1970lyy,Feynman:1969ej}, and found a natural interpretation within the framework of QCD as a consequence of the cascade structure of parton showers in QCD jets~\cite{Konishi:1979cb, Bassetto:1979nt, Dokshitzer:1982ia, Malaza:1984vv, Bassetto:1987fq, Dokshitzer:1991wu, Dokshitzer:1993dc, Vertesi:2020utz, Germano:2024ier, Duan:2025ngi, Dokshitzer:2025owq, Dokshitzer:2025fky}.

Multiplicity distributions are expected to exhibit universality across different high-energy collision systems, including $e^+e^-$, Deep Inelastic Scattering (DIS), and $pp$ collisions, when dominated by high-energy jets. In particular, measurements of hadronic final states in $e^+e^-$ annihilation over a wide energy range, from PETRA to LEP~\cite{TASSO:1983cre,TASSO:1989orr,DELPHI:1990ohs,DELPHI:1991qnt}, showed that charged-particle multiplicity distributions approximately collapse onto a universal KNO curve. Nonetheless, deviations from exact KNO scaling are expected, arising from the superposition of jets with different flavors and topologies~\cite{ALEPH:1995qic,Giovannini:1997ce}, as well as from the running of the strong coupling~\cite{Dokshitzer:2025owq,Dokshitzer:2025fky}. Similar approximate KNO behavior is observed in DIS at HERA~\cite{H1:2020zpd}, where both theoretical predictions~\cite{Liu:2022bru,Liu:2023eve} and experimental data support scaling in the current hemisphere at large momentum transfer. Early low-energy $pp$ collisions, from bubble chamber experiments to ISR data, also reported approximate scaling of multiplicity distributions~\cite{Slattery:1972ni,Ames-Bologna-CERN-Dortmund-Heidelberg-Warsaw:1983cqw}. In high-energy hadronic collisions, however, LHC data~\cite{UA5:1985kkp,UA5:1988gup,CMS:2010qvf,Grosse-Oetringhaus:2009eis,ALICE:2017pcy} show that multiplicity distributions broaden with increasing energy and no longer follow KNO scaling. This breakdown is not surprising, since contributions from mixed event topologies of hard processes, such as quark- and gluon-initiated jets~\cite{Duan:2025ngi,Dokshitzer:2025owq}, as well as soft processes, can lead to scaling violation (for recent discussions, see ref.~\cite{Martins-Fontes:2025iee} and references therein). KNO scaling is suggested to hold within QCD jets in $pp$ collisions~\cite{Vertesi:2020utz,Germano:2024ier}, while such universality is anticipated to reemerge when focusing specifically on high-energy QCD jets with proper quark–gluon discrimination~\cite{Duan:2025ngi,Dokshitzer:2025owq}.

The KNO scaling functions are different for quark- and gluon-initiated jets, referred to as quark and gluon jets below, and have been calculated in perturbative QCD~\cite{Dokshitzer:1982ia, Bassetto:1987fq, Dokshitzer:1991wu, Dokshitzer:1993dc, Duan:2025ngi, Dokshitzer:2025owq, Dokshitzer:2025fky}. In the \emph{double logarithmic approximation} (DLA), where the dominant soft and collinear emissions are resummed with strong ordering in both energy and angle, the forms of the scaling functions $\Psi(x)$ for gluon and quark jets can be evaluated explicitly~\cite{Dokshitzer:1982ia, Bassetto:1987fq, Dokshitzer:1991wu}. These scaling functions are exact in the asymptotic high-energy (high-virtuality) limit, and KNO scaling within DLA has recently been confirmed to hold to a good approximation by numerically solving the multiplicity distributions in QCD jets for a broad jet $p_T$ range relevant to LHC energies~\cite{Duan:2025ngi}.

However, the DLA KNO scaling functions do not provide a reliable description of the measured multiplicity distributions~\cite{Malaza:1984vv, Dokshitzer:1993dc, Duan:2025ngi, Dokshitzer:2025owq, Dokshitzer:2025fky}. 
To remedy this, the DLA for gluon jets was first improved by enforcing energy conservation in the parton cascade~\cite{Dokshitzer:1993dc}. This \emph{modified double logarithmic approximation} (MDLA) has recently been generalized to jets initiated by any source, including quarks and antiquarks, achieving notable phenomenological success~\cite{Dokshitzer:2025owq,Dokshitzer:2025fky}. Meanwhile, the KNO scaling functions have also been extracted using {\tt PYTHIA}~\cite{Vertesi:2020utz,Duan:2025ngi}, showing good agreement with LHC measurements within experimental uncertainties~\cite{ATLAS:2019rqw,Duan:2025ngi}. One of the main goals of this work is to compare quark and gluon KNO scaling functions obtained within MDLA and {\tt PYTHIA}, and to explore their phenomenological implications.

Recent advances in jet physics and jet substructure at the LHC~\cite{Marzani:2019hun,Larkoski:2024uoc} provide an unprecedented opportunity to investigate multiplicity fluctuations in QCD jets at very high transverse momenta, both theoretically and experimentally. A variety of methods for quark–gluon discrimination have been proposed~\cite{Gallicchio:2011xq,Gallicchio:2012ez,Larkoski:2013eya,Larkoski:2014pca,Gras:2017jty,Metodiev:2017vrx,Metodiev:2018ftz,Komiske:2018vkc,Dreyer:2021hhr} and applied in LHC analyses~\cite{CMS:2013kfa,ATLAS:2014vax,ATLAS:2016wzt,ATLAS:2019rqw}. If proven effective for multiplicity distributions, these techniques would enable direct experimental tests of the universality of the KNO scaling functions for quark and gluon jets, as illustrated using energy–correlation functions~\cite{Larkoski:2013eya} in ref.~\cite{Duan:2025ngi}. Another central motivation of this paper is to investigate KNO scaling in QCD jets using jet topics~\cite{Metodiev:2018ftz,Komiske:2018vkc}, which is well suited to extracting multiplicity distributions and can be implemented without additional theoretical input.

The rest of this paper is structured as follows. In \autoref{sec:qgjet}, we start with a review of the generating function method for multiplicity distributions within QCD jets, as well as a summary of the DLA results for parton multiplicity distributions and mean multiplicities. We then calculate the KNO scaling function for both quark and gluon jets within the MDLA, i.e., by incorporating energy conservation into the DLA, following refs.~\cite{Dokshitzer:1982ia, Bassetto:1987fq, Dokshitzer:1993dc}, as an alternative to the recent calculations in refs.~\cite{Dokshitzer:2025owq, Dokshitzer:2025fky}. In \autoref{sec:incjet}, we present a phenomenological study of inclusive charged-particle multiplicity distributions of two leading jets in $pp$ collisions at $\sqrt{s}=13$ TeV, combining the MDLA KNO scaling functions with leading-order cross sections and the calculations of mean multiplicities, and compare our theoretical results with ATLAS data from ref.~\cite{ATLAS:2019rqw}. As a consistency check, we also directly compare the MDLA KNO scaling functions for quark and gluon jets with those extracted from ATLAS measurements using topic modeling~\cite{Metodiev:2018ftz, Komiske:2018vkc}, which is presented in \autoref{sec:topic}. We conclude in \autoref{sec:concl} with a summary of results and outlook for future directions. Finally, for completeness, we collect the next-to-next-to-next-to-leading-order (N$^3$LO) results for mean parton multiplicities in QCD jets from refs.~\cite{Dremin:1999ji,Capella:1999ms} in \autoref{app:nb_N3LO}, which are used in our phenomenological studies in \autoref{sec:incjet}.

\section{Multiplicity distributions and KNO scaling in QCD jets}
\label{sec:qgjet}

In this section, we begin with a brief review, for the sole purpose of completeness in our subsequent discussion, of the method of \emph{generating functions} (GFs)~\cite{Konishi:1979cb, Bassetto:1979nt, Dokshitzer:1982ia, Dokshitzer:1991wu, Ellis:1996mzs,Dremin:2000ep} and the DLA results for parton multiplicity distributions within QCD jets~\cite{Dokshitzer:1982xr,Dokshitzer:1982ia,Dokshitzer:1982fh, Duan:2025ngi}. We then evaluate the KNO scaling functions within MDLA for both quark and gluon jets by building on the treatment of gluon jets presented in ref.~\cite{Dokshitzer:1993dc}, providing an alternative to the recent calculations in refs.~\cite{Dokshitzer:2025owq,Dokshitzer:2025fky}.

\subsection{Review of the generating function method}

The GF method serves as a powerful and convenient tool for describing the evolution of QCD jets produced in high-energy hard scatterings.  To support the forthcoming discussion, we outline the basics of the method and summarize the DLA results of parton multiplicities within QCD jets here. For a more comprehensive discussion, interested readers are referred to refs.~\cite{Dokshitzer:1991wu, Ellis:1996mzs, Dremin:2000ep}.

\subsubsection{The definition of generating functions}

For a QCD jet initiated by a parton $a$ at initial scale $Q$, the GF is defined as  
\begin{align}\label{eq:GFdef}
    Z_a(u,Q) \equiv \sum_{n=0}^{\infty} u^n P_a(n,Q) \qquad \text{with } Z_a(1,Q)=1,
\end{align}
where the \emph{multiplicity probability distribution} $P_a(n, Q)$ denotes the probability of finding $n$ particles. 
By successive differentiation of the GF, one can recover the probability distributions:
\begin{align}\label{eq:Pndef}
    P_a(n,Q) = \frac{1}{n!} \frac{\partial^n}{\partial u^n} Z_a(u,Q) \bigg|_{u=0}.
\end{align}

Analogous to the integral form of the DGLAP equation, the GFs at next-to-leading logarithmic (NLL) order obey~\cite{Ellis:1996mzs}
\begin{align}\label{eq:ZNLL}  
    Z_a(u, Q) &= Z_a(u, Q_0) \Delta_a(Q, Q_0) + \int_{Q_0}^Q \frac{d\bar{Q}}{\bar{Q}}\frac{\Delta_a(Q, Q_0)}{\Delta_a(\bar{Q}, Q_0)} \nonumber\\  
    &\times\int dz \frac{\alpha_s}{\pi}\hat{P}_{a\to bc}(z) Z_{b}(u, z \bar{Q}) Z_{c}(u, (1-z) \bar{Q}),  
\end{align}
where $\alpha_s$ is evaluated at the scale $z(1-z)\bar{Q}$ if the running of the strong coupling is taken into account, the repeated indices $b$ and $c$ run over all parton species, $Q_0 < Q$ denotes another momentum scale, the splitting functions are given by  
\begin{align}\label{eq:Pa2b} 
    \hat{P}_{g\to gg}(z) &= 2 C_A\bigg[\frac{1-z}{z}+\frac{z}{1-z}+z(1-z)\bigg], \nonumber\\ 
    \hat{P}_{g \to q\bar{q}}(z) &= T_F[z^2+(1-z)^2], \nonumber\\ 
    \hat{P}_{q \to gq}(z) &= C_F\bigg[\frac{1+(1-z)^2}{z}\bigg],
\end{align}
and the Sudakov form factors are defined as
\begin{align}
    \Delta_a(Q, Q_0) = \exp\bigg[-\sum_{bc}\int_{Q_0}^Q \frac{d\bar{Q}}{\bar{Q}}\int dz \frac{\alpha_s}{\pi}\hat{P}_{a\to bc}(z)\bigg].
\end{align}
Equivalently, by differentiating eq.~\eqref{eq:ZNLL} with respect to $Q$, one obtains, at $O(\alpha_s)$~\cite{Dremin:2000ep},
\begin{align}\label{eq:ZNLLDiff}  
    \frac{\partial}{\partial\ln Q}Z_a(u, Q) = \int dz \frac{\alpha_s}{\pi}\hat{P}_{a\to bc}(z) [Z_{b}(u, z Q) Z_{c}(u, (1-z) Q) - Z_{a}(u, Q)].
\end{align}

The first term on the right-hand side of eq.~\eqref{eq:ZNLL}, involving the form factor $\Delta_a$, corresponds to the process in which the initial parton $a$ remains unchanged, while the second term accounts for the splitting process $a \to bc$. 
When \emph{coherent branching (i.e., angular ordering)}~\cite{Dokshitzer:1987nm, Dokshitzer:1991wu} is incorporated into the second term, the formalism properly accounts for both soft and collinear enhancements. 
Denote the momenta of the daughter parton $b$ and its parent $a$ by $k$ and $p$, respectively. 
In this case, the evolution variable is most conveniently defined as~\cite{Ellis:1996mzs}
\begin{align}\label{eq:evoltuionscale}
     Q^2 
     = \frac{p^2}{z(1 - z)} 
     = \frac{k_\perp^2}{z^2 (1 - z)^2},
\end{align}
where the daughter parton $b$ carries a longitudinal momentum fraction $z$, and $k_\perp$ represents the transverse momentum of the daughter parton.

\subsubsection{Parton multiplicities in QCD jets within DLA}

Focusing solely on soft and collinear logarithmic contributions, the splitting function simplifies to $\hat{P}_{a \to g a}(z) \approx 2 C_a / z$. And the condition for coherent branching can be expressed as~\cite{Dokshitzer:1991wu}
\begin{align}\label{eq:DLAphasespace}
    z Q = k^0 \theta> k_\perp > Q_0,
\end{align}
where $k$ is the momentum of the emitted soft gluon, $\theta$ is the angle relative to the preceding branching, and the momentum scale of the parent parton is given by $Q = p \theta$, with $p$ denoting the parent parton’s momentum.

Within DLA, eq.~\eqref{eq:ZNLLDiff} becomes
\begin{align}\label{eq:ZDLA}  
    \frac{\partial}{\partial\ln Q} Z_a(u, Q) = Z_a(u, Q) c_a\int \frac{dz}{z} \gamma_0^2 \left[Z_g(u, z Q) - 1\right] \quad \text{with $\gamma_0 = \sqrt{\frac{2N_c\alpha_s}{\pi}}$}, 
\end{align}
where $c_g = 1$ for gluon jets and $c_q \equiv {C_F}/{C_A}$ for quark jets with the Casimir factors $C_A = N_c$ and $C_F = (N_c^2-1)/(2N_c)$. 
It has the formal solution
\begin{align}
    Z_a(u, Q) = Z_a(u, Q_0)\exp\left\{\int_{Q_0}^Q \frac{d\bar{Q}}{\bar{Q}} \int \frac{dz}{z} \frac{2\alpha_s(z\bar{Q}) C_a}{\pi} \left[Z_g(u, z\bar{Q}) - 1\right]\right\}.
\end{align}
For hadron multiplicities, the GF necessarily includes non-perturbative contributions. 
In this section, we instead focus on parton multiplicities at the reference virtuality $Q_0$, where the system contains a single parton. This sets the initial condition: $Z_a(u, Q_0) = u$.\footnote{
Note that in refs.~\cite{Dokshitzer:1982xr,Dokshitzer:1982ia,Dokshitzer:1982fh}, the GFs account only for the radiated soft gluons with $Z_a(u, Q_0) = 1$. In contrast, our convention follows ref.~\cite{Dokshitzer:1991wu}, where the original leading parton is counted in $P_a(n)$.
}
Restricting to the DLA phase space in eq.~\eqref{eq:DLAphasespace}, in terms of $k_\perp = z\bar{Q}$ and $z$, one obtains~\cite{Dokshitzer:1991wu}
\begin{align}\label{eq:GF_DLA}
    Z_a(u, y) = u \exp \left\{ c_a \int_0^y d\bar{y}~(y - \bar{y}) \gamma_0^2 [Z_g(u,\bar{y})-1] \right\},
\end{align}
where $y \equiv \ln(Q/Q_0)$ and $\bar{y} \equiv \ln(k_\perp/Q_0)$. From the GFs given in eq.~\eqref{eq:GF_DLA}, we obtain the following recursive relation between $P_a(n)$'s~\cite{Duan:2025ngi}:
\begin{align}\label{eq:Pqg}
    P_a(1,Q)
    &= \exp \left\{-c_a \int_0^y d\bar{y}~(y-\bar{y}) \gamma_0^2 \right\}, \nonumber\\
    P_a(n+1,Q)
    &= c_a \sum_{k=1}^{n} \frac{k}{n}P_a(n+1-k,Q) \int_0^y  d\bar{y}~(y-\bar{y}) \gamma_0^2 P_g(k,\bar{y}).
\end{align}

The mean multiplicity, denoted by $\bar{n}_a$, can also be derived by taking the first derivative of the GFs with respect to $u$ at $u = 1$:
\begin{align}\label{eq:nbar}
    \bar{n}_a(Q)
    = 1 + c_a \int_0^y d\bar{y}~(y-\bar{y}) \gamma_0^2 \bar{n}_g(\bar{y}),
\end{align}
where the normalization condition of the GFs has been applied in the final step. 
Accordingly, one has
\begin{align}\label{eq:nbarQG}
    \bar{n}_q(Q) - 1 = c_a [\bar{n}_g(Q) - 1].
\end{align}
Differentiating eq.~\eqref{eq:nbar} twice with respect to the evolution variable $y$ for gluon jets gives
\begin{align}
    \frac{\partial^2}{\partial y^2}\bar{n}_g
    &= \gamma_0^2 \bar{n}_g
\end{align}
with the boundary conditions $\bar{n}_g'(0) = 0$ and $\bar{n}_g(0) = 1$. This second-order ordinary differential equation admits the analytic solution:~\cite{Dokshitzer:1982fh, Dokshitzer:1982xr, Dokshitzer:1991wu}
\begin{align}\label{eq:mean2}
    \bar{n}_g = \left\{
    \begin{array}{ll}
         \cosh(\gamma_0 y)&\text{for fixed coupling} \\
         z_1 \left[ I_1(z_1) K_0(z_2) + K_1(z_1) I_0(z_2) \right]&\text{for running coupling} 
    \end{array}
    \right.
    ,
\end{align}
where $z_1 \equiv A \sqrt{y+\lambda}$ and $z_2 \equiv A \sqrt{\lambda}$, with $\lambda \equiv \ln (Q_0/\Lambda)$, $A \equiv \sqrt{{16N_c}/{\beta_0}}$, and $\beta_0=({11}N_c-{2}n_f)/{3}$. Alternatively, given $P_a(n, Q)$ numerically evaluated according to eq.~\eqref{eq:Pqg}, the mean parton multiplicities in quark and gluon jets can be obtained via:
\begin{align}\label{eq:mean1}
    \bar{n}_a(Q) = \sum_{n=1}^\infty nP_a(n,Q).
\end{align}
The agreement of these two approaches has been verified in ref.~\cite{Duan:2025ngi}.

\subsection{KNO scaling within QCD jets}

In the asymptotic limit $Q\to\infty$, KNO scaling in multiplicity distributions within QCD jets naturally follows the definition of the GFs~\cite{Konishi:1979cb, Bassetto:1979nt, Dokshitzer:1982ia, Dokshitzer:1991wu, Dremin:2000ep}. As previously verified in ref.~\cite{Duan:2025ngi}, $\bar{n}_a(Q)P_a(n, Q)$ for partons within DLA, given in eq.~\eqref{eq:Pqg}, indeed converge to universal scaling functions for quark and gluon jets respectively for $Q\gg Q_0$. However, they are quite different from those for hadrons in \texttt{PYTHIA} simulations and the LHC data. In this subsection, we evaluate the KNO scaling functions within MDLA for both quark and gluon jets, improving the DLA results~\cite{Dokshitzer:1982ia, Bassetto:1987fq, Dokshitzer:1991wu} by imposing energy conservation~\cite{Dokshitzer:1993dc, Dokshitzer:2025owq, Dokshitzer:2025fky}.

\subsubsection{Definition of KNO scaling functions for quark and gluon jets}
\label{ssec:KNO}

Following refs.~\cite{Dokshitzer:1982ia, Dokshitzer:1991wu}, we first outline the definition of, and the method for evaluating, the KNO scaling functions for QCD jets, which holds exactly in the asymptotic limit $Q\to\infty$. Setting $u = \exp\{-\beta/\bar{n}_a\}$ in the definition of GFs in eq.~\eqref{eq:GFdef} and taking the limit $Q \to \infty$ in $Z_a$ yields
\begin{align}\label{eq:Phi}
    \Phi_a(\beta)
    &\equiv \lim\limits_{Q\to \infty} Z_a(e^{-\frac{\beta}{\bar{n}_a}},Q) = \lim\limits_{Q\to \infty} \sum_{n=0}^{\infty} \frac{e^{-\beta \frac{n}{\bar{n}_a}}}{\bar{n}_a} [\bar{n}_a P_a(n, Q)]\nonumber\\
    &\equiv \int_0^{\infty} dx~ \Psi_a(x) e^{-\beta x} = \sum\limits_{k=0}^\infty \frac{(-\beta)^k}{k!} f_a^{(k)}
\end{align}
where $x \equiv n/\bar{n}_a$, and 
\begin{align}
    f_a^{(k)} \equiv \int_0^{\infty} dx~x^k \Psi_a(x)
\end{align}
with
\begin{align}
    f_a^{(0)} = 1 = f_a^{(1)}.
\end{align} 
Accordingly, the asymptotic KNO scaling functions $\Psi_a(x)$ are given by the inverse Laplace transform of $\Phi_a(\beta)$
\begin{align}
    \Psi_a(n/\bar{n}_a) \equiv \lim\limits_{Q\to \infty}[\bar{n}_a P_a(n, Q)] = \int \frac{d\beta}{2\pi i}\Phi_a(\beta) e^{\beta \frac{n}{\bar{n}_a}},
\end{align}
where the integral over $\beta$ runs parallel to the imaginary axis, with all singularities of $\Phi_a$ lying to the left in the complex plane. This definition thus extends the original KNO scaling in eq.~(\ref{eq:KNO_old}) to jets initiated by specific parton species.

One way to evaluate the above KNO scaling functions is to boil it down to evaluating $f_a^{(k)}$. In terms of \emph{multiplicity correlators}:
\begin{equation}
    n^{(k)}_a(Q) \equiv \left.\frac{\partial^k}{\partial u^k} Z_a(u, Q)\right|_{u=1} = \sum_{n=k}^{\infty}\frac{n!}{(n-k)!}P_a(n, Q),
\end{equation}
one has
\begin{equation}\label{eq:GF-taylor}
    Z_a(u, Q) = \sum_{k=0}^{\infty}\frac{(u-1)^k}{k!}n^{(k)}_a(Q).
\end{equation}
From this expression of $Z_a$, one obtains
\begin{align}
     \Phi_a(\beta) 
     = \lim\limits_{Q\to \infty} \sum_{k=0}^{\infty}\frac{(e^{- \frac{\beta}{\bar{n}_a(Q)}}-1)^k}{k!}n^{(k)}_a(Q) 
     = \sum\limits_{k=0}^\infty \frac{(-\beta)^k}{k!} \lim\limits_{Q\to \infty} \frac{n^{(k)}_a(Q)}{[\bar{n}_a(Q)]^k}.
\end{align}
Comparing with the power series expansion in eq.~\eqref{eq:Phi}
which yields
\begin{align}\label{eq:fak}
    f_a^{(k)} = \lim\limits_{Q\to \infty} \frac{n^{(k)}_a(Q)}{[\bar{n}_a(Q)]^k}.
\end{align}

\subsubsection{KNO scaling functions within MDLA}
\label{sec:MDLA}

Similar to the evolution equation for the GFs within DLA in eq.~\eqref{eq:ZDLA}, incorporating the approximation $\hat{P}_{a \to g a}(z) \approx 2 C_a / z$ along with energy conservation in the NLL evolution equation in eq.~\eqref{eq:ZNLLDiff} yields
\begin{align}\label{eq:ZMDLA}
    \frac{\partial}{\partial\ln Q} Z_a(u, Q) = c_a\int \frac{dz}{z} \gamma_0^2 \left[Z_g(u, z Q) Z_a(u, (1-z)Q) - Z_a(u, Q)\right].
\end{align}
By differentiating both sides of this equation at $u=1$, one has
\begin{align}\label{eq:nmaMDLA}
    \frac{\partial n^{(m)}_a(Q)}{\partial\ln Q} = c_a\int \frac{dz}{z} \gamma_0^2 \left[ \sum\limits_{k=0}^m C_m^k n_g^{(k)}(z Q) n_a^{(m-k)}((1-z)Q) - n^{(m)}_a(Q)\right],
\end{align}
where the binomial coefficient is denoted as
\begin{align}
    C_m^k\equiv\frac{m!}{k!(m-k)!}.
\end{align}
As we are interested in the large $Q$ limit, $n^{(m)}_a(Q)$ can be expressed in terms of $f_a^{(m)}$ and $\bar{n}_a(Q)$ according to eq.~\eqref{eq:fak} in the above equation. Following ref.~\cite{Dokshitzer:1993dc}, we consider only the corrections arising from energy conservation, while neglecting the effects of the running coupling and higher-order corrections to the anomalous dimension $\gamma_0$. In this case, one can approximate the mean multiplicities as
\begin{align}
    \bar{n}_a(Q) \approx c_a (Q/Q_0)^{\gamma_0}/2, 
\end{align}
following their fixed-coupling expressions in eq.~\eqref{eq:mean2}. 

Under the above approximation, for $m\geq2$, after integrating out $z$ on the right-hand side of eq.~\eqref{eq:nmaMDLA}, one obtains
\begin{align}
   c_a^{m-1} \{ m +& c_a \gamma_0[\psi(\gamma_0 m + 1) - \psi(1)]\}f^{(m)}_a - \frac{f^{(m)}_g}{m} \nonumber\\
   &= \gamma_0 \sum\limits_{k=1}^{m-1} 
   C_m^k
   \frac{\Gamma(\gamma_0 k) \Gamma(\gamma_0 (m-k)+1)}{\Gamma(\gamma_0 m + 1)} c_a^{m-k} f^{(k)}_g f^{(m-k)}_a,
\end{align}
where the digamma function $\psi(z)\equiv \Gamma'(z)/\Gamma(z)$. Following ref.~\cite{Dokshitzer:1993dc}, we neglect the $\psi$ terms. Then, starting with $f_a^{(0)} = f_a^{(1)}=1$, one can iteratively obtain, for $m\geq 2$,
\begin{align}
    f^{(m)}_g &= \frac{\gamma_0 m}{m^2 - 1}\sum\limits_{k=1}^{m-1} C_m^k\frac{\Gamma(\gamma_0 k) \Gamma(\gamma_0 (m-k)+1)}{\Gamma(\gamma_0 m + 1)} f^{(k)}_g f^{(m-k)}_g, \nonumber\\
    f^{(m)}_q &= \frac{c_q^{1-m}}{m^2} f^{(m)}_g + \gamma_0 \sum\limits_{k=1}^{m-1} \frac{1}{m}  C_m^k \frac{\Gamma(\gamma_0 k) \Gamma(\gamma_0 (m-k)+1)}{\Gamma(\gamma_0 m + 1)} c_q^{1-k} f^{(k)}_g f^{(m-k)}_q.
\end{align}
The DLA results~\cite{Dokshitzer:1982ia, Dokshitzer:1991wu} are recovered in the limit $\gamma_0 \to 0$.

Given the expressions $f_a^{(m)}$ above, the KNO scaling functions $\Psi_a(x)$ can be presented using an expansion in Laguerre polynomials~\cite{Bassetto:1987fq}, which obey the relations
\begin{align}
\label{eq:Lm}
    L_m(x) = \sum_{k=0}^m C^k_m \frac{(-1)^k}{k!} x^k \qquad \text{and} \qquad \int_0^\infty e^{-x} L_n(x) L_m(x) = \delta_{mn}.
\end{align}
That is, $\Psi_a(x)$ can be approximated as
\begin{align}\label{eq:Psia_Laguerre}
    \Psi_a(x) \approx e^{-x}\sum_{m=0}^{N} \hat{f}_a^{(m)} L_m(x),
\end{align}
where $N$ denotes the truncation order of the expansion and according to the relations in eq.~\eqref{eq:Lm}, the expansion coefficients are given by
\begin{align}
    \hat{f}^{(m)}_a = \int_0^\infty dx L_m(x) \Psi_a(x) = \sum_{k=0}^m C^k_m \frac{(-1)^k}{k!} f^{(m)}_a.
\end{align}
We observe that for $N \geq 30$, the series converges reliably for both quark and gluon jets. All results presented below are evaluated with $N=40$.

\begin{figure}[htbp]
    \centering
    \includegraphics[height=0.24\textheight]{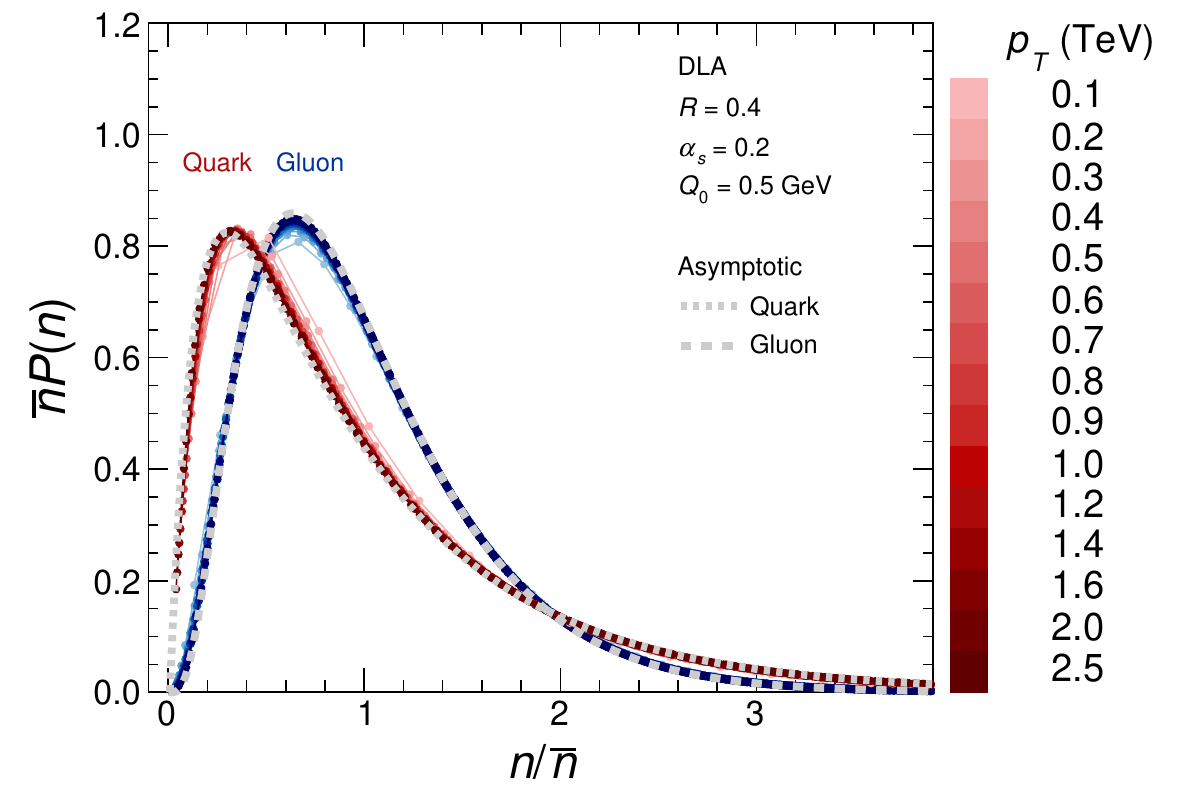}
    \includegraphics[height=0.24\textheight]{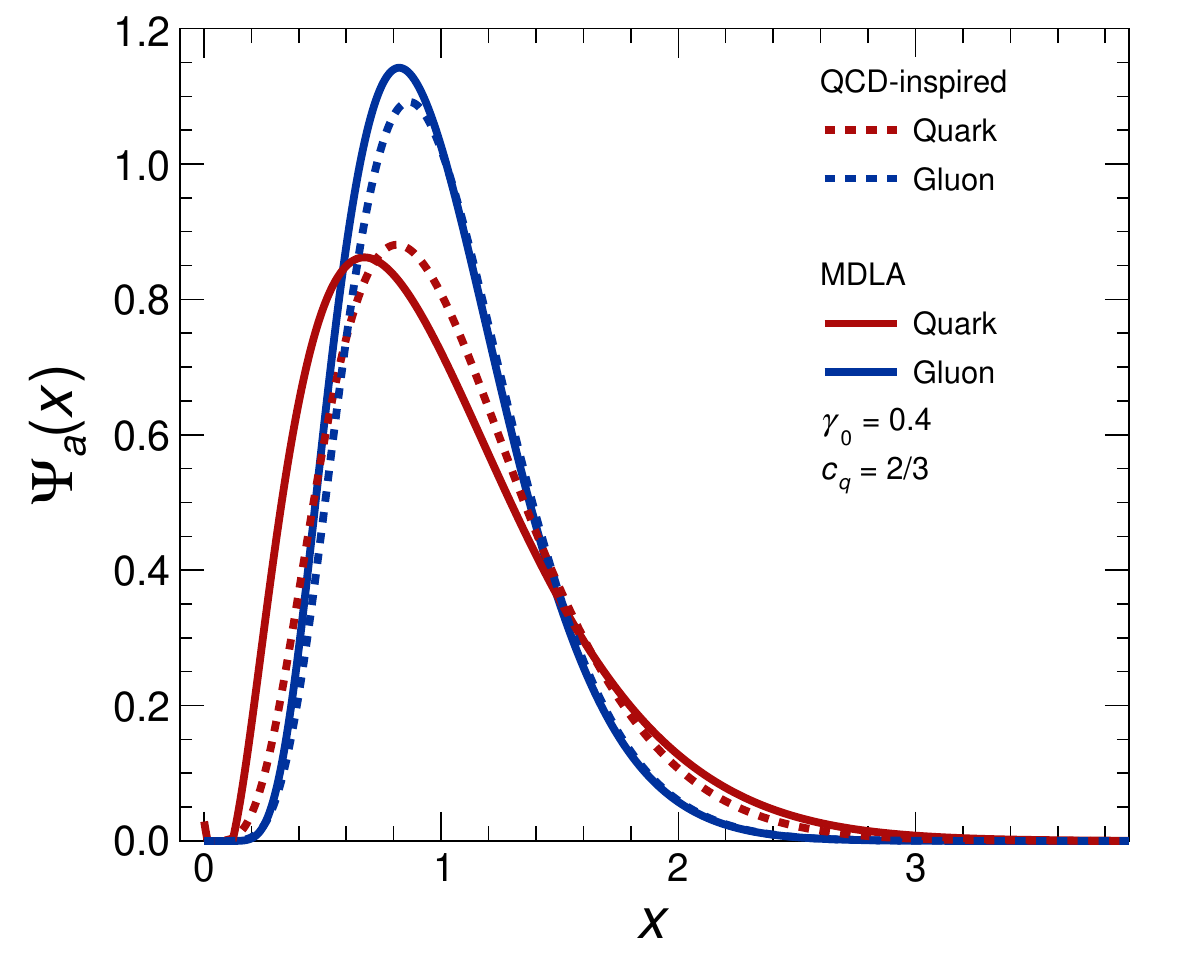}
    \caption{KNO scaling functions for quark and gluon jets. Left: Verification of KNO scaling within DLA by directly evaluating $P_a(n)$ according to eq.~\eqref{eq:Pqg} with $Q_0 = 0.5$ GeV, using a fixed coupling $\alpha_s = 0.2$, over a $p_T$ range from 0.1 to 2.5 TeV (solid), and comparing them to the DLA KNO scaling functions derived in the asymptotic limit $Q\to\infty$ (dashed). Within this $p_T$ range, we observe that the curves of $\bar{n}_a P_a(n)$ for different $p_T$ still show some dependence on the value of $\alpha_s$. With the chosen value, they all converge to the DLA KNO scaling functions. Right: Comparison of the MDLA KNO scaling functions evaluated from eq.~(\ref{eq:nmaMDLA}) (solid) with the QCD-inspired forms in ref.~\cite{Dokshitzer:2025fky} (dashed) for $\gamma_0 = 0.4=\gamma$ and $c_q=2/3$.}
    \label{fig:KNO_ADLA_MDLA}
\end{figure}

The left panel of figure~\ref{fig:KNO_ADLA_MDLA} verifies the validity of KNO scaling in DLA with a fixed $\alpha_s$, which is assumed in the preceding MDLA derivation. This is verified by directly evaluating $P_a(n)$, as given in eq.~\eqref{eq:Pqg} with $Q_0 = 0.5$ GeV, over a broad $p_T$ range from 0.1 to 2.5 TeV. With running coupling, the resulting curves of $\bar{n}_a P_a(n)$ all converge to some universal functions for quark and gluon jets, respectively, though with slight deviations from the asymptotic DLA scaling functions~\cite{Duan:2025ngi}. With fixed coupling, we find that $\bar{n}_a P_a(n)$ shows some dependence on the value of $\alpha_s$ and the results with $\alpha_s = 0.2$ converge to the DLA scaling functions, as shown in this plot. Here, the asymptotic KNO functions within DLA are given by eq.~(\ref{eq:Psia_Laguerre}) with $f_a^{(n)}$ evaluated in the limit $\gamma_0\to 0$, which agree with the results of refs.~\citep{Bassetto:1987fq, Dokshitzer:1993dc}.

The right panel of figure~\ref{fig:KNO_ADLA_MDLA} compares the MDLA scaling functions evaluated above with the QCD-inspired expressions recently proposed in ref.~\cite{Dokshitzer:2025fky}. The comparison is shown for $\gamma_0 = 0.4$~\cite{Dokshitzer:1993dc} (with $\gamma = 0.4$ in the results of ref.\cite{Dokshitzer:2025fky}) and $c_q = 2/3$~\cite{Dokshitzer:2025fky}. For both quark and gluon jets, including energy conservation significantly shifts the peak of the scaling functions toward larger values of $x$. These significant discrepancies highlight the need to go beyond DLA to obtain reliable KNO scaling functions. Our results, obtained using Laguerre polynomial expansions, agree with the expression for gluon jets given in ref.~\cite{Dokshitzer:1993dc}, except in the $x \ll 1$ region, where our method may yield negative values, like the DLA results in ref.~\cite{Bassetto:1987fq}. Hence, for the remainder of the paper, we adopt the analytic form from ref.~\cite{Dokshitzer:1993dc} for gluon jets. In comparison with recently suggested forms in ref.~\cite{Dokshitzer:2025fky}, our results for gluon jets are slightly different from that at $x\lesssim1$, starting around the peak region, while the differences for quark jets are more pronounced. We find that this difference can be traced back to the fact that the coefficients $f_q^{(m)}$, calculated iteratively above, no longer satisfy the relation within DLA: $\Phi_q(\beta)=[\Phi_g(\beta/c_q)]^{c_q}$, which is used to obtain the quark-jet KNO functions in refs.~\cite{Dokshitzer:2025owq, Dokshitzer:2025fky}.

\subsection{Comparison with KNO scaling functions using {\tt PYTHIA}}
\label{ssec:KNOPYTHIA}

As shown by the results above, while the mean parton multiplicity and the parton multiplicity distribution functions exhibit a logarithmic dependence on $Q_0$, the asymptotic KNO scaling functions remain infrared safe. Consequently, one may not expect logarithmically enhanced corrections when transitioning from the parton level to the hadron level. In this subsection, we compare the above results at the parton level with those for charged particles extracted in \texttt{PYTHIA} simulations~\cite{Bierlich:2022pfr}, which exhibit KNO scaling to a very reasonable level in both quark and gluon jets~\cite{Duan:2025ngi}.

In our \texttt{PYTHIA} simulations of $pp$ collisions~\cite{Duan:2025ngi}, the event selection procedure follows the methodology in ref.~\cite{ATLAS:2019rqw}. Particles satisfying $p_T > 0.5$ GeV and $|\eta| < 2.5$ are used to reconstruct the two leading jets via the anti-$k_t$ algorithm~\cite{Cacciari:2008gp} with a radius parameter $R = 0.4$, implemented in the \texttt{FastJet} package~\cite{Cacciari:2011ma}. The jets are required to satisfy $|\eta| < 2.1$, and the leading and subleading jets must have a transverse momentum ratio $p_T^{\text{lead}}/p_T^{\text{sublead}} < 1.5$. The \texttt{PYTHIA} samples are generated using the A14 tune, which has been shown to provide a good description for charged-particle multiplicity distribution observed by ATLAS~\cite{ATLAS:2019rqw}. In our analysis, the charged particles from jet constituents are used to define the multiplicity distributions, while jet $p_T$ is reconstructed from all particles inside the jet. In addition, quark and gluon jets are identified based on their spatial proximity to the initiating parton from the underlying $2\rightarrow2$ hard scattering event.

\begin{figure}[htbp]
    \centering
    \includegraphics[height=0.24\textheight]{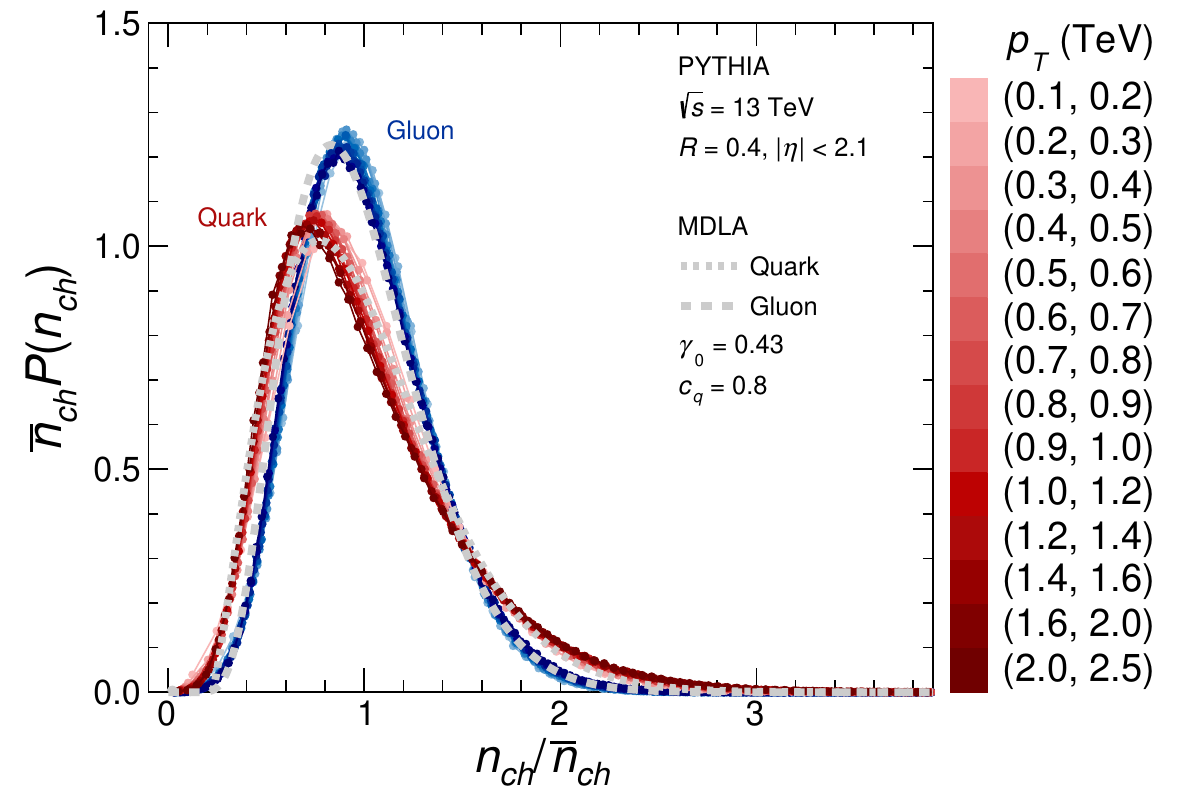}
    \includegraphics[height=0.24\textheight]{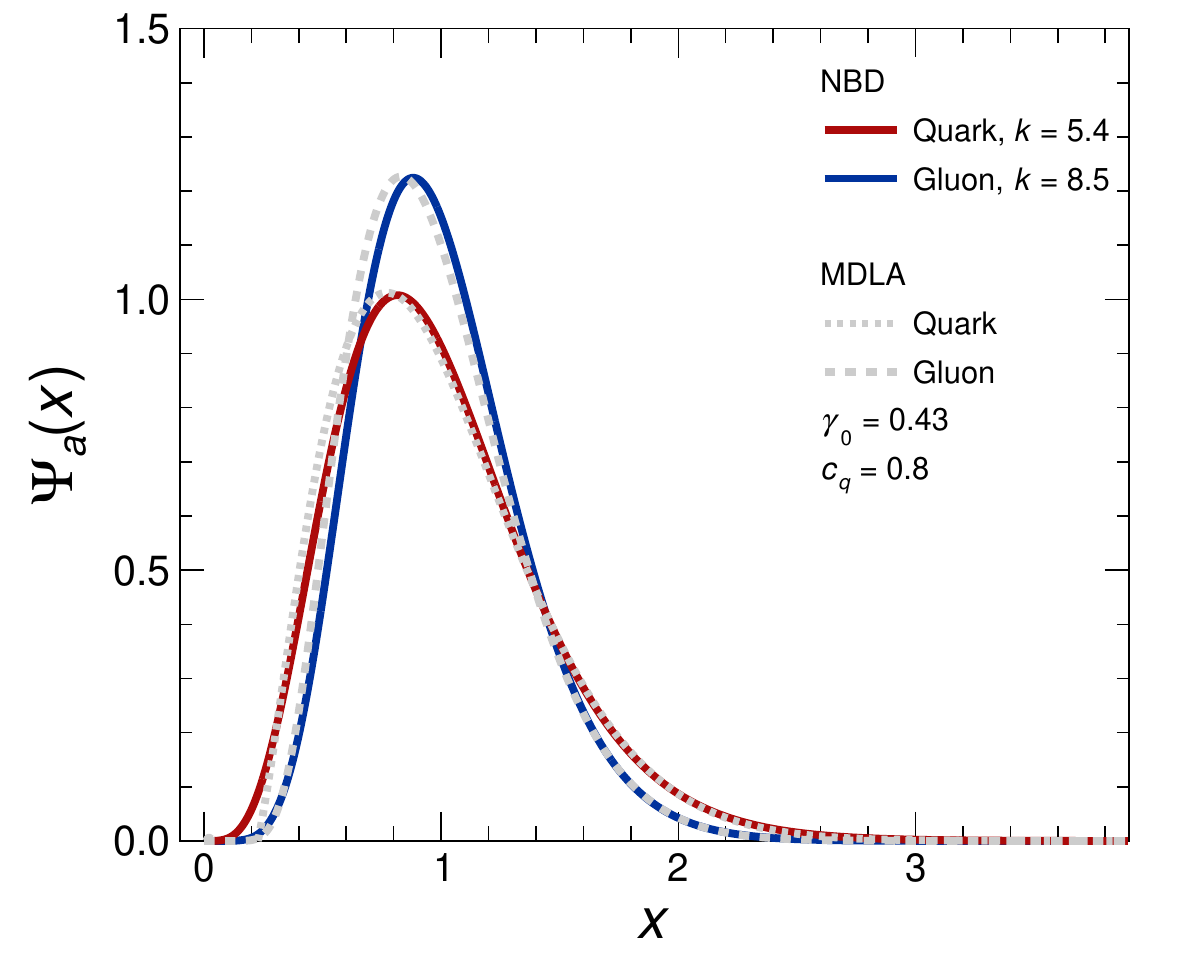}
    \caption{KNO scaling functions for quark and gluon jets in \texttt{PYTHIA} and MDLA. Left: Charged-particle multiplicity distributions for quark and gluon jets in \texttt{PYTHIA}~\cite{Duan:2025ngi} (solid) versus MDLA results (dashed) with $\gamma_0 = 0.43$ and $c_q = 0.8$. Right: Negative binomial distributions (NBD, solid), with $k = 5.4$ for quark jets and $k = 8.5$ for gluon jets (see eq.~\eqref{eq:NBD}) compared to the same MDLA results (dashed).}
    \label{fig:KNO_MDLA_PYTHIA_QG}
\end{figure}

The left panel of  figure~\ref{fig:KNO_MDLA_PYTHIA_QG} compares our MDLA results with {\tt PYTHIA} simulations that include both MPI and hadronization, facilitating comparison with ATLAS data in the next sections. We find that incorporating energy conservation indeed improves significantly the agreement between the KNO scaling functions and the \texttt{PYTHIA} results for both quark and gluon jets. Inspired by refs.~\cite{Dokshitzer:2025owq, Dokshitzer:2025fky}, we take $c_q$ as a fitting parameter rather than fixing it to the DLA value $C_F/C_A$. With $\gamma_0 = 0.43$ and $c_q = 0.8$, the MDLA predictions achieve good agreement not only with \texttt{PYTHIA} simulations but also, as will be shown later, with the ATLAS measurements reported in ref.~\cite{ATLAS:2019rqw}. In details, we first determine $\gamma_0$ by fitting the \texttt{PYTHIA} results over the entire $p_T$ range for gluon jets. Keeping this value of $\gamma_0$ fixed, we then extract $c_q$ by fitting the \texttt{PYTHIA} results over the entire $p_T$ range for quark jets. The fitted $c_q$ turns out larger than the chosen value of $2/3$ in ref.~\cite{Dokshitzer:2025fky}, reflecting quantitative differences between the two results.

Note that setting the momentum scale in $\alpha_s$ to $Q=Rp_T$, using {\tt LHAPDF}~\cite{Buckley:2014ana}, we obtain $\gamma_0$ values in the range 0.41–0.51 across the plotted $p_T$ interval.\footnote{In refs.~\cite{Dokshitzer:2025owq, Dokshitzer:2025fky}, the two-loop anomalous dimension $\gamma(\alpha_s)$ is employed in place of $\gamma_0$. It is defined as
$\gamma(\alpha_s) = \gamma_0 - \left[{\beta_0}/{4} + {10 n_f}/{(3 N_c^2)}\right]{\alpha_s}/{(2\pi)}$, which yields values in the range 0.36–0.43 for $n_f=5$. 
} Our choice of $\gamma_0=0.43$ indeed lies in this range although allowing for this running produces a broader spread than is seen in the \texttt{PYTHIA} results. Since the MDLA scaling functions with $\gamma_0=0.43$ and $c_q=0.8$ can very well describe the experimental data in ref.~\cite{ATLAS:2019rqw}, we proceed with this phenomenological approach and defer a more refined analysis beyond DLA to future work.

Similar to the DLA results (see, e.g., ref.~\cite{Bassetto:1987fq}), we find that the MDLA KNO functions are well approximated by negative binomial distributions (NBDs) of the form
\begin{align}\label{eq:NBD}
    \Psi^{NB}(x) = x^{k-1} e^{-k x} \frac{k^k}{\Gamma(k)},
\end{align}
where $k$ serves as a fit parameter (see also ref.~\cite{Dokshitzer:1993dc}). The right panel of figure~\ref{fig:KNO_MDLA_PYTHIA_QG} demonstrates this agreement, where the MDLA results are reproduced using $k = 5.4$ for quark jets and $k = 8.5$ for gluon jets, exhibiting only minor deviations.

\section{Inclusive multiplicity distributions in $pp$ collisions at the LHC}
\label{sec:incjet}

In $pp$ collisions, distinguishing quark and gluon jets to study their multiplicity distributions is a nontrivial challenge~\cite{Marzani:2019hun, Larkoski:2024uoc}, which we defer to the next section. As an alternative, the validity of KNO scaling can be inferred through inclusive multiplicity measurements of QCD jets~\cite{Germano:2024ier, Duan:2025ngi, Dokshitzer:2025owq}. In this section, we compare our theoretical calculations, based on the MDLA scaling functions evaluated above, with ATLAS measurements of the inclusive multiplicity distributions of the two leading jets at $\sqrt{s} = 13$ TeV~\cite{ATLAS:2019rqw}.

\subsection{Theoretical calculations}

Inclusive multiplicity distributions inside jets in $pp$ collisions have been measured at the LHC~\cite{ATLAS:2011eid, ATLAS:2019rqw}. These inclusive distributions can be represented as a combination of contributions from quark and gluon jets~\cite{Duan:2025ngi, Dokshitzer:2025owq}:
\begin{align}\label{eq:Pall}
    P(n) &= r_q P_q(n) + r_g P_g(n), \nonumber\\
    \bar{n} &= r_q \bar{n}_q + r_g \bar{n}_g,
\end{align}
where the coefficient
\begin{align}\label{eq:ratioqg}
    r_a \equiv \frac{d\sigma_a / dp_T}{d\sigma_q / dp_T + d\sigma_g / dp_T}
\end{align}
denotes the fraction of jets initiated by a parton of type $a$ at a given jet transverse momentum $p_T$, and $d\sigma_a / dp_T$ represents the corresponding differential cross section.

To establish a theoretical baseline, we begin with the leading-order (LO) differential cross section in $pp$ collisions, which is factorized as a convolution of parton distribution functions (PDFs) with the partonic hard-scattering cross sections, summed over all contributing subprocesses $ab \rightarrow cd$:
\begin{align}\label{eq:sigma}
    \frac{d\sigma}{dp_{T}} = 2 p_{T} \sum_{a,b,c,d} \int dy_c dy_d x_a f_{a/p}(x_a,\mu^2) x_b f_{b/p}(x_b,\mu^2) \frac{d\hat{\sigma}_{ab \rightarrow cd}}{d\hat{t}},
\end{align}
where $y_c$ and $y_d$ are the rapidities of the final-state partons $c$ and $d$ that initiate the jets, and the longitudinal momentum fractions $x_a$ and $x_b$ of the incoming partons are determined by
\begin{align}
    x_a = \frac{1}{2} x_{T} (e^{y_c} + e^{y_d}),
    \qquad
    x_b = \frac{1}{2} x_{T} (e^{-y_c} + e^{-y_d})
\end{align}
with $x_{T} \equiv 2p_{T} / \sqrt{s}$. The PDFs $f_{a/p}(x_a, \mu^2)$ and $f_{b/p}(x_b, \mu^2)$ are evaluated at the factorization scale $\mu^2 = p_{T}^2$. The partonic cross sections, denoted as $d\hat{\sigma}_{ab \rightarrow cd} / d\hat{t}$ and found in the standard textbooks such as ref.~\cite{Peskin:1995ev}, are expressed in terms of the Mandelstam variables:
\begin{align}
    \hat{s} = x_a x_b s,\qquad
    \hat{t} = -x_a p_{T} \sqrt{s} e^{-y_c},\qquad
    \hat{u} = -x_b p_{T} \sqrt{s} e^{y_c}.
\end{align}

In our analysis, we explicitly separate the contributions from quark and gluon jet channels by examining each partonic subprocess $ab \rightarrow cd$ and evaluating their corresponding partonic cross sections. This enables us to compute the differential cross sections $d\sigma_q / dp_T$ for quark jets and $d\sigma_g / dp_T$ for gluon jets, which in turn allows us to extract the quark and gluon fractions $r_q$ and $r_g$, as defined in eq.~\eqref{eq:ratioqg}. For multiplicity distributions, we utilize the MDLA results as presented in the previous section. Based on the theoretical framework outlined above, we investigate the mean multiplicities and multiplicity probability distributions within jets across a broad jet transverse momentum range of $p_T = 0.1$--$2.5$ TeV in $pp$ collisions at $\sqrt{s} = 13$ TeV. Our theoretical predictions are to be compared with results from \texttt{PYTHIA} simulations and experimental measurements from ATLAS~\cite{ATLAS:2019rqw}.

\subsection{Mean charged-particle multiplicity}

\begin{figure}[htbp]
    \centering
    \includegraphics[height=0.25\textheight]{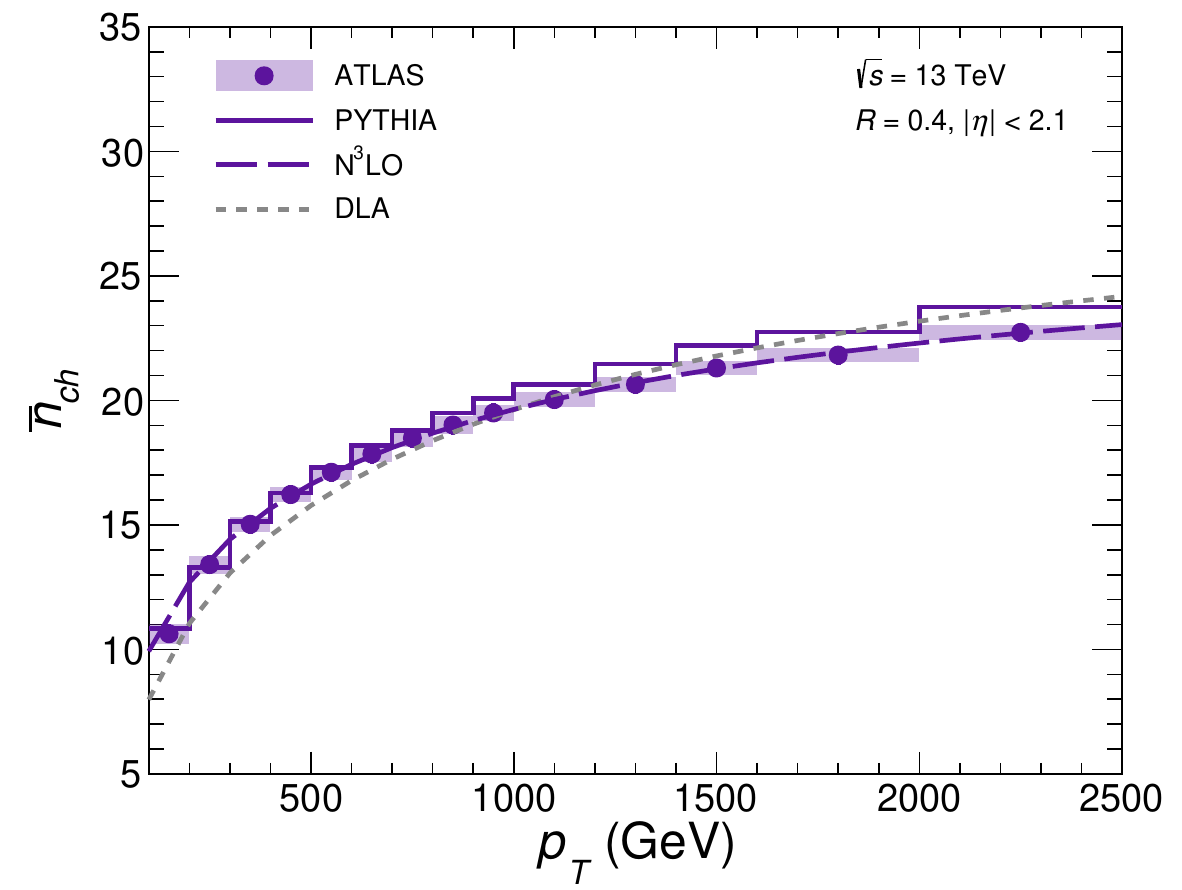}
    \includegraphics[height=0.25\textheight]{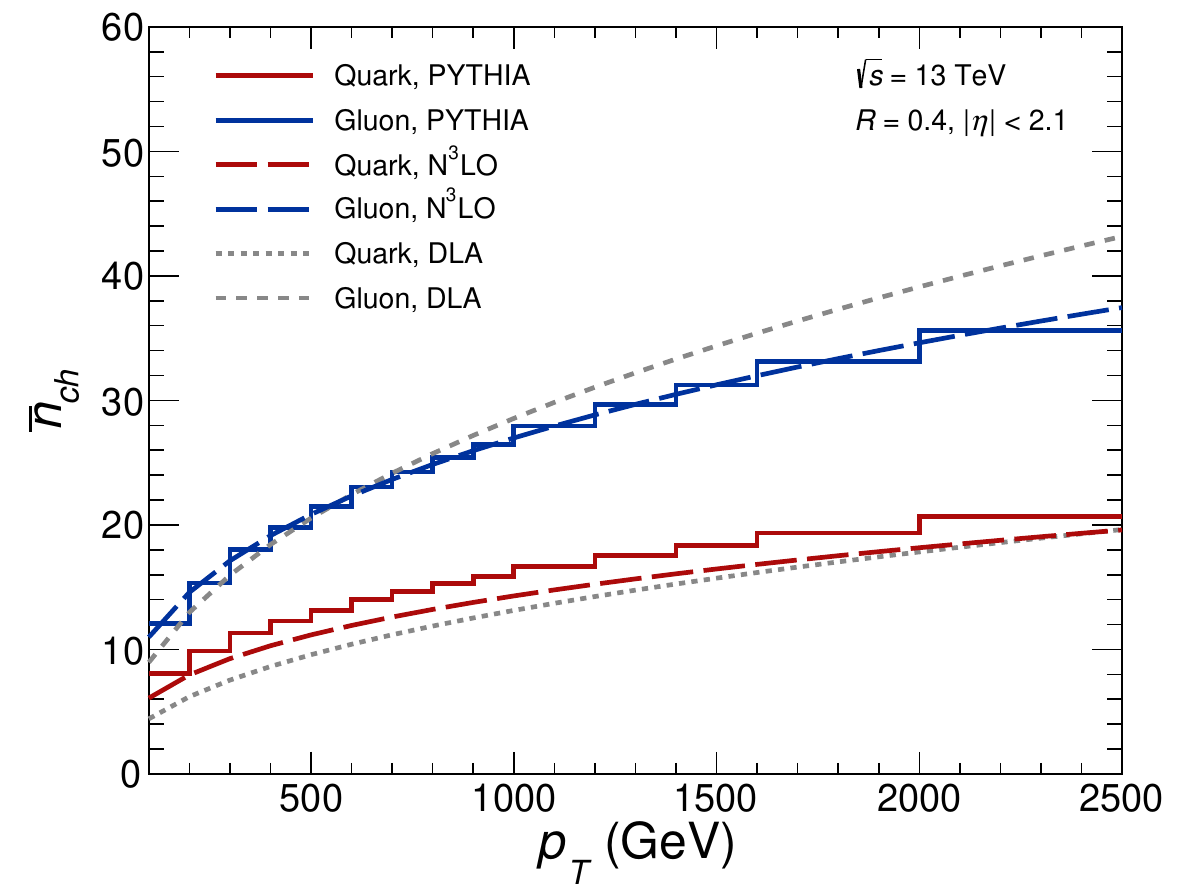}
    \caption{Mean charged-particle multiplicity within QCD jets at the LHC. Left: Inclusive mean charged-particle multiplicity as a function of jet transverse momentum $p_T$, comparing ATLAS data~\cite{ATLAS:2019rqw}, \texttt{PYTHIA} simulations, DLA, and N$^3$LO results.
    Right: Mean charged-particle multiplicity in quark and gluon jets from \texttt{PYTHIA}, DLA, and N$^3$LO predictions. For the DLA results, $K_{\text{LPHD}}=0.80$ is obtained by fitting the inclusive mean multiplicity data shown in the left panel with $Q_0=0.5$~GeV held fixed, while for the N$^3$LO results $K=0.0353$ and $Q_0=0.053$~GeV are both determined from fits to the same data.}
    \label{fig:mean}
\end{figure}

In this subsection, we briefly summarize the results of mean charged-particle multiplicities used in our studies of inclusive multiplicity distributions, as given in eq.~(\ref{eq:Pall}). Specifically, we employ the DLA results with running coupling, given in eqs.~\eqref{eq:nbarQG} and \eqref{eq:mean2}, as well as the next-to-next-to-next-to-leading-order (N$^3$LO) results from refs.~\cite{Dremin:1999ji,Capella:1999ms}, which are summarized in appendix~\ref{app:nb_N3LO} (see also ref.~\cite{Dremin:2000ep} for a review). To facilitate a meaningful comparison between parton-level theoretical predictions and experimental charged-particle data in ref.~\cite{ATLAS:2019rqw}, we make use of the Local Parton–Hadron Duality (LPHD) hypothesis~\cite{Azimov:1984np,Dokshitzer:1995ev}. Within this approach, we introduce a normalization factor $K_{\text{LPHD}}$ for the DLA results, defined as the ratio of the mean charged-particle multiplicity to the mean parton multiplicity. The corresponding normalization parameter in the N$^3$LO results is denoted by $K$. 

To confront the theoretical calculations for the inclusive mean charged-particle multiplicity with the measurements in ref.~\cite{ATLAS:2019rqw}, we first calculate the fractions of quark and gluon jets at LO using eqs.~\eqref{eq:ratioqg} and~\eqref{eq:sigma}. Here, we consider two leading jets with $|\eta| < 2.1$ over the $p_T$ range $0.1$--$2.5$ TeV in $pp$ collisions at $\sqrt{s} = 13$ TeV. The PDFs are taken from the CT18NLO via \texttt{LHAPDF}~\cite{Buckley:2014ana}. Using these extracted jet fractions, we then fix the parameters in the theoretical calculations by fitting to the experimental data for the inclusive mean charged-particle multiplicity in jets with radius $R = 0.4$.

The left panel of figure~\ref{fig:mean} compares the theoretical results of the inclusive mean multiplicity with ATLAS data and \texttt{PYTHIA} simulations. The DLA mean multiplicities depend on two parameters, $Q_0$ and $K_{\text{LPHD}}$. We fix $Q_0=0.5$~GeV, to be consistent with the discussions in sec.~\ref{sec:MDLA} and ref.~\cite{Duan:2025ngi}, and determine $K_{\text{LPHD}}$ through a $\chi^2$ fit to the ATLAS data~\cite{ATLAS:2019rqw}, obtaining $K_{\text{LPHD}}=0.80$ with $\chi^2/\text{d.o.f.}=23.1/13=1.78$.\footnote{
We have also tried a $\chi^2$ fit to the experimental data to determine both parameters. However, this yields a value of $Q_0$ very close to $\Lambda$, due to the infrared divergence in the limit $Q_0 \to \Lambda$ (see eq.~\eqref{eq:mean2}), without improving the $\chi^2$. In such fits, the global minimum of $\chi^2$ does exist, but it occurs at a highly perturbative value of $Q_0 > 20$~GeV.
}
For the N$^3$LO results, the two parameters $Q_0$ and $K$ are both determined via fits to the ATLAS data, yielding $K=0.0353$ and $Q_0=0.053$~GeV with $\chi^2/\text{d.o.f.}=0.904/12=0.075$. As shown in the plot, the N$^3$LO predictions provide an impressively accurate description of the data, while the DLA results remain reasonable but tend to underestimate $\bar{n}_{ch}$ at low $p_T$ and overestimate it at high $p_T$. The \texttt{PYTHIA} simulations, with MPI and hadronization included, also reproduce the measurements well, aside from a mild overshoot at high $p_T$.

\begin{figure}[htbp]
    \centering
    \includegraphics[height=0.25\textheight]{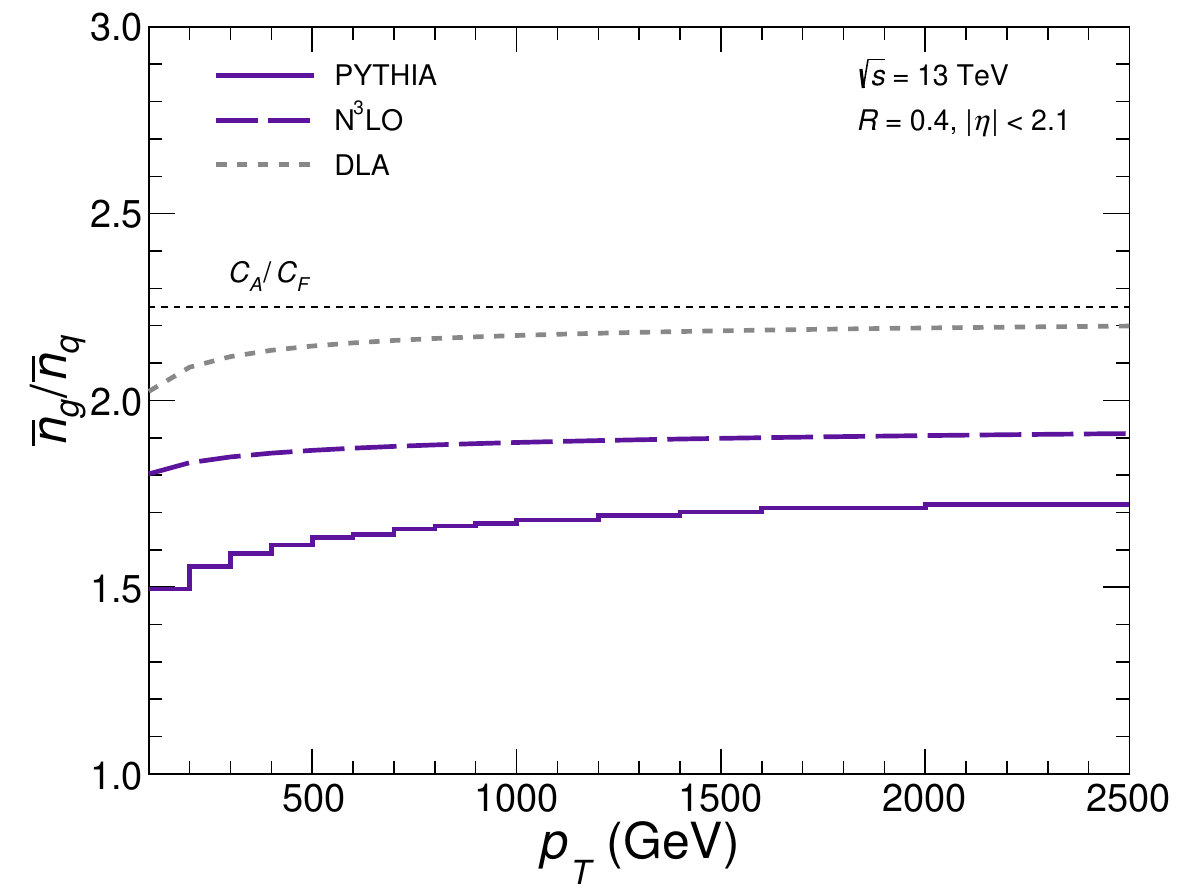}
    \caption{Ratio of mean multiplicities between gluon and quark jets. In the case of DLA, this ratio is not constant at the expected value of $C_A/C_F = 9/4$, indicated by the horizontal dashed line, because the leading partons are included in the mean multiplicity given in eq.~(\ref{eq:nbarQG}).
    }
    \label{fig:ratioQG}
\end{figure}

With the parameters fixed as above, the right panel of figure~\ref{fig:mean} compares the mean charged-particle multiplicities in quark and gluon jets from DLA, N$^3$LO, and \texttt{PYTHIA}. Overall, in comparison with the DLA curves, the N$^3$LO results get closer to the \texttt{PYTHIA} results (the corresponding experimental results are discussed in the next section). Figure~\ref{fig:ratioQG} shows the corresponding ratio of mean multiplicities in gluon and quark jets. Consistent with the behavior observed in the right panel of figure~\ref{fig:mean}, the N$^3$LO ratio lies noticeably below the DLA prediction. In the next subsection, we employ both the DLA and N$^3$LO mean multiplicities to investigate the sensitivity of inclusive multiplicity distributions to these mean multiplicities.

\subsection{Inclusive multiplicity distributions}

In this subsection, we present a detailed investigation of inclusive multiplicity distributions in two leading jets produced in $pp$ collisions at $\sqrt{s} = 13$ TeV. Based on eqs.~\eqref{eq:Pall} and \eqref{eq:ratioqg}, and on the assumption that KNO scaling holds separately for quark and gluon jets, the inclusive multiplicity distributions in KNO form are expressed as
\begin{align}\label{eq:KNO_inclusive}
    \bar{n}P(n)
    &= \bar{n}[r_qP_q(n) + r_gP_g(n)]\notag\\
    &= r_q\frac{\bar{n}}{\bar{n}_q}\Psi_q(x{\bar{n}}/{\bar{n}_q}) + r_g\frac{\bar{n}}{\bar{n}_g}\Psi_g(x{\bar{n}}/{\bar{n}_g}),
\end{align}
where we have used the relation $n = x\bar{n} = x_q\bar{n}_q = x_g\bar{n}_g$. Here, the fractions $r_q$ and $r_g$ are determined from the LO cross section, as in the mean multiplicity calculations. For the KNO scaling functions $\Psi_a$, we adopt the MDLA results with $\gamma_0 = 0.43$ and $c_q = 0.8$, as described in sec.~\ref{sec:MDLA}, since the DLA predictions clearly fail to describe the measurements reported in ref.~\cite{ATLAS:2019rqw}. The remaining relevant quantity is the ratio $\bar{n}_g/\bar{n}_q$ appearing in $\bar{n}/\bar{n}_a$.

\begin{figure}[htbp]
    \centering
    \includegraphics[height=0.23\textheight]{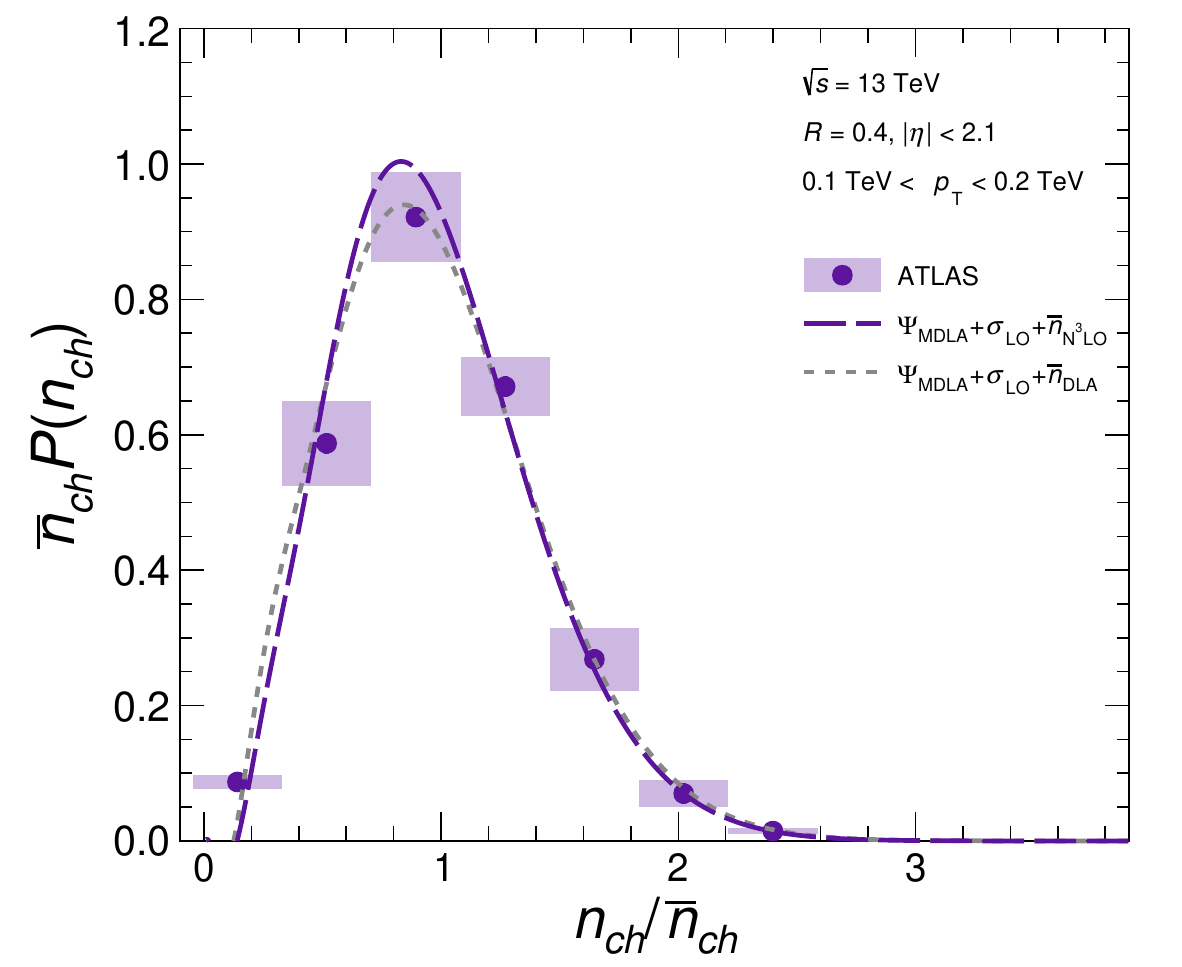}
    \includegraphics[height=0.23\textheight]{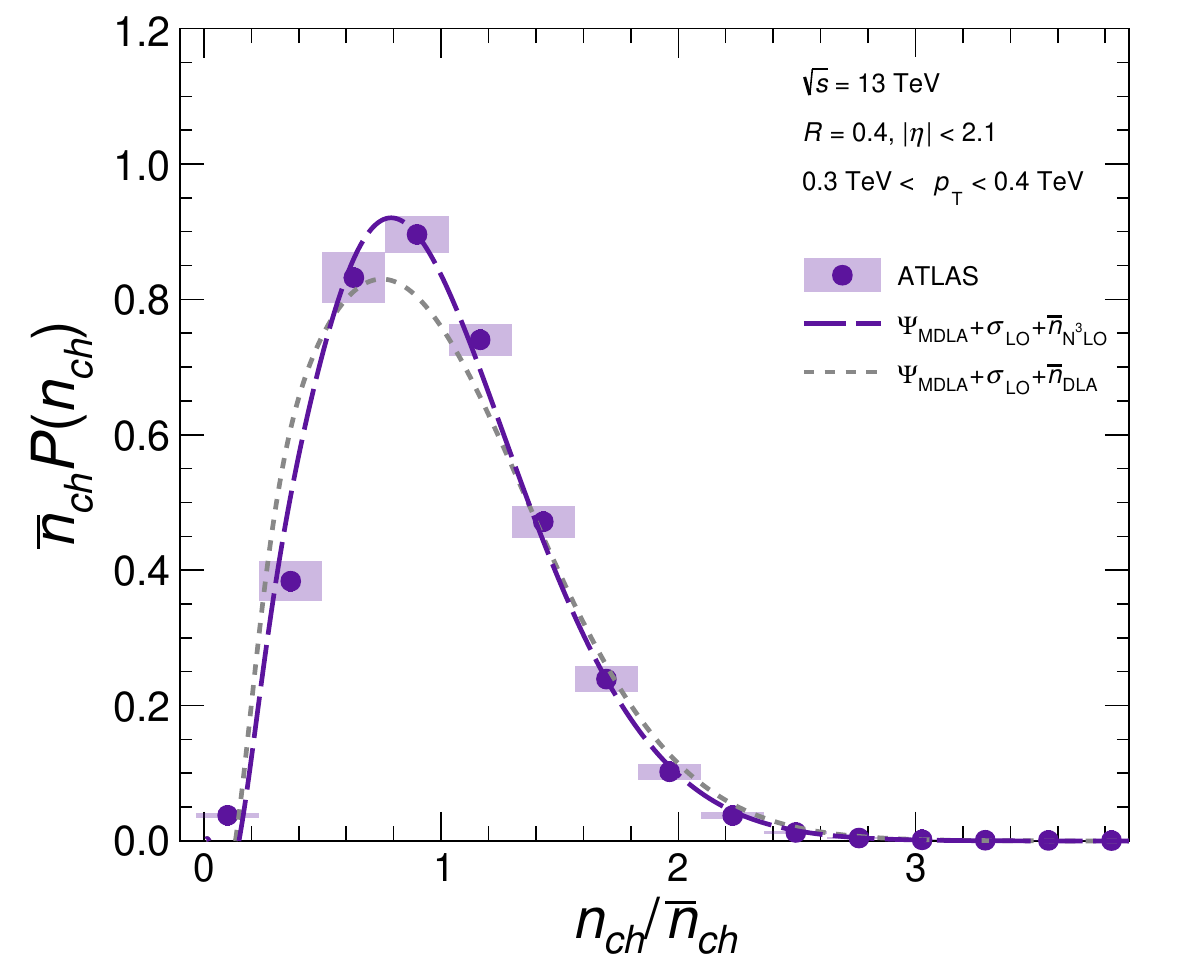} \\
    \includegraphics[height=0.23\textheight]{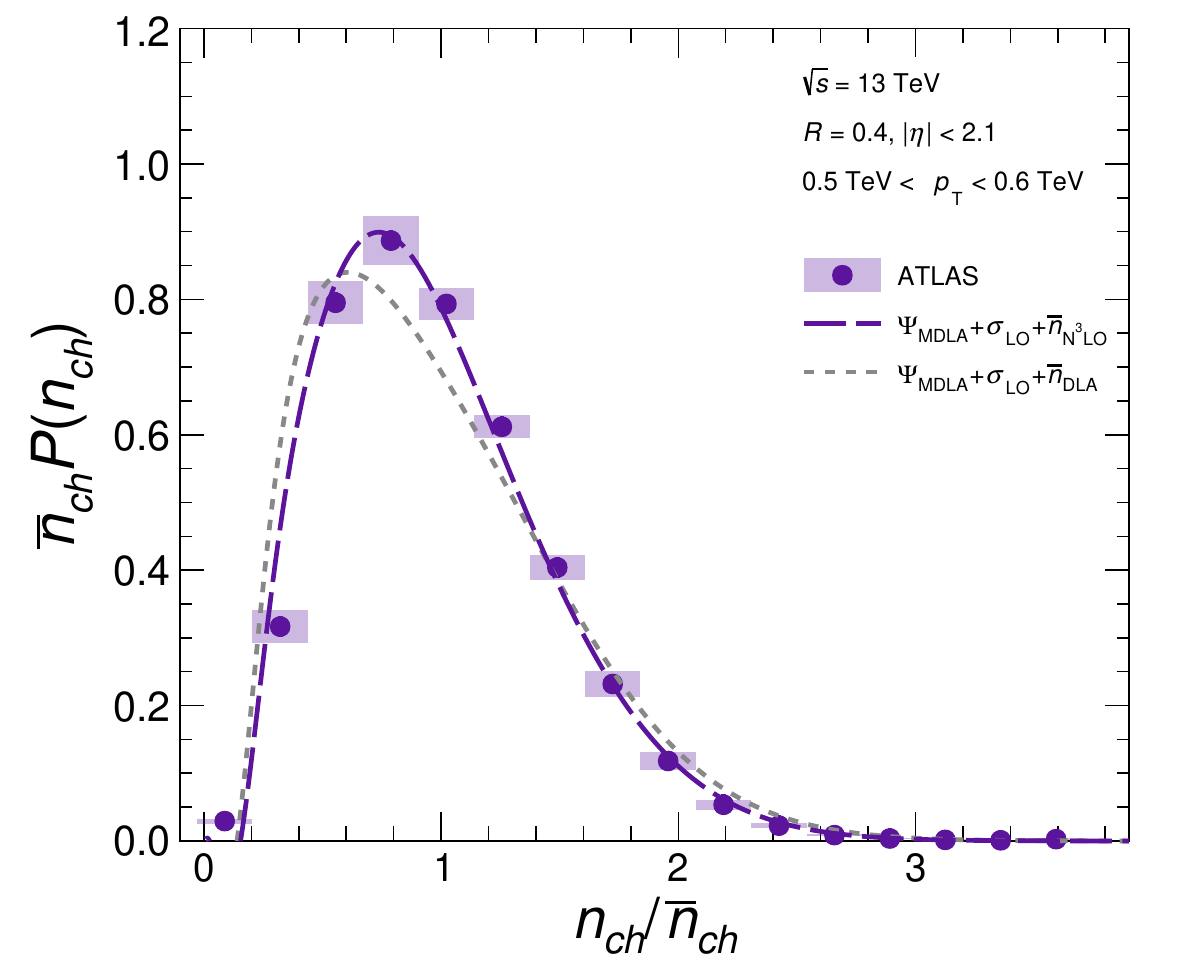}
    \includegraphics[height=0.23\textheight]{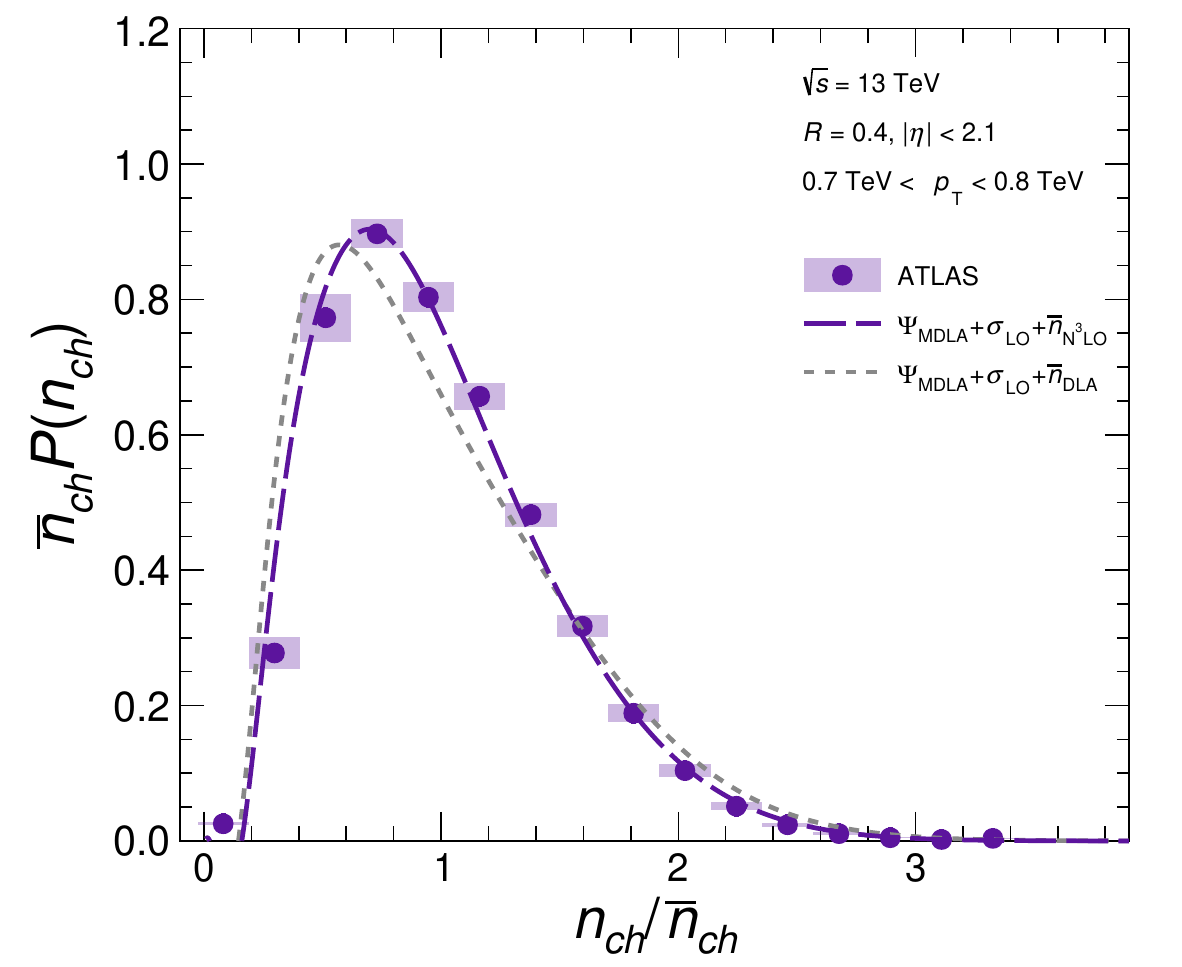} \\
    \includegraphics[height=0.23\textheight]{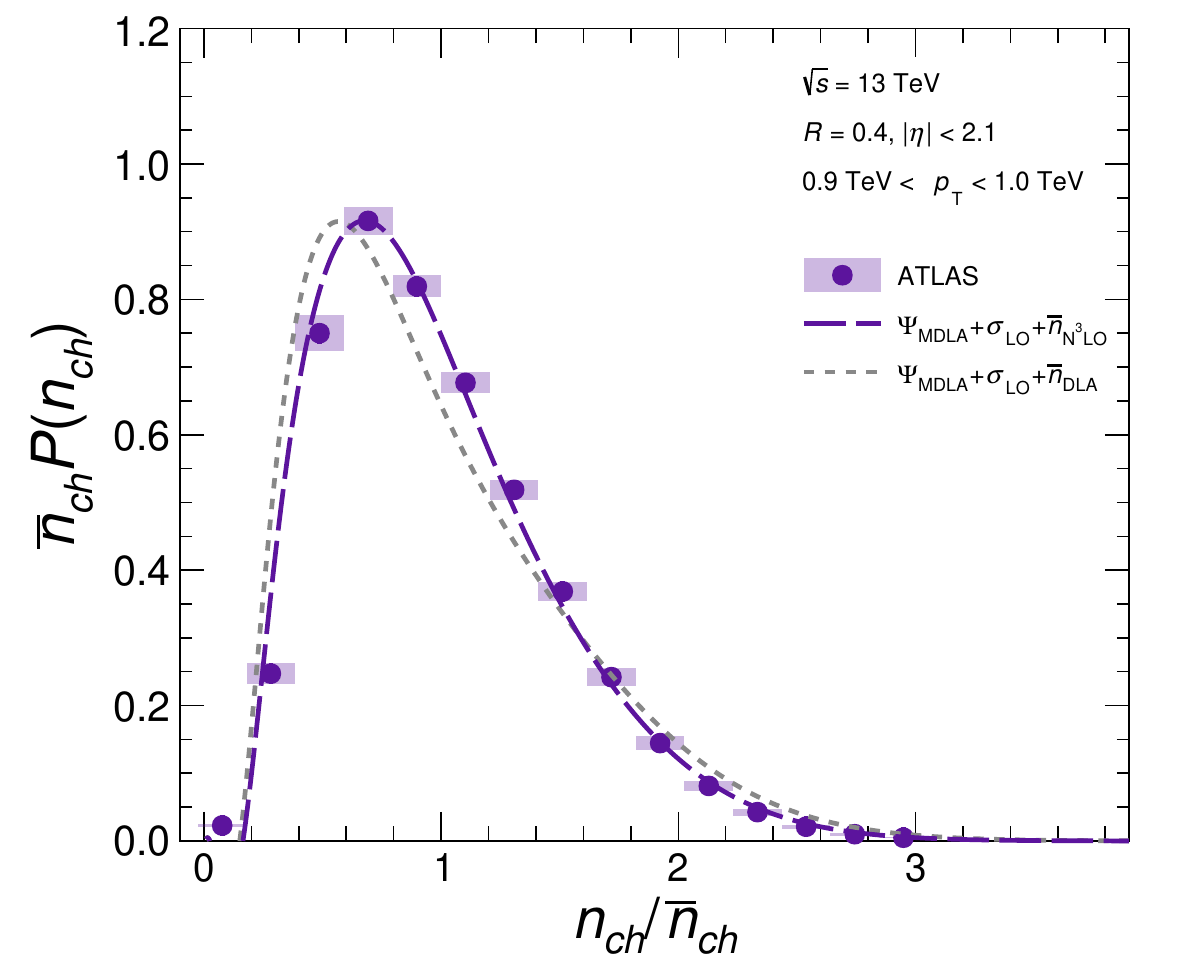}
    \includegraphics[height=0.23\textheight]{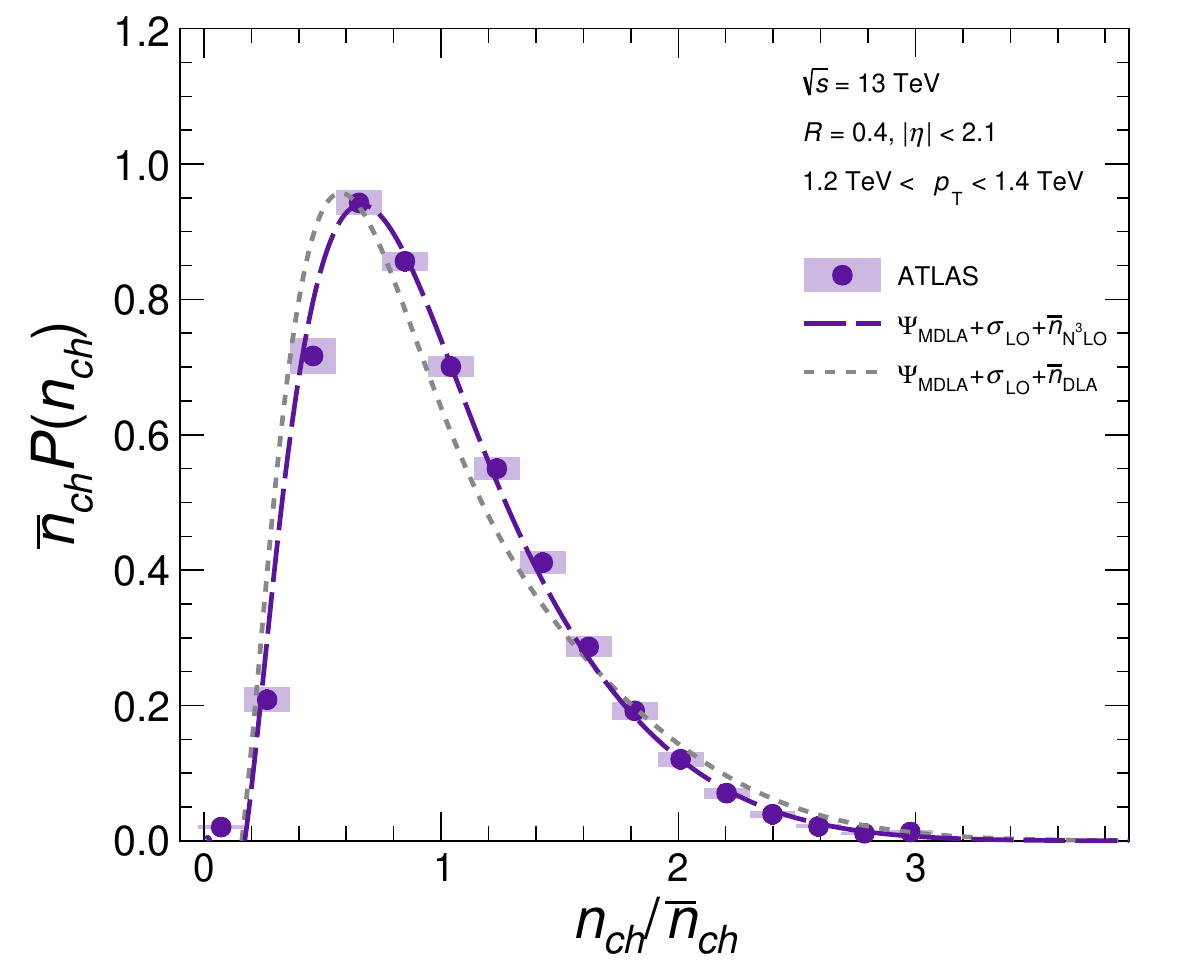} \\
    \includegraphics[height=0.23\textheight]{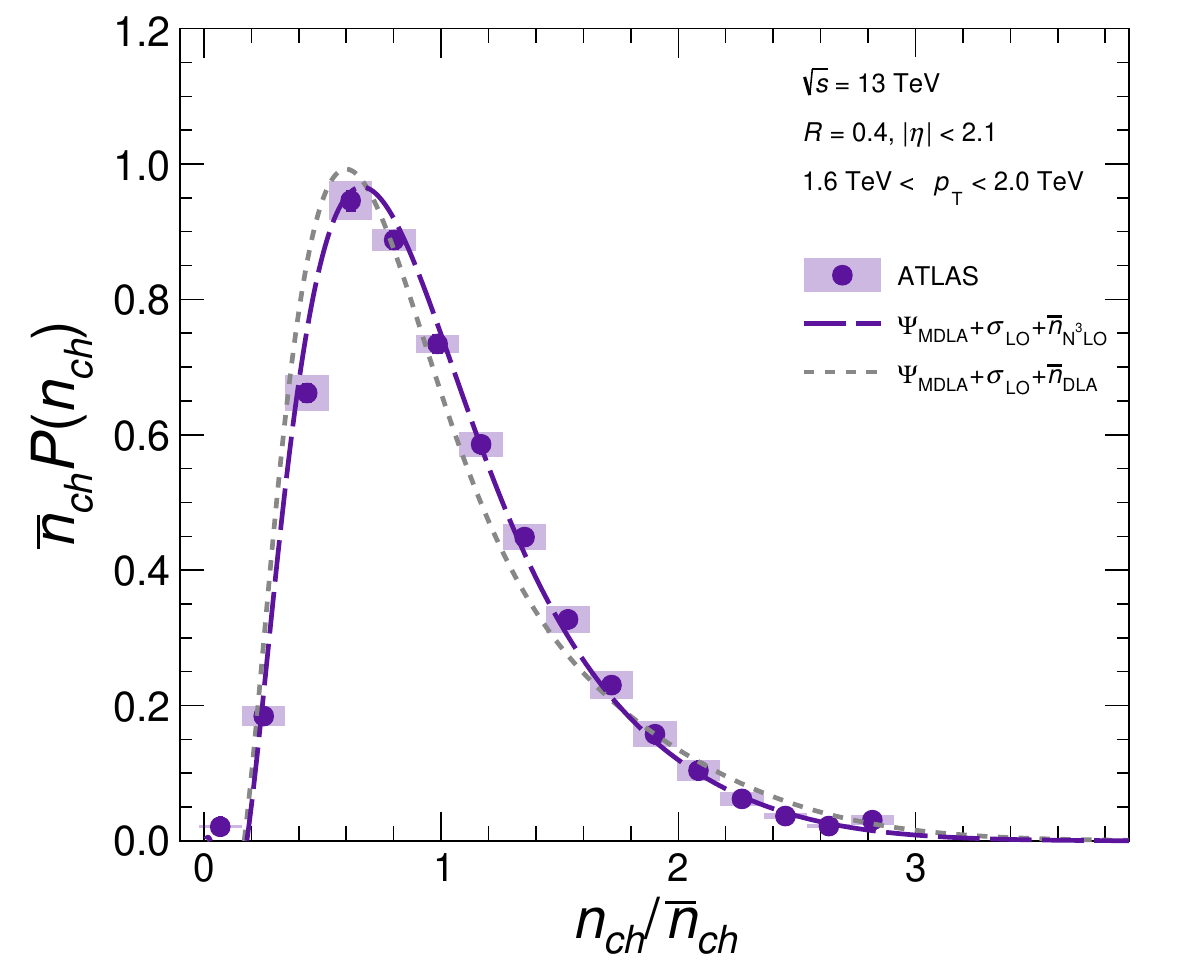}
    \includegraphics[height=0.23\textheight]{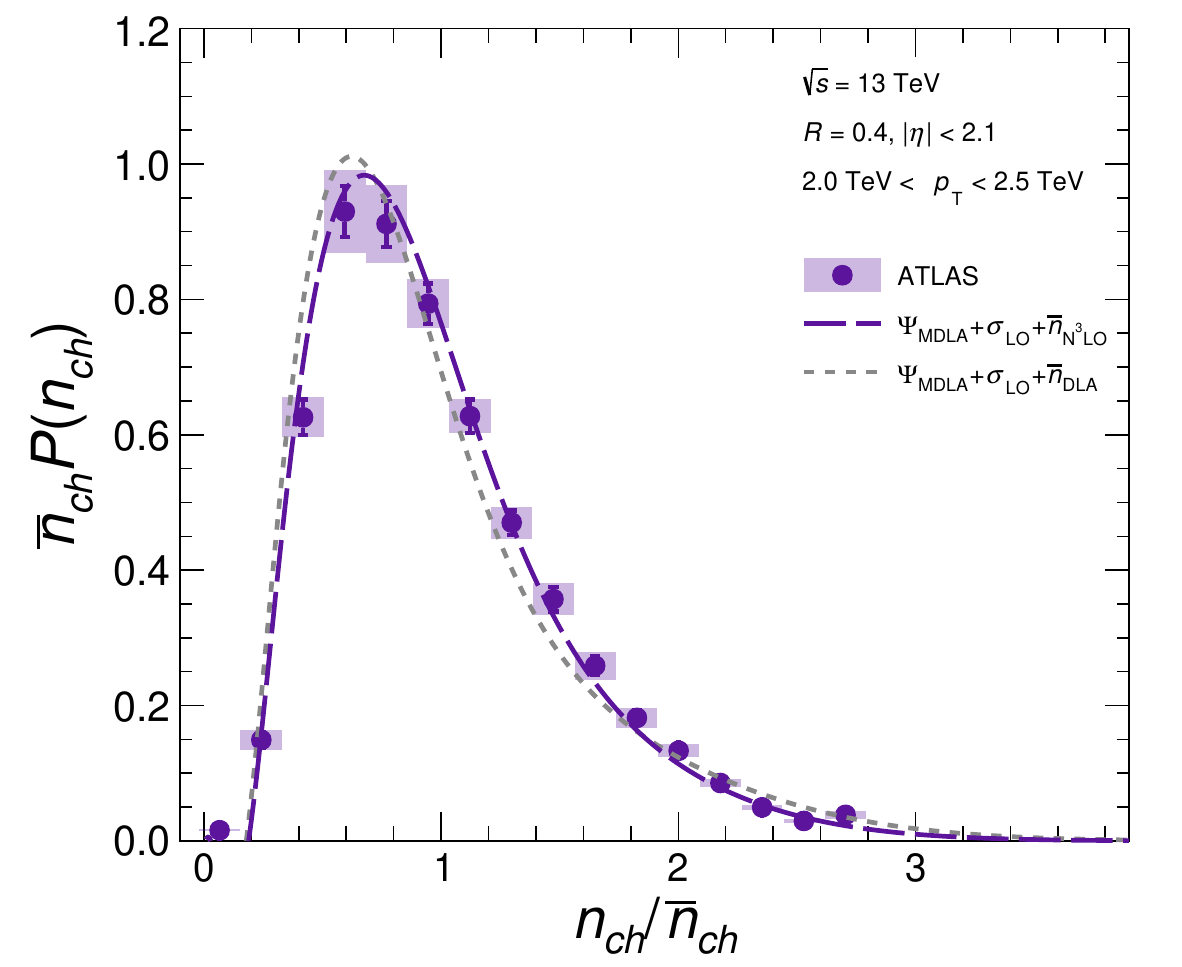} \\
    \caption{Investigation of KNO scaling via inclusive multiplicity distributions in $pp$ collisions. Shown here are our theoretical results using the MDLA KNO scaling functions ($\Psi_{\text{MDLA}}$), the LO cross section ($\sigma_{\text{LO}}$), and the ratios of mean multiplicities between gluon and quark jets in DLA ($\bar{n}_{\text{DLA}}$) and N$^3$LO ($\bar{n}_{\text{N$^3$LO}}$), compared with the ATLAS measurements of $\bar{n}_{ch} P(n_{ch})$ reported in ref.~\cite{ATLAS:2019rqw} across eight $p_T$ bins.}
    \label{fig:KNO_DLA_MDLA_PYTHIA_ATLAS}
\end{figure}

Figure~\ref{fig:KNO_DLA_MDLA_PYTHIA_ATLAS} presents our theoretical predictions for charged-particle multiplicity distributions in KNO form: $\bar{n}_{ch}P(n_{ch})$. For the ratio of mean multiplicities in gluon and quark jets, we employ both the DLA and N$^3$LO results, shown in figure~\ref{fig:ratioQG}. The differences between these predictions can be understood from eq.~(\ref{eq:KNO_inclusive}) and from the fact that the N$^3$LO multiplicity ratio is smaller than that in DLA across the entire $p_T$ range. At low $p_T$, gluon jets dominate. When the contribution from quark jets can be neglected, the DLA curve is expected to be broader, lower, and shifted to the right relative to the N$^3$LO curve. This is because the coefficient in front of $\Psi_g$ in the last line of eq.~(\ref{eq:KNO_inclusive}), proportional to $\bar{n}/\bar{n}_g = r_g + r_q \bar{n}_q/\bar{n}_g$, which also enters the argument of $\Psi_g$, decreases as $\bar{n}_g/\bar{n}_q$ increases. Conversely, at high $p_T$, quark jets dominate. When gluon-jet contributions are negligible, the DLA curve is expected to become narrower, higher, and shifted to the left relative to the N$^3$LO curve, since the coefficient of the quark-jet KNO scaling function, proportional to $\bar{n}/\bar{n}_q = r_q + r_g \bar{n}_g/\bar{n}_q$, increases with $\bar{n}_g/\bar{n}_q$, which also affects the argument of $\Psi_q$. The eight panels of the figure, arranged from top left with $0.1~\text{TeV}<p_T<0.2~\text{TeV}$ to bottom right with $2.0~\text{TeV}<p_T<2.5~\text{TeV}$, quantitatively illustrate the evolution between these two extremes.

In figure~\ref{fig:KNO_DLA_MDLA_PYTHIA_ATLAS}, our theoretical predictions are also compared with the measurements reported in ref.~\cite{ATLAS:2019rqw}. The results obtained with the N$^3$LO mean multiplicities show very good agreement with the experimental data across all reported $p_T$ bins. In contrast, predictions using DLA mean multiplicities provide a good description only in the range $0.1~\text{TeV}<p_T<0.2~\text{TeV}$, as shown in the top left panel. For the remaining $p_T$ bins, clear discrepancies appear, even when accounting for the relatively large uncertainties and coarse binning ($\Delta n_{ch}=4$).

\begin{figure}[htbp]
    \centering
    \includegraphics[height=0.23\textheight]{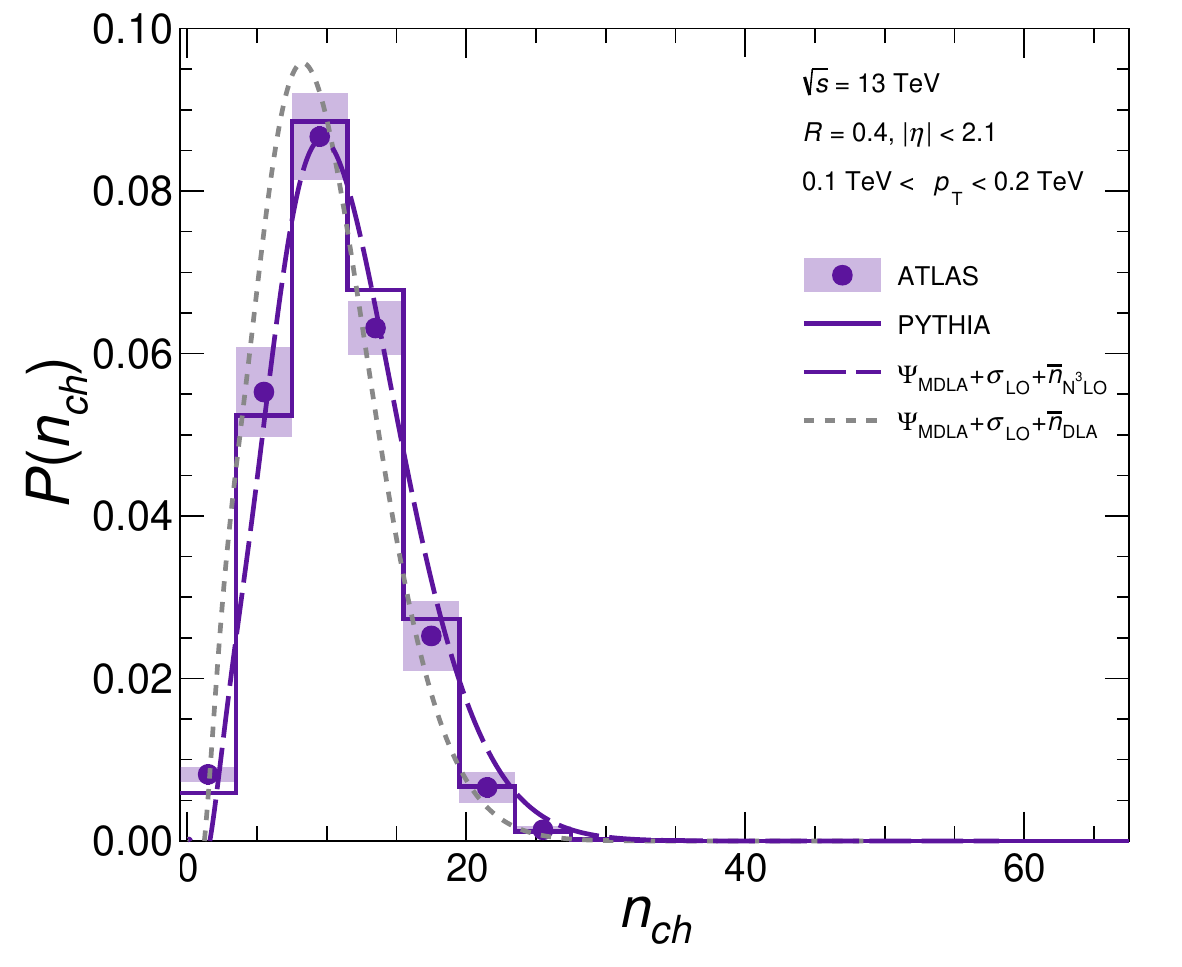}
    \includegraphics[height=0.23\textheight]{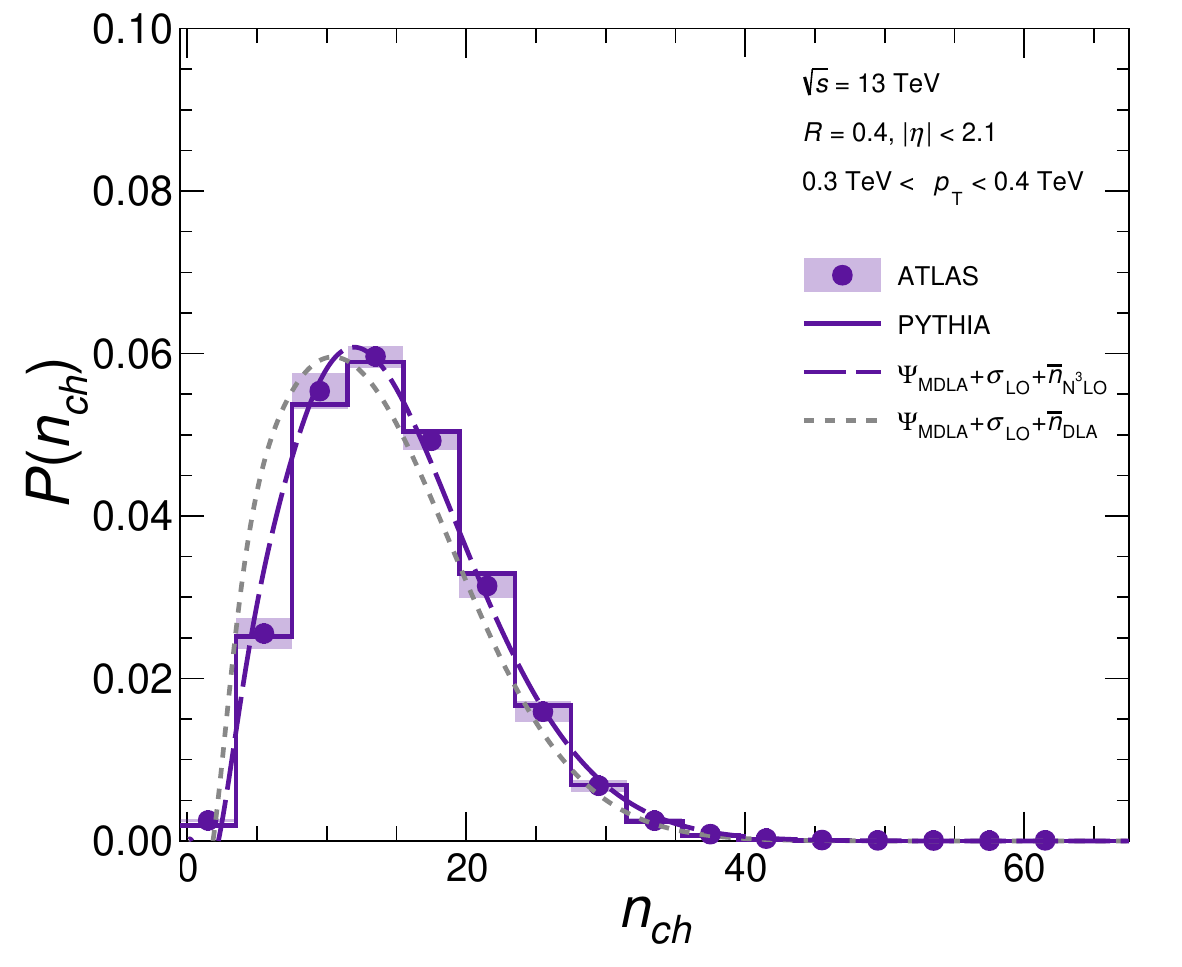} \\
    \includegraphics[height=0.23\textheight]{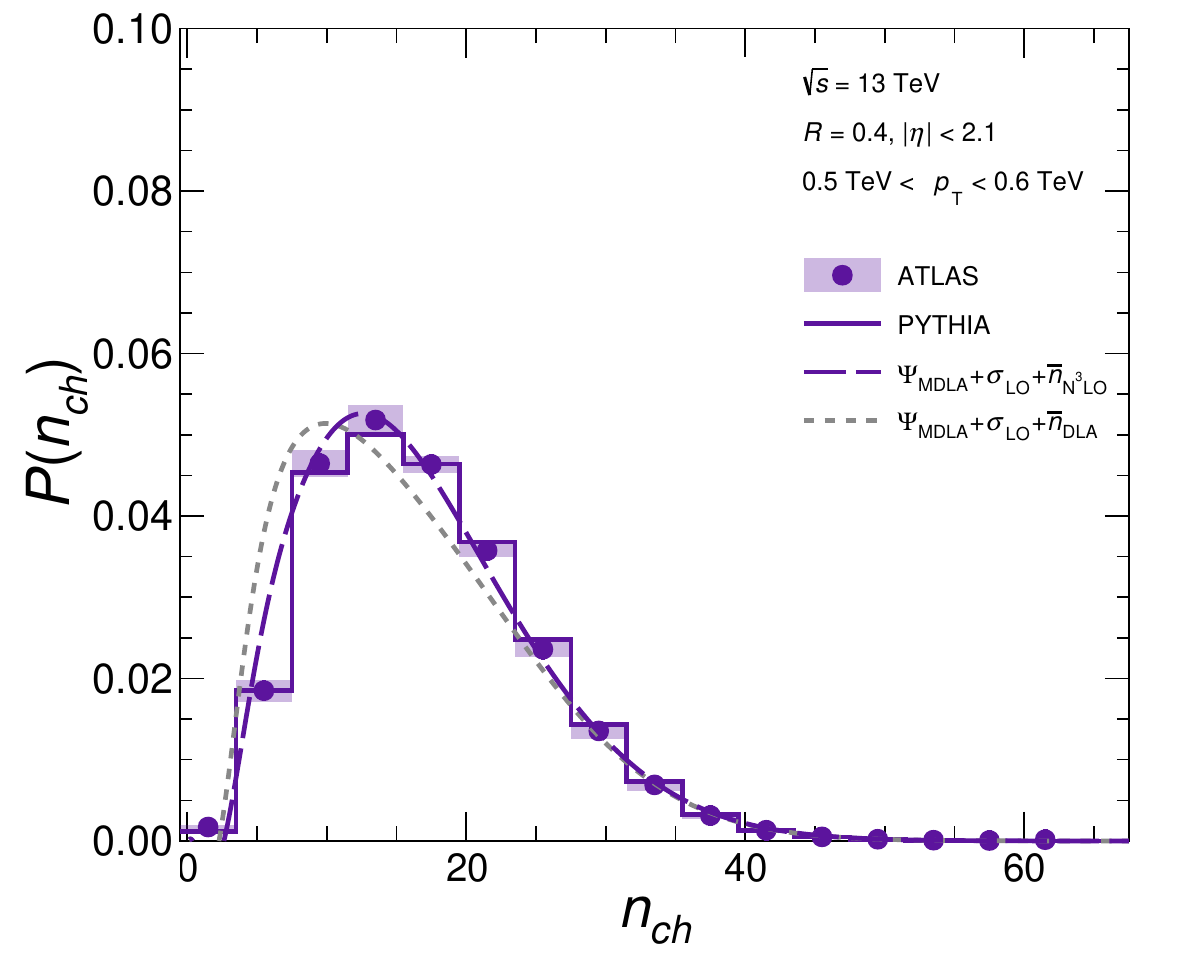}
    \includegraphics[height=0.23\textheight]{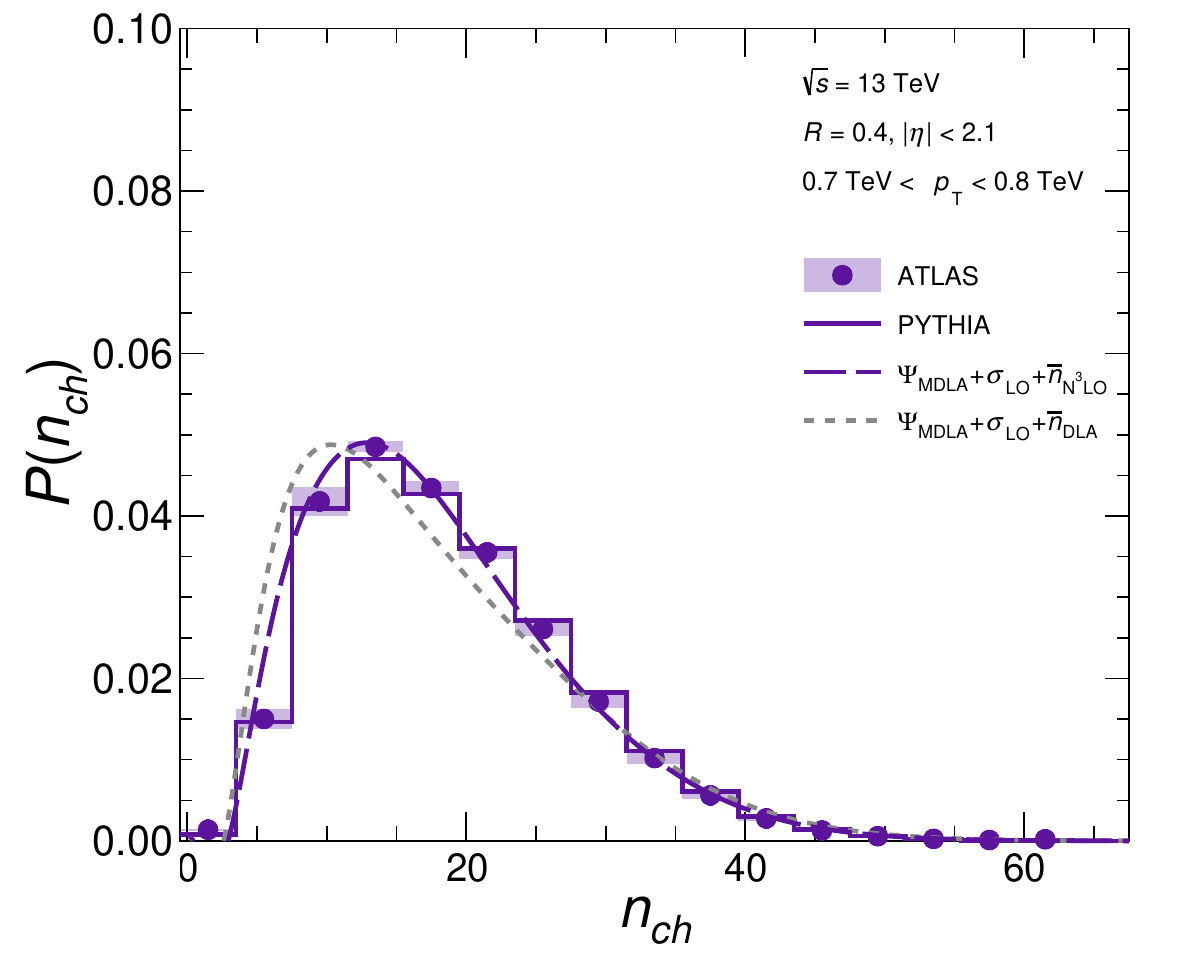} \\
    \includegraphics[height=0.23\textheight]{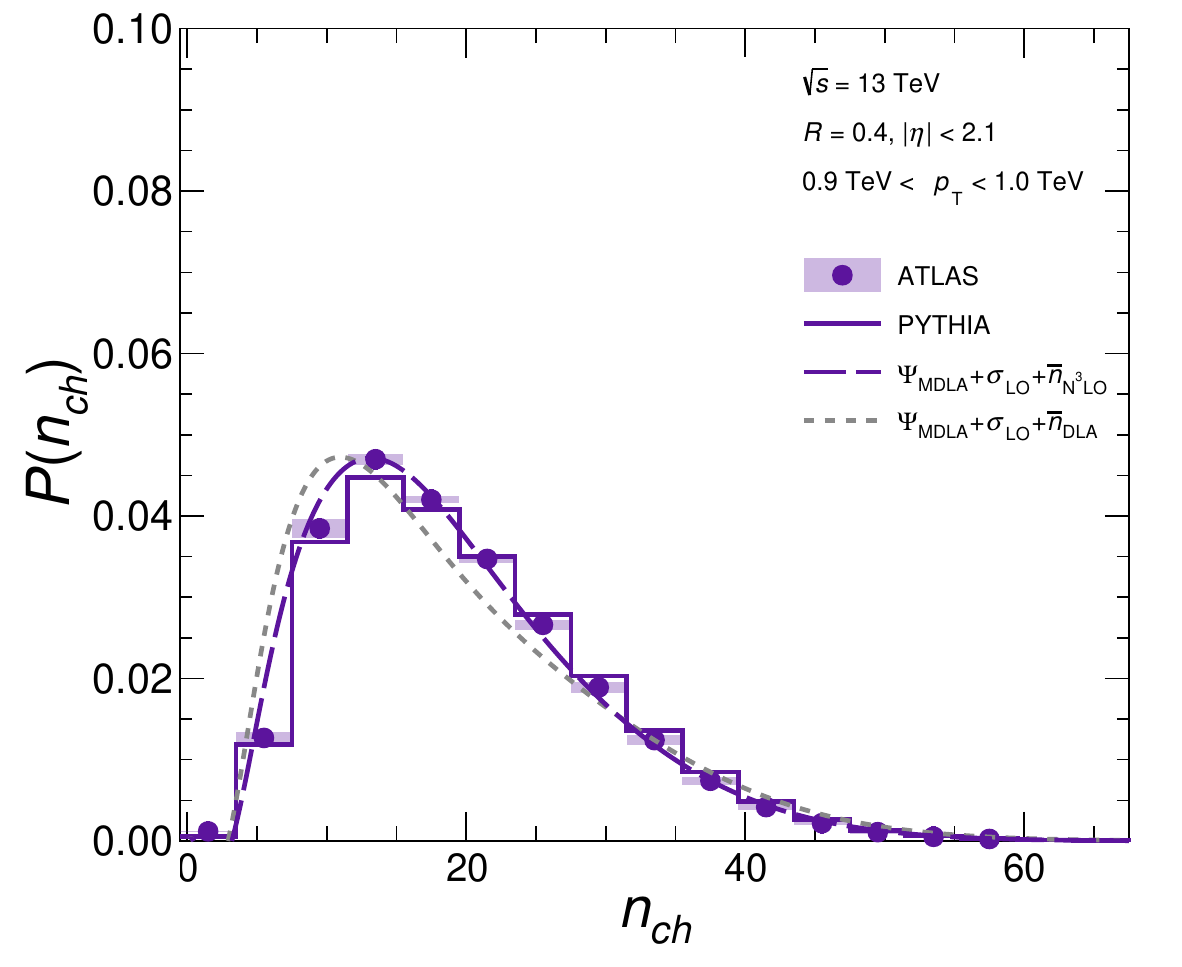}
    \includegraphics[height=0.23\textheight]{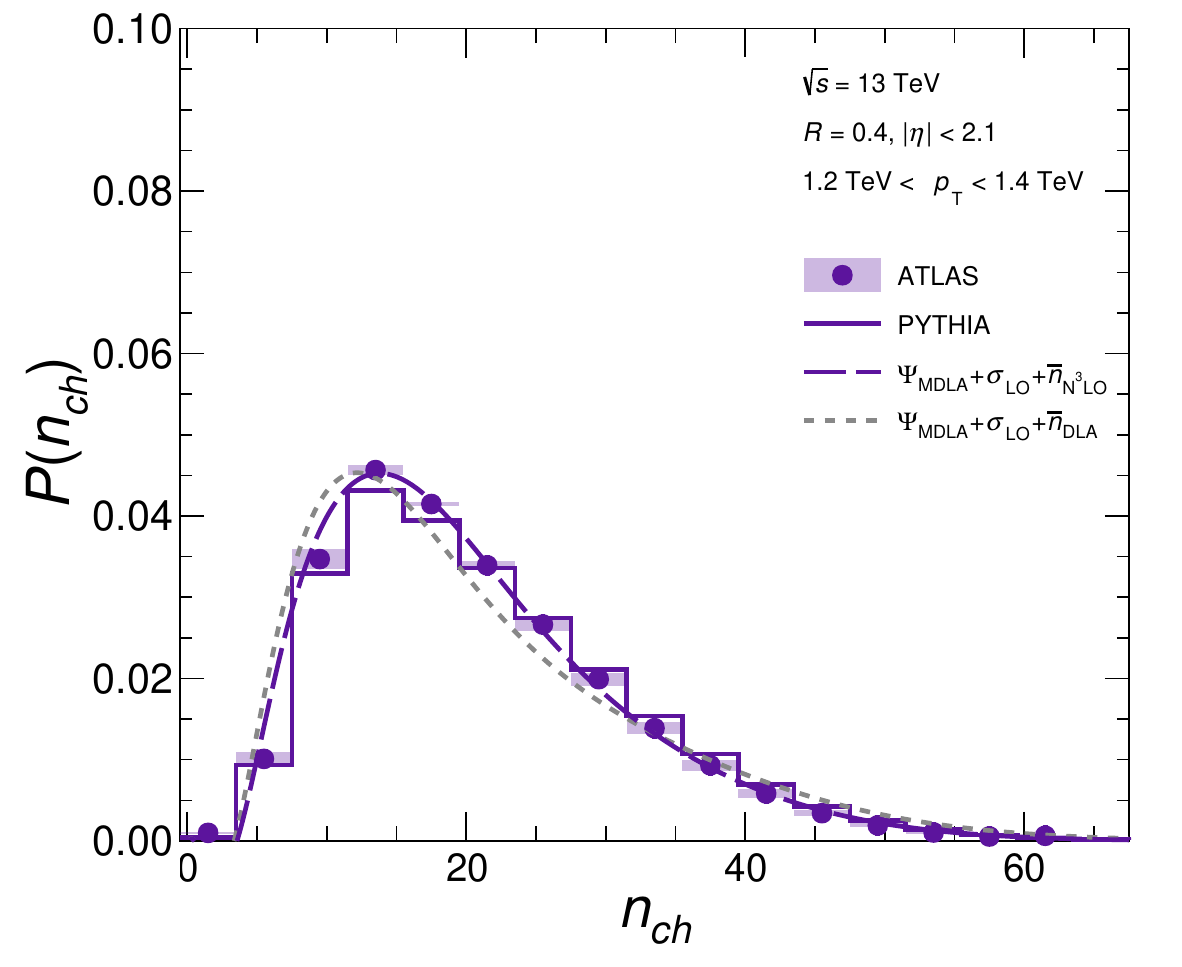} \\
    \includegraphics[height=0.23\textheight]{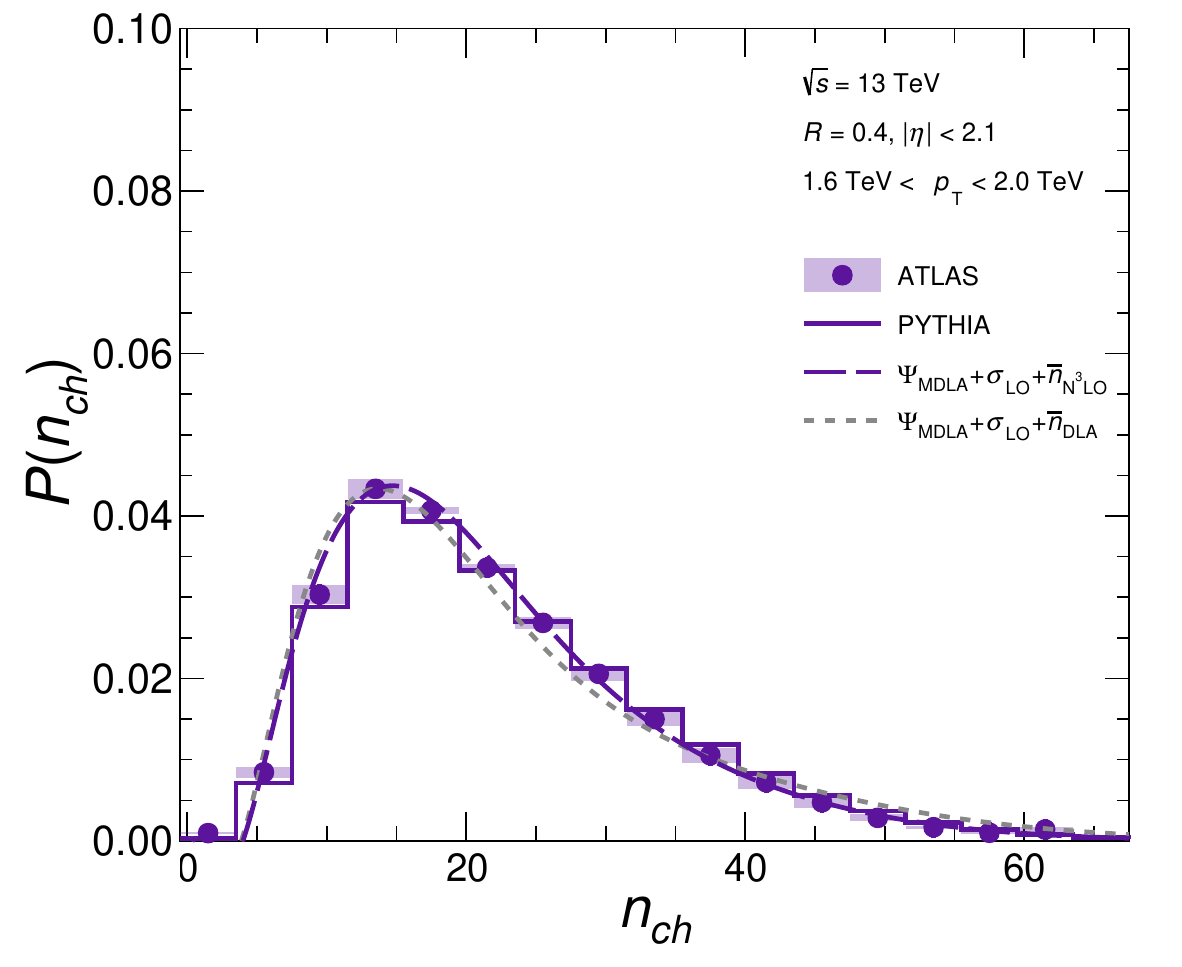}
    \includegraphics[height=0.23\textheight]{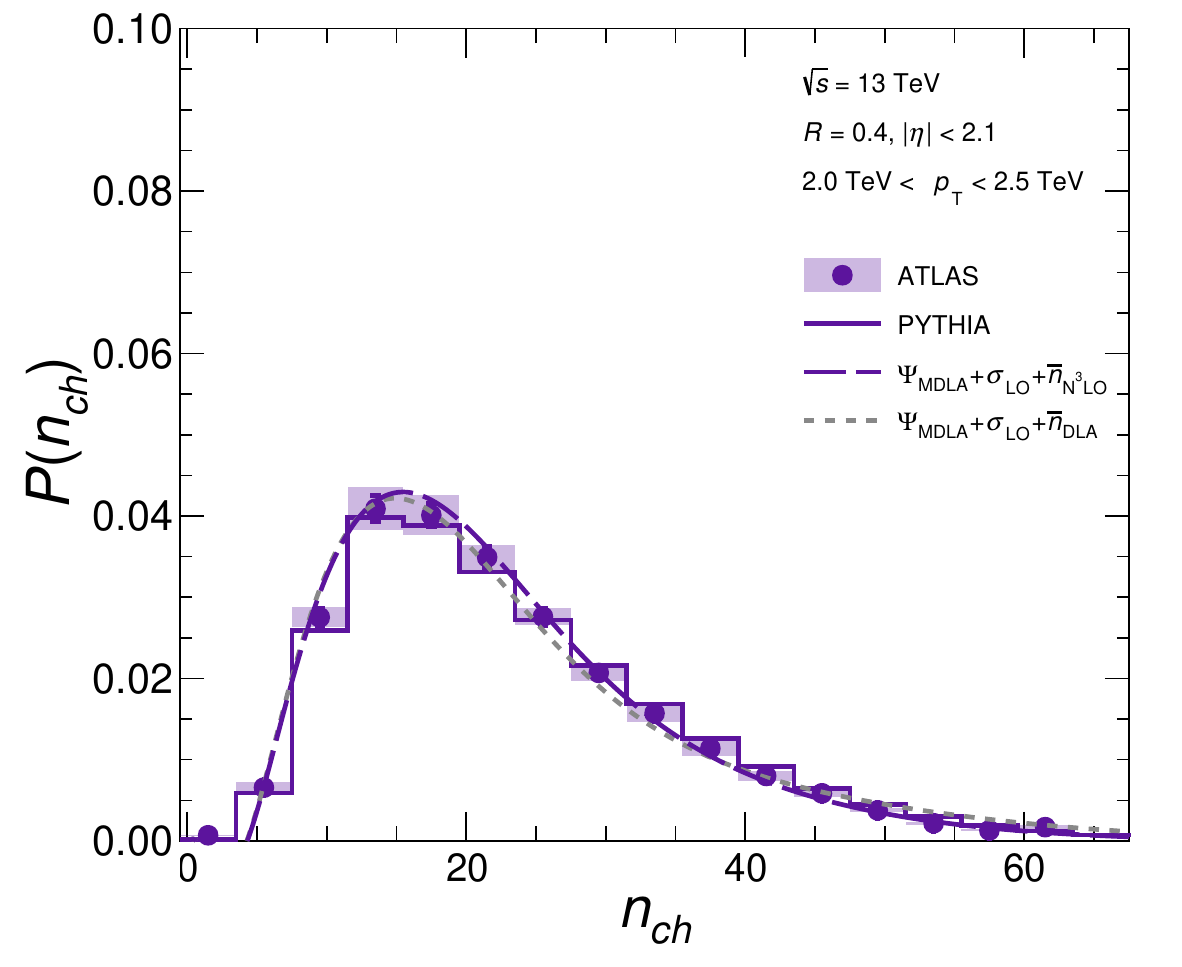} \\
    \caption{Inclusive multiplicity distributions in $pp$ collisions. Our theoretical predictions, shown in comparison to the $P(n_{ch})$ measurements reported in ref.~\cite{ATLAS:2019rqw}, are the same as those in figure~\ref{fig:KNO_DLA_MDLA_PYTHIA_ATLAS}, divided by the corresponding theoretical results of the inclusive mean multiplicity. The figure also displays the {\tt PYTHIA} results, with a bin width of $\Delta n_{ch} = 4$ in each panel.}
    \label{fig:P_MDLA_ATLAS_PYTHIA}
\end{figure}

Figure~\ref{fig:P_MDLA_ATLAS_PYTHIA} presents the same theoretical results as in figure~\ref{fig:KNO_DLA_MDLA_PYTHIA_ATLAS}, now divided by the corresponding theoretical inclusive mean multiplicities and compared directly with the $P(n_{ch})$ measurements from ref.~\cite{ATLAS:2019rqw}. The predictions with N$^3$LO mean multiplicities continue to agree very well with the experimental data across the entire $p_T$ range, as the N$^3$LO inclusive mean multiplicity provides a very good description of the ATLAS measurements, as shown in the left panel of figure~\ref{fig:mean}. In contrast, results obtained with DLA mean multiplicities again show less accurate agreement, though with a noticeable improvement for $2.0~\text{TeV}<p_T<2.5~\text{TeV}$ (and a deterioration for $0.1~\text{TeV}<p_T<0.2~\text{TeV}$). This behavior reflects the deviations of the DLA inclusive mean multiplicity from the data in the low- and high-$p_T$ bins.

The same figure also shows comparisons with \texttt{PYTHIA} simulations. The \texttt{PYTHIA} $P(n_{ch})$ distributions are presented using the same bin width of $\Delta n_{ch} = 4$ as in the ATLAS measurements. As shown in this figure, our theoretical predictions are consistent with the \texttt{PYTHIA} results across all $p_T$ bins. The agreement with \texttt{PYTHIA} is expected, as the values of $\gamma_0$ and $c_q$ were determined by fitting to \texttt{PYTHIA} results in ref.~\cite{Duan:2025ngi}, as shown in figure~\ref{fig:KNO_MDLA_PYTHIA_QG}. In turn, the consistency between our theoretical predictions and experimental measurements validates our approach of fixing $\gamma_0$ and $c_q$ based on \texttt{PYTHIA} simulations, although a more sophisticated determination of these parameters could be achieved by fitting directly to experimental data.

Thus, the agreement of our theoretical results with the ATLAS measurements provides further evidence that the measurements of ref.~\cite{ATLAS:2019rqw} can be consistently described in terms of two distinct KNO scaling functions, as presented within MDLA in sec.~\ref{sec:MDLA}, in line with our earlier observation using {\tt PYTHIA} in ref.~\citep{Duan:2025ngi}. Nonetheless, our theoretical calculations can be further refined, as it relies on the LO cross section, neglects KNO scaling violations induced by the running coupling (see ref.~\cite{Dokshitzer:2025owq} for a discussion at large $n_{ch}/\bar{n}_{ch}$) and treats $c_q$ as a free fitting parameter. A more systematic analysis, which would require improved theoretical inputs for each quantity involved, as discussed at the beginning of this subsection, is deferred to future work.

\section{Multiplicity distributions in quark and gluon jets versus jet topics}
\label{sec:topic}

As discussed above, the analysis of inclusive multiplicity distributions requires additional theoretical inputs, such as $r_a$ and $\bar{n}_a$, besides the KNO scaling functions, which makes establishing KNO scaling less straightforward. A more direct investigation of KNO scaling in QCD jets can be carried out alongside quark–gluon discrimination using jet substructure techniques (see refs.\cite{Marzani:2019hun, Larkoski:2024uoc} for recent reviews). The feasibility of this approach has been demonstrated with \texttt{PYTHIA} simulations in ref.~\cite{Duan:2025ngi}, where energy–correlation functions (ECFs)~\cite{Larkoski:2013eya} are employed to probe KNO scaling. However, because ECF-based studies select jets within a constrained phase space, the MDLA results presented in sec.~\ref{sec:qgjet} are not directly applicable. In this section, we instead compare the MDLA KNO scaling functions for quark and gluon jets with results obtained via the topic modeling method~\cite{Metodiev:2018ftz, Komiske:2018vkc} applied to the two jet samples measured by ATLAS in ref.~\cite{ATLAS:2019rqw}.

In ref.~\cite{ATLAS:2019rqw}, the two selected leading jets are separated into two samples based on the pseudorapidities of the two jets: the more forward jet sample and the more central jet sample. The charged-particle multiplicity distributions with a bin width of $\Delta n_{ch} = 4$ are measured for both jet samples, denoted as $h_i^f$ and $h_i^c$, where $i$ labels the $n_{ch}$ bins. According to topic modeling~\cite{Metodiev:2018ftz, Komiske:2018vkc}, the multiplicity distributions in quark and gluon jets can be approximated by the multiplicity distributions for topics 1 ($T_1$) and 2 ($T_2$):
\begin{align}\label{eq:topic}
    h_i^{T_1} = \frac{h_i^f - \min_j\{ h_j^f / h_j^c \} \, h_i^c}{1 - \min_j\{ h_j^f / h_j^c \}}, \qquad
    h_i^{T_2} = \frac{h_i^c - \min_j\{ h_j^c / h_j^f \} \, h_i^f}{1 - \min_j\{ h_j^c / h_j^f \}}.
\end{align}
As a test of this method for discriminating the quark and gluon jets, we also generate two jet samples in the same way using {\tt PYTHIA} simulations.

In our data analysis, we do not retain all $n_{ch}$ bins. For the \texttt{PYTHIA} samples, we impose the condition $h_j^{f,c} > \sqrt{N}/N$ (with $N$ the sample size) to exclude bins dominated by statistical fluctuations. For the experimental data, we discard bins with large uncertainties. Defining $r_{a/b} \equiv h_j^a / h_j^b$, where $a$ and $b$ correspond to two different jet samples, the propagated uncertainty of $r_{a/b}$ from the statistical and systematic errors in $h_j^a$ and $h_j^b$ is given by
\begin{align}
    \sigma_{r_{a/b}}^2 = \frac{1}{(h_j^b)^2}\sigma_{h_j^a}^2 + \frac{(h_j^a)^2}{(h_j^b)^4}\sigma_{h_j^b}^2,
\end{align}
where $\sigma_{o_j}^2$ represents the variance (squared uncertainty) associated with a data point $o_j$. Throughout our discussion, uncertainties are treated as uncorrelated. When evaluating $\kappa_{a/b} \equiv \min_j\{ h_j^a / h_j^b \}$, only those $n_{ch}$ bins are retained for which the total relative uncertainty, calculated as the square root of the sum of the squares of the propagated statistical and systematic errors, sufficiently small. We find that one obtains stable results when requiring:
\begin{align}\label{eq:Topic_ATLAS_select}
    \sigma_{r_{a/b}} / r_{a/b} < 25\%.
\end{align}
This threshold can be lowered, for example, to 15\%, but using an even smaller value may lead to deviations from the results presented below in certain $p_T$ bins, highlighting the need for improved experimental precision. Conversely, using a larger threshold can result in $\kappa_{a/b}$ being determined randomly by $n_{ch}$ bins with very small probabilities ($<0.005$), producing unreliable results. 

\begin{figure}[htbp]
    \centering
    \includegraphics[height=0.25\textheight]{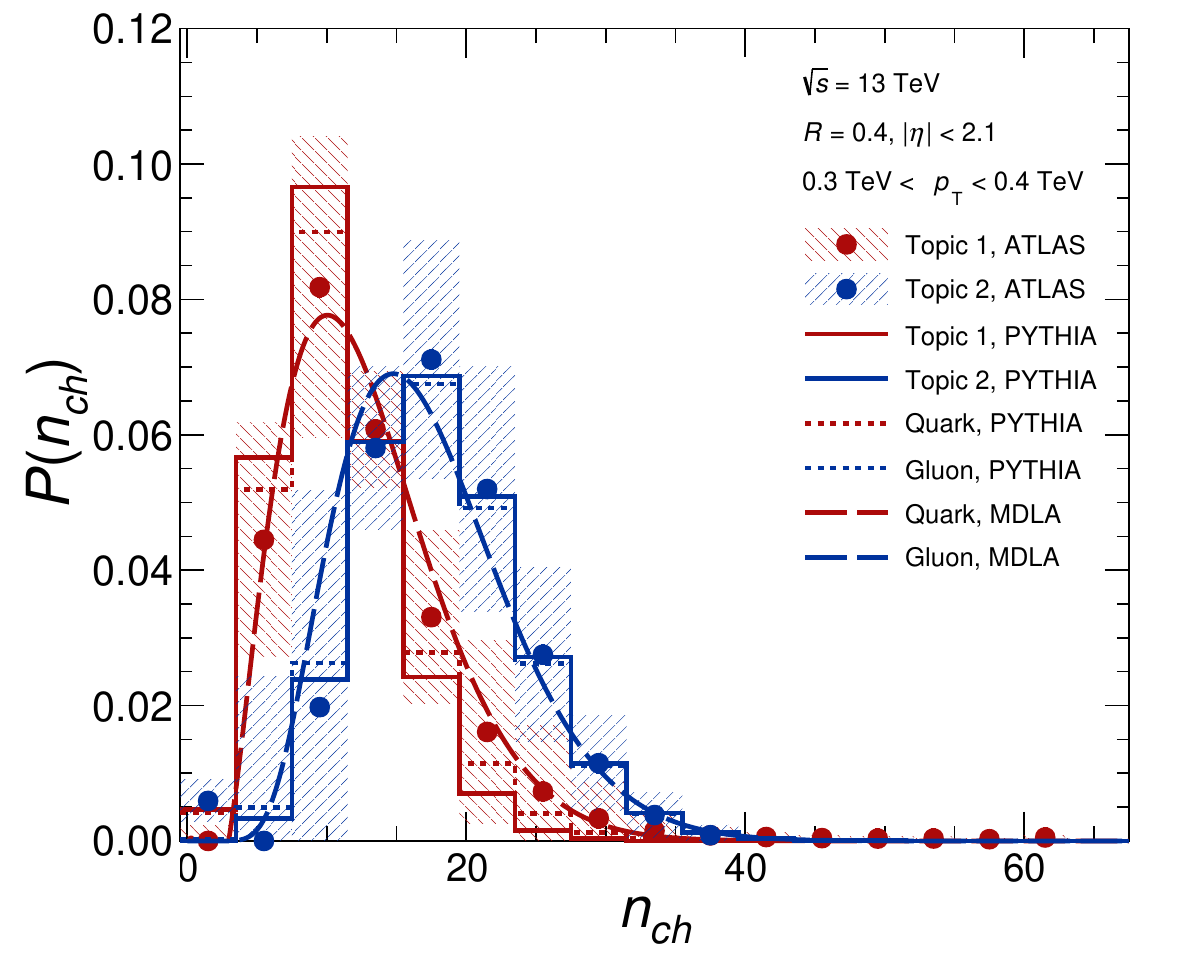}
    \includegraphics[height=0.25\textheight]{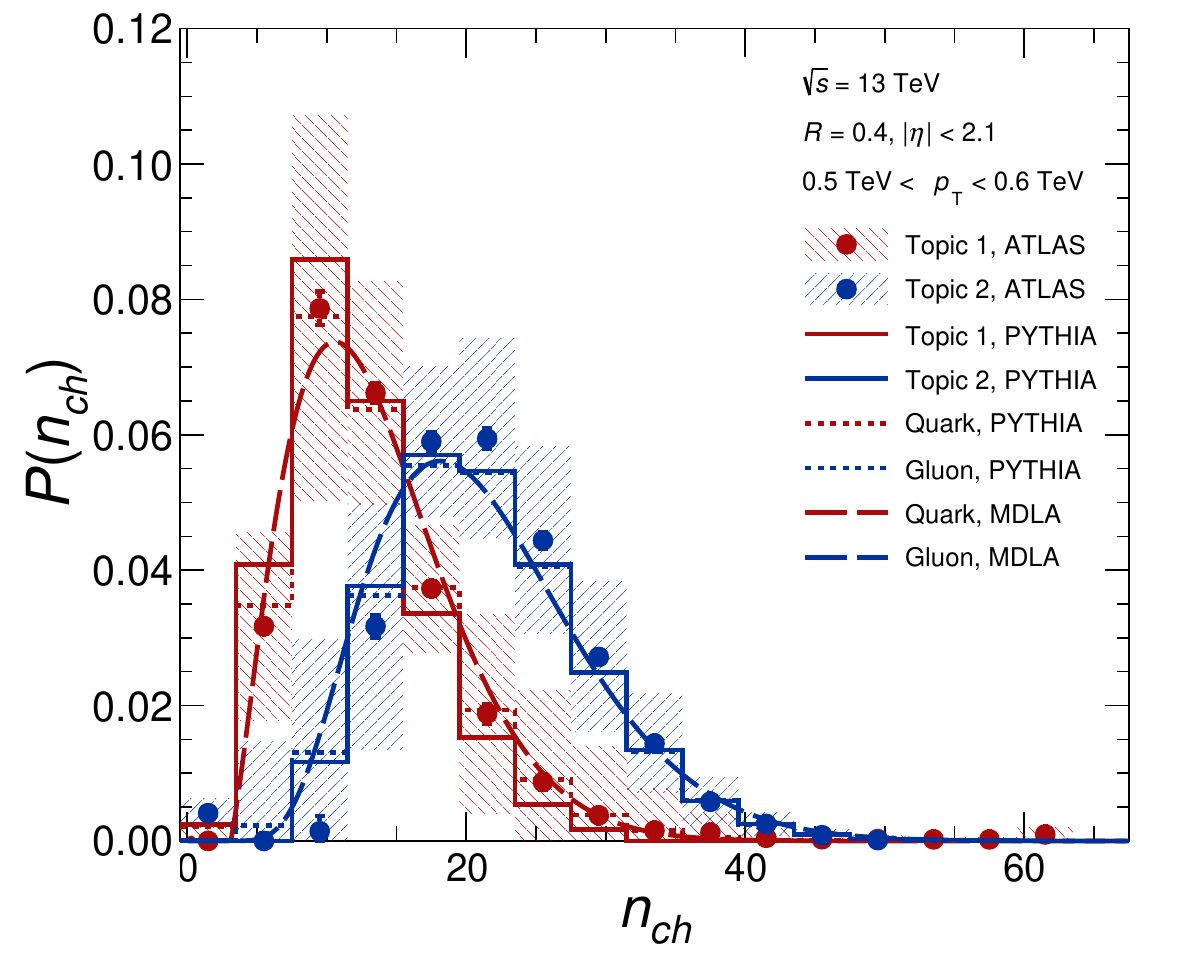} \\
    \includegraphics[height=0.25\textheight]{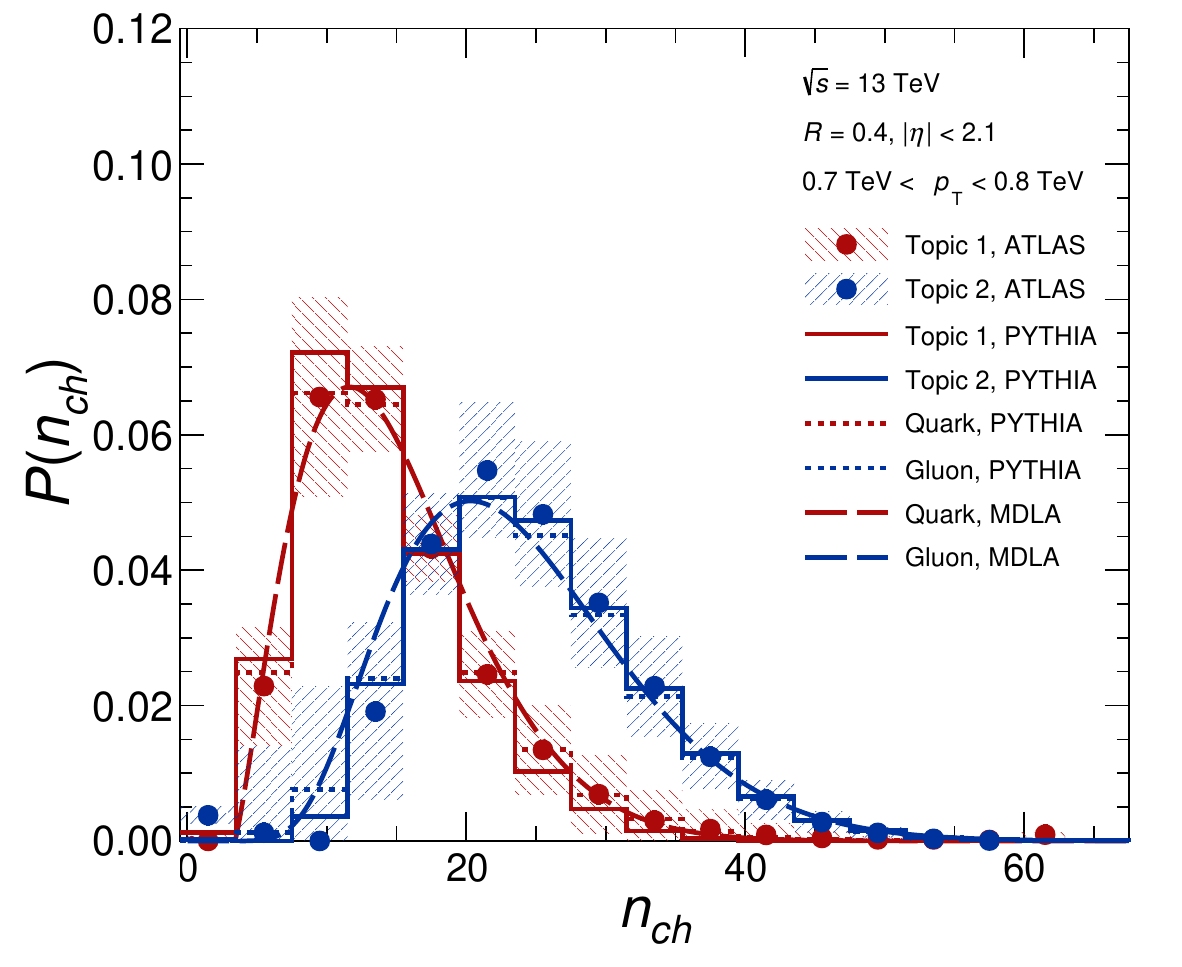}
    \includegraphics[height=0.25\textheight]{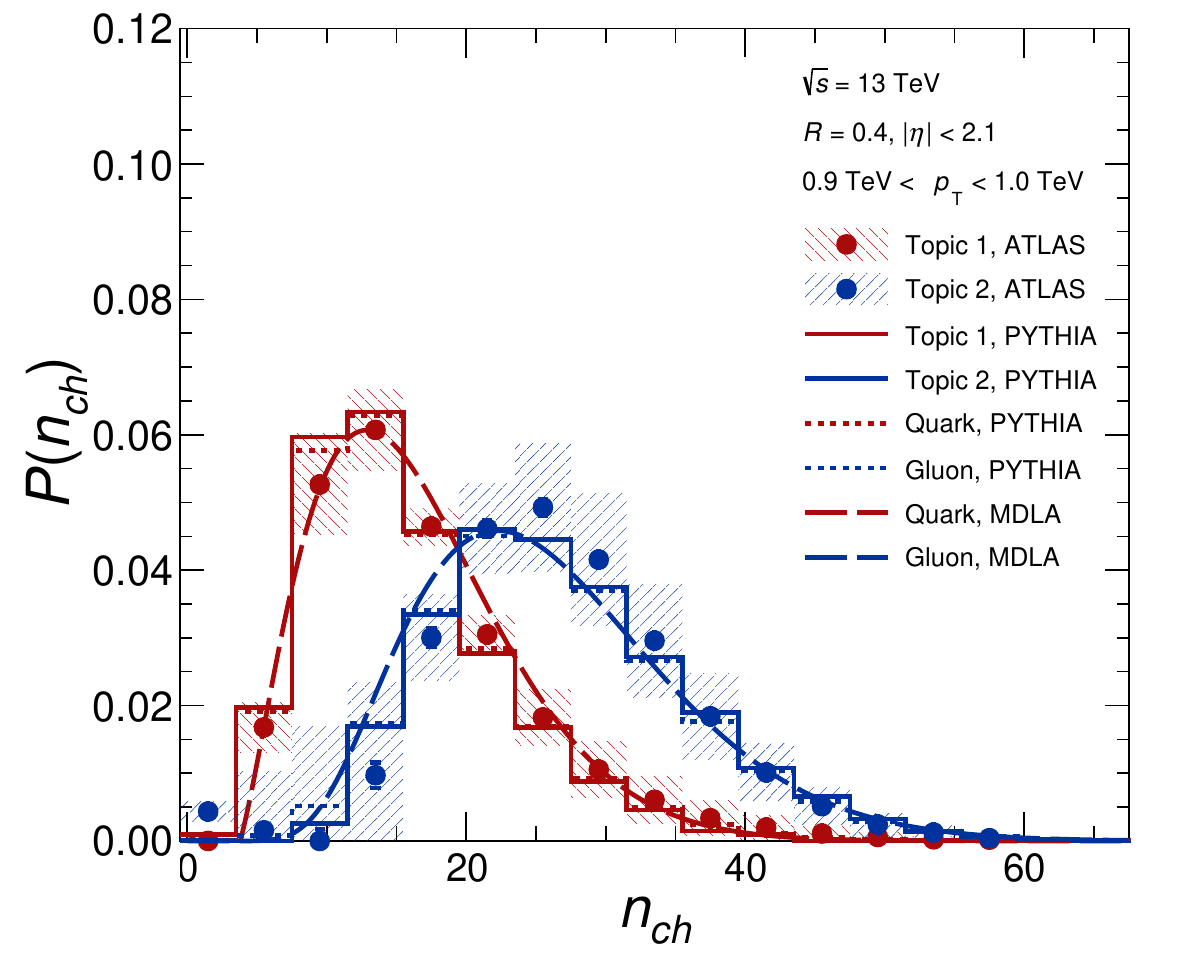} \\
    \includegraphics[height=0.25\textheight]{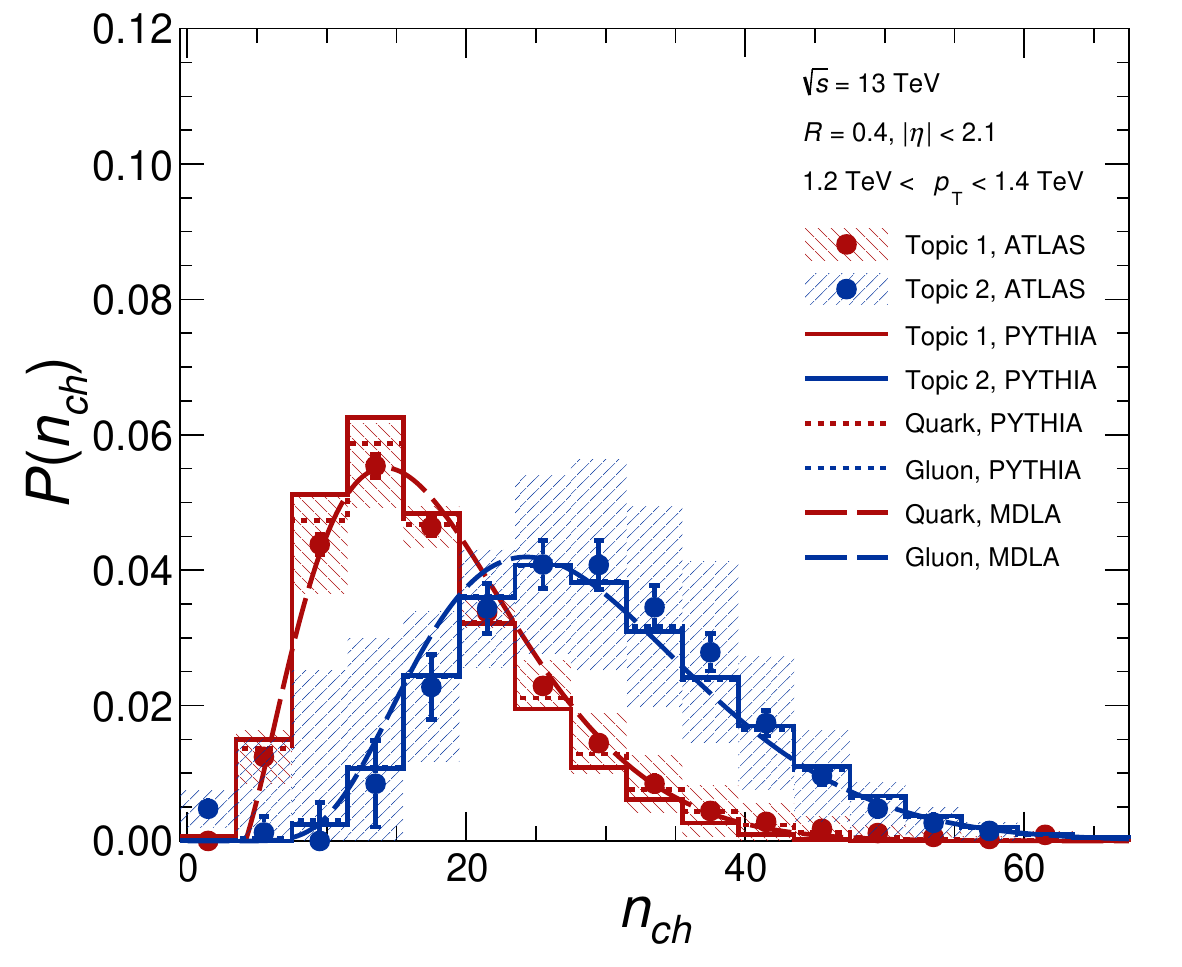}
    \includegraphics[height=0.25\textheight]{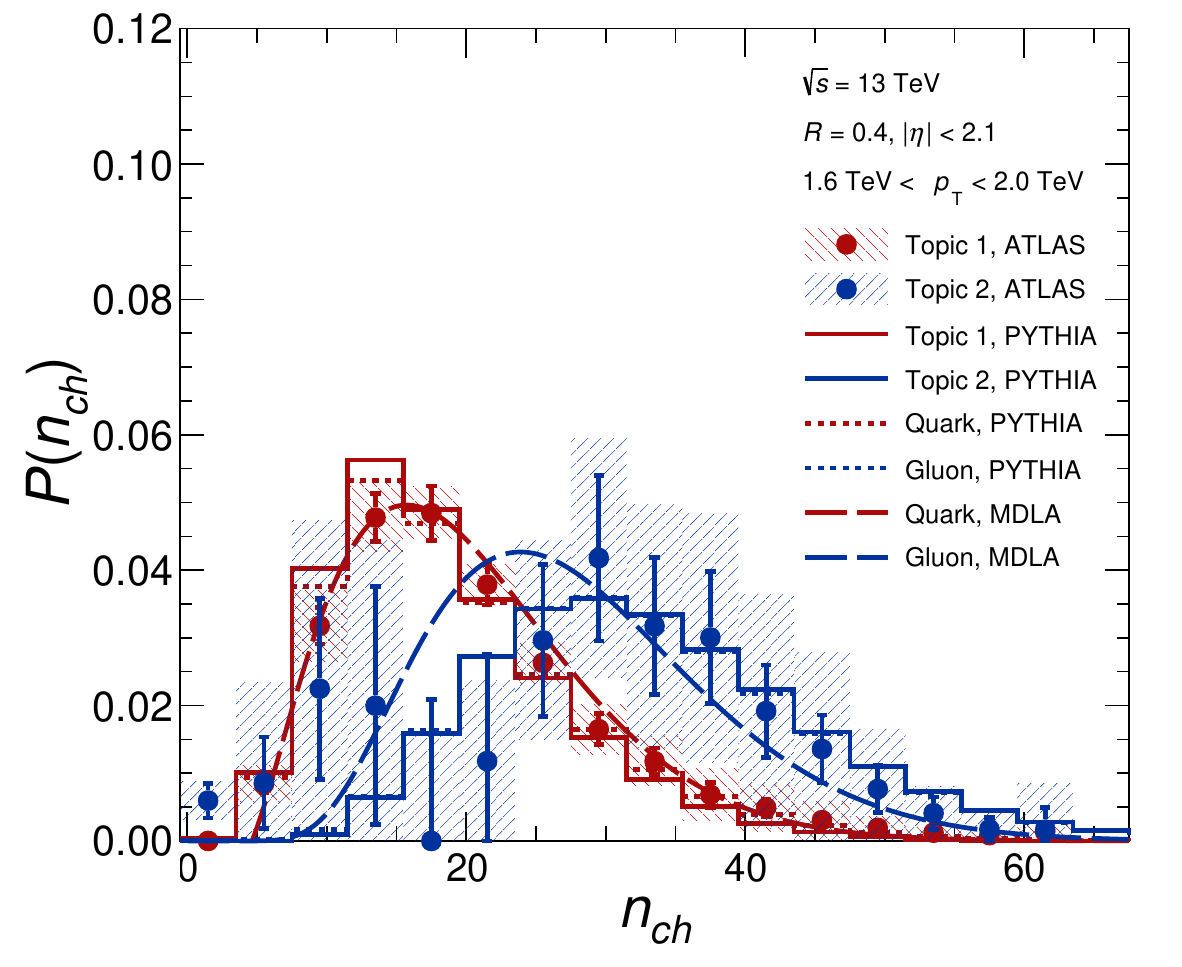} \\
    \caption{Charge-particle multiplicity distributions in quark (topic 1) and gluon (topic 2) jets. The plots compare the MDLA results (dashed) with the multiplicity distributions for topics 1 and 2 extracted from ATLAS data~\cite{ATLAS:2019rqw} (dots). In the experimental results, the error bars indicate statistical uncertainties, while the shaded bands correspond to systematic uncertainties. For comparison, predictions from {\tt PYTHIA} are also shown, where quark and gluon jets are identified according to the underlying hard processes~\cite{Duan:2025ngi} (dotted) and via the topic modeling approach~\cite{Metodiev:2018ftz, Komiske:2018vkc} (solid).}
    \label{fig:MFC-P-topic}
\end{figure}

\begin{figure}[htbp]
    \centering
    \includegraphics[height=0.25\textheight]{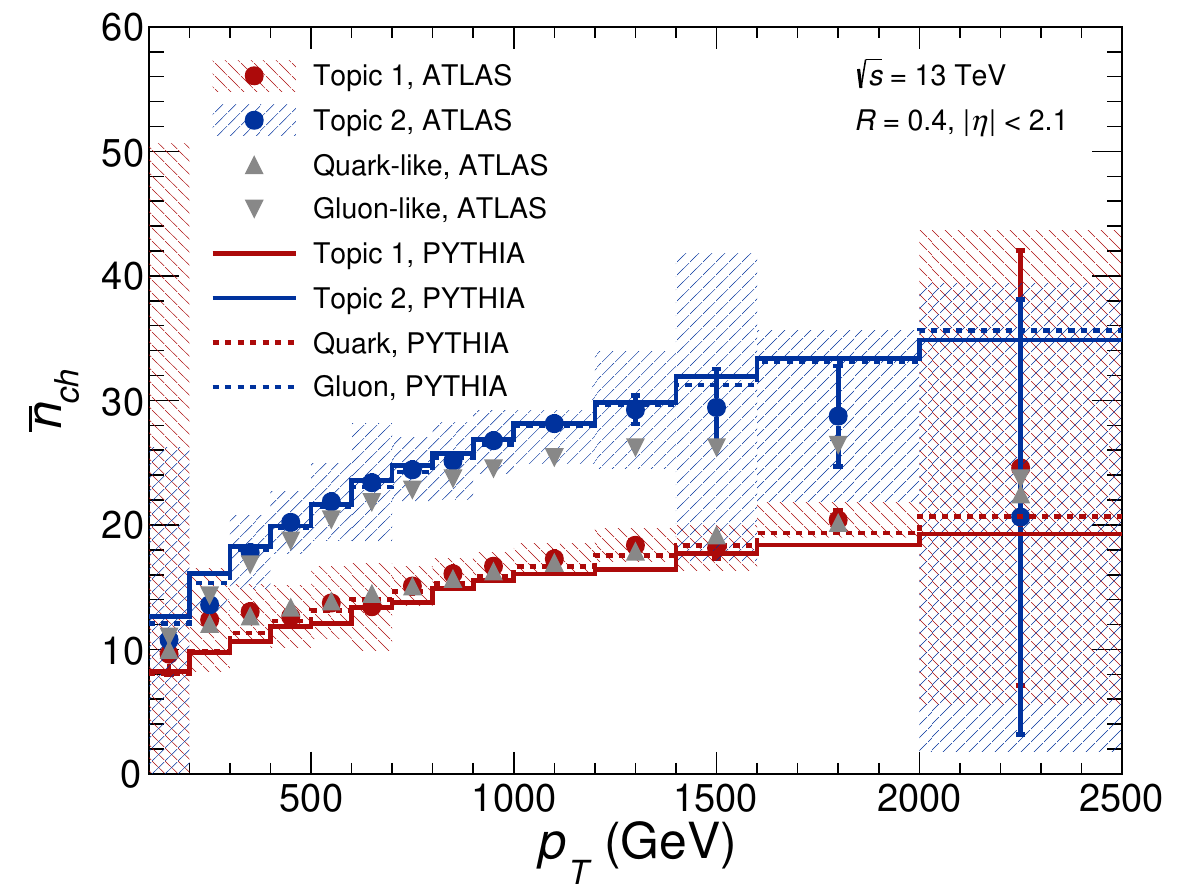}
    \caption{Mean charged-particle multiplicities in quark (topic 1) and gluon (topic 2) jets. The figure shows the average multiplicities for topics 1 and 2 extracted by applying topic modeling to the ATLAS more-forward and more-central jet samples (dots; statistical errors shown as error bars, systematic uncertainties as shaded bands), compared with the ATLAS results using the nominal quark and gluon fractions from the \texttt{PYTHIA} simulation (triangles) in ref.~\cite{ATLAS:2019rqw}, with experimental uncertainties omitted, as well as with \texttt{PYTHIA} results from ref.~\cite{Duan:2025ngi} (dotted) and \texttt{PYTHIA} results obtained via topic modeling (solid).
    }
    \label{fig:MFC-mean}
\end{figure}

Figure~\ref{fig:MFC-P-topic} compares the MDLA results, divided by $\bar{n}_a$, with the multiplicity distributions for $T_1$ and $T_2$ extracted from ATLAS data, as well as the corresponding {\tt PYTHIA} results. Here, $\bar{n}_q$ and $\bar{n}_g$ are taken as the mean multiplicities for $T_1$ and $T_2$, respectively. To validate the topic modeling method, we first compare the charged-particle distributions of quark and gluon jets identified through the hard processes in {\tt PYTHIA}, as implemented in ref.~\cite{Duan:2025ngi}, with those obtained via topic modeling applied to {\tt PYTHIA}-generated jet samples. Overall, the two results are consistent across all $p_T$ bins, though the degree of agreement varies among different $p_T$ ranges. For example, in the range $0.3~\mathrm{TeV} < p_T < 0.4~\mathrm{TeV}$, as shown in the top left panel, when restricting to $n_{ch}$ bins with probability above 0.01 to eliminate those dominated by statistical uncertainties, the gluon jet results agree at the percent level, and for quark jets only a single $n_{ch}$ bin shows a relative deviation exceeding 10\%.

Applying the same procedure to the two jet samples from ATLAS by imposing the selection criteria in eq.~(\ref{eq:Topic_ATLAS_select}), we obtain the multiplicity distributions for the jet topics, shown as data points with both statistical errors (error bars) and systematical errors (shaded bands) in figure~\ref{fig:MFC-P-topic}. The zeros of the extracted distributions for the topics indicate which $n_{ch}$ bin determines the minimum value of $\kappa_{a/b}$. For $T_2$, these zeros do not occur in the outermost $n_{ch}$ bins, reflecting the impact of experimental uncertainties and coarse binning ($\Delta n_{ch}=4$), which can be expected to account for the rise of the central values of $P(n_{ch})$ in the smaller $n_{ch}$ bins. Here, error propagation is performed using the standard formula; for example, for the $T_1$ distribution,
\begin{align}
    \sigma_{h_i^{T_1}}^2
    = \frac{1}{(1 - \kappa_{f/c})^2}\sigma_{h_i^f}^2
    + \frac{\kappa_{f/c}^2}{(1 - \kappa_{f/c})^2}\sigma_{h_i^c}^2
    + \frac{(h_i^f - h_i^c)^2}{(1 - \kappa_{f/c})^4}\sigma_{\kappa_{f/c}}^2.
\end{align} 
Large uncertainties appear in the very low- and high-$p_T$ regions, where the two jet samples, dominated by gluon and quark jets, respectively, become increasingly similar (i.e., $\kappa_{f/c} \to 1$), which amplifies the propagated errors according to the above expression.

As shown in figure~\ref{fig:MFC-P-topic}, the extracted jet-topic distributions from the ATLAS samples are consistent with the {\tt PYTHIA} results once uncertainties are taken into account. Notably, for $0.9~\mathrm{TeV} < p_T < 1.0~\mathrm{TeV}$, the extracted distributions from experimental data and {\tt PYTHIA}, whether topic modeling is used or not, are consistent, differing from the results exemplified in ref.~\cite{ATLAS:2019rqw}, which we attribute to our exclusion of $n_{ch}$ bins with large uncertainties. At higher $p_T$ (see the bottom right panel), the agreement between our experimental extractions and the {\tt PYTHIA} results deteriorates, as the central values of the two ATLAS jet samples become more alike than those in {\tt PYTHIA}.

As shown in figure~\ref{fig:MFC-mean}, the mean multiplicities extracted from ATLAS jet samples via topic modeling differ from those reported in ref.~\cite{ATLAS:2019rqw} using the same method, due to the exclusion of $n_{ch}$ bins with large uncertainties. In contrast, they agree with the results obtained via the alternative method using the nominal quark and gluon fractions from the \texttt{PYTHIA} simulation provided by ATLAS in ref.~\cite{ATLAS:2019rqw} and are consistent with the \texttt{PYTHIA} predictions within propagated uncertainties across the full $p_T$ range, although the central values lie below the \texttt{PYTHIA} results at higher $p_T$ for gluon jets. Here, the large uncertainties for topics 1 and 2 at both low and high $p_T$ arise from the near-similarity of the two jet samples, as mentioned above.

Overall, the MDLA scaling functions (with $\gamma_0=0.43$ and $c_q=0.8$) presented in sec.~\ref{sec:MDLA} show good agreement with the multiplicity distributions extracted from ATLAS data using jet topics, within the propagated experimental uncertainties. A notable discrepancy appears for gluon jets at low $n_{ch}$ in the $1.6~\mathrm{TeV} < p_T < 2.0~\mathrm{TeV}$ bin (see the bottom right panel of figure~\ref{fig:MFC-P-topic}). This is consistent with the mean multiplicities in figure~\ref{fig:MFC-mean}, where the central values of $\bar{n}_g$ at high $p_T$ are smaller than those predicted by {\tt PYTHIA}. Since $P_g(n) = \Psi_g(n/\bar{n}_g)/\bar{n}_g$, this shift explains why the MDLA curve lies higher and more to the left relative to {\tt PYTHIA} and the experimental results. We have also verified that, when $\bar{n}_g$ from {\tt PYTHIA} is used, the MDLA results agree very well with both the ATLAS extractions at large $n_{ch}$ and the {\tt PYTHIA} predictions.

\begin{figure}
    \centering
    \includegraphics[height=0.22\textheight]{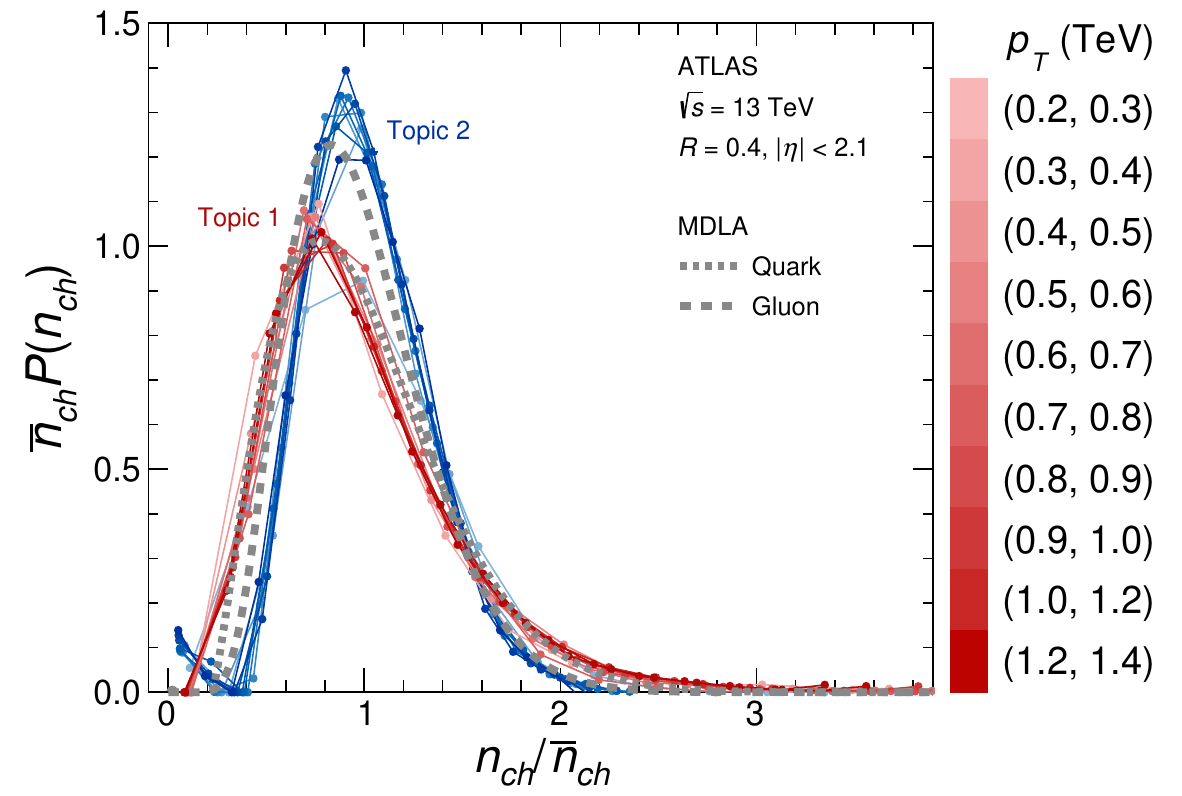}
    \includegraphics[height=0.22\textheight]{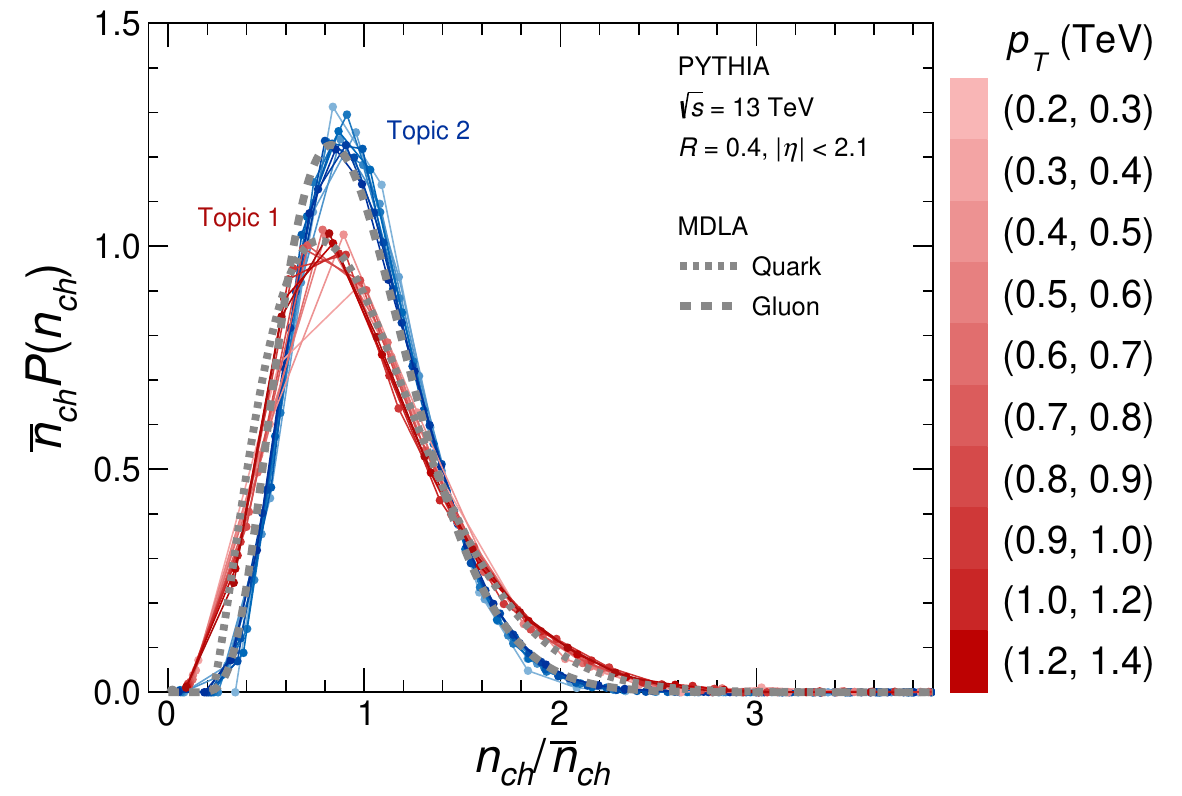}
    \caption{KNO scaling in jet topics. Left: Multiplicity distributions extracted from the two ATLAS jet samples~\cite{ATLAS:2019rqw} using topic modeling, with propagated experimental uncertainties (see figure~\ref{fig:MFC-P-topic}) omitted for clarity. Right: Corresponding distributions from {\tt PYTHIA} jet samples. In both panels, the bin width is $n_{ch}=4$, and the results are compared with the MDLA scaling functions (dashed).}
    \label{fig:MFC-KNO}
\end{figure}

As summarized in figure~\ref{fig:MFC-KNO}, the extracted multiplicity distributions for the two jet topics exhibit reasonably distinct scaling behavior. For the {\tt PYTHIA} results, they are consistent with the distributions extracted in ref.~\cite{Duan:2025ngi}, and the MDLA predictions provide a good description when the slight spread in {\tt PYTHIA} results is neglected. For the extracted ATLAS results, the central values also display similar scaling, with quark jets (Topic 1) agreeing better with the MDLA results than gluon jets (Topic 2). While caution is required when drawing final conclusions due to experimental uncertainties and coarse binning as explicitly shown in figures~\ref{fig:MFC-P-topic} and \ref{fig:MFC-mean}, these observations demonstrate that jet topic modeling offers a promising approach for extracting multiplicity distributions and supporting studies of KNO scaling.

\section{Summary and outlook}
\label{sec:concl}

Using an expansion in Laguerre polynomials~\cite{Bassetto:1987fq} and following the calculations for gluon jets in ref.~\cite{Dokshitzer:1993dc}, we obtain the KNO scaling functions for quark and gluon jets by incorporating energy conservation into the Double Logarithmic Approximation (DLA) of the generating function evolution equation. The resulting modified DLA (MDLA) expressions deviate markedly from the DLA results, supporting earlier conclusions that the latter fail to provide a reliable description of experimental data~\cite{Malaza:1984vv, Dokshitzer:1993dc, Duan:2025ngi, Dokshitzer:2025owq, Dokshitzer:2025fky}, although recent DLA-based studies have successfully explained the observed KNO scaling in DIS~\cite{Liu:2022bru, Liu:2023eve}. For gluon jets, our results are consistent with ref.~\cite{Dokshitzer:1993dc}, while for both quark and gluon jets they exhibit quantitative differences compared with the recently suggested QCD-inspired expressions in ref.~\cite{Dokshitzer:2025fky}.

The MDLA KNO scaling function for gluon jets depends solely on the anomalous dimension $\gamma_0$, while the quark KNO scaling function additionally depends on the ratio of mean multiplicities of quark and gluon jets in the asymptotic high-virtuality limit, $c_q$. Inspired by recent studies in refs.~\cite{Dokshitzer:2025owq, Dokshitzer:2025fky}, we treat $c_q$ as a fitting parameter in our analysis of jets in $pp$ collisions, rather than using the DLA value of $C_F/C_A$. We find that by fixing $\gamma_0$ and $c_q$, the MDLA results can very well reproduce the approximate KNO scaling functions for both quark and gluon jets in the two leading jets of $pp$ collisions at $\sqrt{s}=13$ TeV, as extracted using {\tt PYTHIA} for a broad jet $p_T$ range of $0.1$–$2.5$ TeV in ref.~\cite{Duan:2025ngi}. Furthermore, we observe that combining the MDLA scaling functions with the leading-order cross section and the N$^3$LO mean multiplicities in refs.~\cite{Dremin:1999ji,Capella:1999ms} provides a very good description of the inclusive charged-particle multiplicity distributions measured by ATLAS~\cite{ATLAS:2019rqw}.

Describing the inclusive multiplicity distributions requires additional theoretical input, such as the fraction of jets initiated by each parton type and mean multiplicities. This extra dependence may introduce uncertainties when evaluating the validity of KNO scaling. To address this and as a consistency check, we utilize jet substructure techniques to distinguish quark and gluon jets. Specifically, we apply the topic modeling method~\cite{Metodiev:2018ftz, Komiske:2018vkc} to two jet samples provided by ATLAS~\cite{ATLAS:2019rqw} to extract multiplicity  distributions for jet topics, serving as an approximation for quark and gluon jets. We find that these results are consistent with our MDLA results, although the experimental uncertainties remain significant.

Several aspects of our study could be further improved. First, a complete calculation of the KNO scaling functions for quark and gluon jets beyond the DLA approximation is still lacking. This is important because, although MDLA only incorporates energy conservation compared to DLA, it already introduces significant modifications. Moreover, describing the high-multiplicity tail observed by ATLAS~\cite{ATLAS:2019rqw}, where hadronic multiplicities exceed the average by a factor of 3 or more, would require detailed knowledge of hard jet constituents~\cite{Dokshitzer:2025owq}, which lies beyond both DLA and MDLA. A key issue to address beyond DLA is how to perform a direct calculation using the same jet definition (e.g., the anti-$k_t$ algorithm) employed in experiments. For inclusive multiplicity distributions, higher-order dijet cross sections are available~\cite{Currie:2017eqf}. In addition, the discrepancies between theoretical predictions and the measured mean multiplicities need to be resolved to achieve better agreement. One promising approach for high-precision calculations involves infrared-safe multiplicity definitions~\cite{Medves:2022ccw,Medves:2022uii}, although multiplicity distributions within this approach are yet to be studied. Notably, KNO scaling functions are infrared safe within both DLA and MDLA, and parton-level results have been used for comparisons with experimental data in this work (see also refs.~\cite{Dokshitzer:2025owq, Dokshitzer:2025fky}). At the same time, non-perturbative effects are observed in {\tt PYTHIA} simulations~\cite{Vertesi:2020utz, Duan:2025ngi}, highlighting the importance of clarifying the role of hadronization effects~\cite{Azimov:1984np,Kang:2023zdx}. Another promising avenue is to explore how studies of multiplicity distributions in quark and gluon jets can benefit from complementary methods of quark–gluon discrimination~\cite{Gallicchio:2011xq,Gallicchio:2012ez,Larkoski:2014pca,Gras:2017jty,Metodiev:2017vrx,Dreyer:2021hhr}. On the experimental side, more precise measurements are needed, and the $n_{ch}$ binning should be refined.

\acknowledgments

X.D. thanks Dr. Wen-Ya Wu for valuable discussions regarding the treatment of experimental data. This work is supported by the European Research Council under project ERC-2018-ADG-835105 YoctoLHC; by Maria de Maeztu excellence unit grant CEX2023-001318-M and projects PID2020-119632GB-I00 and PID2023-152762NB-I00 funded by MICIU/AEI/10.13039/501100011033; and by ERDF/EU. It has received funding from Xunta de Galicia (CIGUS Network of Research Centres).
X.D. is supported by the China Scholarship Council under Grant No. 202306100165. 
L.C. is supported by the Marie Sk\l{}odowska-Curie Actions Postdoctoral Fellowships under Grant No.~101210595. 
X.D. and G.M. are supported by the National Natural Science Foundation of China under Grants No.12147101, No. 12547102, No. 12325507, the National Key Research and Development Program of China under Grant No. 2022YFA1604900, and the Guangdong Major Project of Basic and Applied Basic Research under Grant No. 2020B0301030008. 
B.W. acknowledges the support of the Ram\'{o}n y Cajal program with the Grant No. RYC2021-032271-I and the support of Xunta de Galicia under the ED431F 2023/10 project.

\appendix
\section{Mean multiplicities at N$^3$LO}
\label{app:nb_N3LO}

To clarify the role of the parameters appearing in the N$^3$LO results~\cite{Dremin:1999ji,Capella:1999ms}, we provide a summary in this appendix. The mean multiplicity for gluon jets takes the form
\begin{align}\label{eq:GN3LOg}
    \bar{n}_g = K y^{-a_1C^2} \exp[2C\sqrt{y}+\delta_g(y)],
\end{align}
where $K$ is the normalization constant, $y \equiv \ln(Q/Q_0)$ (as defined in eq.~\eqref{eq:GF_DLA}), $C=\sqrt{4N_c/\beta_0}$, and
\begin{align}
    \delta_g(y) = \frac{C}{\sqrt{y}}\left\{2a_2C^2 + \frac{\beta_1}{\beta_0^2}[\ln(2y)+2]\right\} + \frac{C^2}{y}\left\{a_3C^2-\frac{a_1\beta_1}{\beta_0^2}[\ln(2y)+1]\right\}.
\end{align}
Here, the QCD $\beta$-function coefficients are given by $\beta_0=(11N_c-2n_f)/3$, $\beta_1=[17N_c^2-n_f(5N_c+3C_F)]/3$. Analogously, the mean multiplicity for quark jets takes the form
\begin{align}\label{eq:GN3LOq}
    \bar{n}_q = \frac{K}{r_0} y^{-a_1C^2} \exp[2C\sqrt{y}+\delta_q(y)],
\end{align}
with
\begin{align}
    r_0 = \frac{N_c}{C_F},
    \qquad
    \delta_q(y) = \delta_g(y) + \frac{C}{\sqrt{y}}r_1 + \frac{C^2}{y}(r_2+\frac{r_1^2}{2}).
\end{align}
The coefficients $a_i, r_i$ depend on the number of flavors $n_f$ and are tabulated in refs.~\cite{Capella:1999ms}. In our analysis, these coefficients are evaluated using $n_f = 5$, with $a_1 = 0.314, a_2 = -0.301, a_3 = 0.112, r_1 = 0.198, r_2 = 0.510$, and $r_3 = -0.041$.

\bibliographystyle{JHEP}
\bibliography{ref.bib}

@article{Polyakov:1970lyy,
    author = "Polyakov, A. M.",
    title = "{A Similarity hypothesis in the strong interactions. 1. Multiple hadron production in e+ e- annihilation}",
    journal = "Zh. Eksp. Teor. Fiz.",
    volume = "59",
    pages = "542--552",
    year = "1970"
}

@article{Koba:1972ng,
    author = "Koba, Z. and Nielsen, Holger Bech and Olesen, P.",
    title = "{Scaling of multiplicity distributions in high-energy hadron collisions}",
    doi = "10.1016/0550-3213(72)90551-2",
    journal = "Nucl. Phys. B",
    volume = "40",
    pages = "317--334",
    year = "1972"
}

@article{TASSO:1983cre,
    author = "Althoff, M. and others",
    collaboration = "TASSO",
    title = "{Jet Production and Fragmentation in e+ e- Annihilation at 12-GeV to 43-GeV}",
    reportNumber = "DESY-83-130",
    doi = "10.1007/BF01547419",
    journal = "Z. Phys. C",
    volume = "22",
    pages = "307--340",
    year = "1984"
}

@article{DELPHI:1990ohs,
    author = "Abreu, P. and others",
    collaboration = "DELPHI",
    title = "{Charged particle multiplicity distributions in Z0 hadronic decays}",
    reportNumber = "CERN-PPE-90-173",
    doi = "10.1007/BF01474073",
    journal = "Z. Phys. C",
    volume = "50",
    pages = "185--194",
    year = "1991"
}

@article{DELPHI:1991qnt,
    author = "Abreu, P. and others",
    collaboration = "DELPHI",
    title = "{Charged particle multiplicity distributions in restricted rapidity intervals in Z0 hadronic decays.}",
    doi = "10.1007/BF01560444",
    journal = "Z. Phys. C",
    volume = "52",
    pages = "271--281",
    year = "1991"
}

@article{H1:2020zpd,
    author = "Andreev, V. and others",
    collaboration = "H1",
    title = "{Measurement of charged particle multiplicity distributions in DIS at HERA and its implication to entanglement entropy of partons}",
    eprint = "2011.01812",
    archivePrefix = "arXiv",
    primaryClass = "hep-ex",
    reportNumber = "DESY-20-176",
    doi = "10.1140/epjc/s10052-021-08896-1",
    journal = "Eur. Phys. J. C",
    volume = "81",
    number = "3",
    pages = "212",
    year = "2021"
}

@article{UA5:1985kkp,
    author = "Alner, G. J. and others",
    collaboration = "UA5",
    title = "{A New Empirical Regularity for Multiplicity Distributions in Place of KNO Scaling}",
    reportNumber = "CERN-EP/85-62",
    doi = "10.1016/0370-2693(85)91492-3",
    journal = "Phys. Lett. B",
    volume = "160",
    pages = "199--206",
    year = "1985"
}

@article{UA5:1988gup,
    author = "Ansorge, R. E. and others",
    editor = "Kotthaus, R. and Kuhn, Johann H.",
    collaboration = "UA5",
    title = "{Charged Particle Multiplicity Distributions at 200-GeV and 900-GeV Center-Of-Mass Energy}",
    reportNumber = "CERN-EP-88-172",
    doi = "10.1007/BF01506531",
    journal = "Z. Phys. C",
    volume = "43",
    pages = "357",
    year = "1989"
}

@article{CMS:2010qvf,
    author = "Khachatryan, Vardan and others",
    collaboration = "CMS",
    title = "{Charged Particle Multiplicities in $pp$ Interactions at $\sqrt{s}=0.9$, 2.36, and 7 TeV}",
    eprint = "1011.5531",
    archivePrefix = "arXiv",
    primaryClass = "hep-ex",
    reportNumber = "CERN-PH-EP-2010-048, CMS-QCD-10-004",
    doi = "10.1007/JHEP01(2011)079",
    journal = "JHEP",
    volume = "01",
    pages = "079",
    year = "2011"
}

@article{Grosse-Oetringhaus:2009eis,
    author = "Grosse-Oetringhaus, Jan Fiete and Reygers, Klaus",
    title = "{Charged-Particle Multiplicity in Proton-Proton Collisions}",
    eprint = "0912.0023",
    archivePrefix = "arXiv",
    primaryClass = "hep-ex",
    doi = "10.1088/0954-3899/37/8/083001",
    journal = "J. Phys. G",
    volume = "37",
    pages = "083001",
    year = "2010"
}

@article{ALICE:2017pcy,
    author = "Acharya, S. and others",
    collaboration = "ALICE",
    title = "{Charged-particle multiplicity distributions over a wide pseudorapidity range in proton-proton collisions at $\sqrt{s}=$ 0.9, 7, and 8 TeV}",
    eprint = "1708.01435",
    archivePrefix = "arXiv",
    primaryClass = "hep-ex",
    reportNumber = "CERN-EP-2017-192",
    doi = "10.1140/epjc/s10052-017-5412-6",
    journal = "Eur. Phys. J. C",
    volume = "77",
    number = "12",
    pages = "852",
    year = "2017"
}

@article{Konishi:1979cb,
    author = "Konishi, K. and Ukawa, A. and Veneziano, G.",
    title = "{Jet Calculus: A Simple Algorithm for Resolving QCD Jets}",
    reportNumber = "RL-79-026",
    doi = "10.1016/0550-3213(79)90053-1",
    journal = "Nucl. Phys. B",
    volume = "157",
    pages = "45--107",
    year = "1979"
}

@article{Bassetto:1979nt,
    author = "Bassetto, A. and Ciafaloni, M. and Marchesini, G.",
    title = "{Inelastic Distributions and Color Structure in Perturbative QCD}",
    reportNumber = "UTF-45",
    doi = "10.1016/0550-3213(80)90413-7",
    journal = "Nucl. Phys. B",
    volume = "163",
    pages = "477--518",
    year = "1980"
}

@article{Dokshitzer:1982ia,
    author = "Dokshitzer, Yuri L. and Fadin, Victor S. and Khoze, Valery A.",
    title = "{On the Sensitivity of the Inclusive Distributions in Parton Jets to the Coherence Effects in QCD Gluon Cascades}",
    reportNumber = "LENINGRAD-82-789",
    doi = "10.1007/BF01571703",
    journal = "Z. Phys. C",
    volume = "18",
    pages = "37",
    year = "1983"
}

@article{Bassetto:1987fq,
    author = "Bassetto, A.",
    title = "{KNO SCALING IN QCD JETS AND THE NEGATIVE BINOMIAL DISTRIBUTION}",
    reportNumber = "DFPD-9/87",
    doi = "10.1016/0550-3213(88)90426-9",
    journal = "Nucl. Phys. B",
    volume = "303",
    pages = "703--712",
    year = "1988"
}

@book{Dokshitzer:1991wu,
    author = "Dokshitzer, Yuri L. and Khoze, Valery A. and Mueller, Alfred H. and Troian, S. I.",
    title = "{Basics of perturbative QCD}",
    year = "1991",
    publisher = "Editions Frontieres",
    url = "https://www.lpthe.jussieu.fr/~yuri/BPQCD/cover.html"
}

@article{Dokshitzer:1993dc,
    author = "Dokshitzer, Yuri L.",
    title = "{Improved QCD treatment of the KNO phenomenon}",
    reportNumber = "LU-TP-93-3",
    doi = "10.1016/0370-2693(93)90121-W",
    journal = "Phys. Lett. B",
    volume = "305",
    pages = "295--301",
    year = "1993"
}

@article{Malaza:1984vv,
    author = "Malaza, E. D. and Webber, B. R.",
    title = "{QCD CORRECTIONS TO JET MULTIPLICITY MOMENTS}",
    reportNumber = "HEP 84/3",
    doi = "10.1016/0370-2693(84)90375-7",
    journal = "Phys. Lett. B",
    volume = "149",
    pages = "501--503",
    year = "1984"
}

@article{Dremin:2000ep,
    author = "Dremin, I. M. and Gary, J. W.",
    title = "{Hadron multiplicities}",
    eprint = "hep-ph/0004215",
    archivePrefix = "arXiv",
    reportNumber = "FIAN-TD31-00, UCRHEP-E273",
    doi = "10.1016/S0370-1573(00)00117-4",
    journal = "Phys. Rept.",
    volume = "349",
    pages = "301--393",
    year = "2001"
}

@book{Marzani:2019hun,
    author = "Marzani, Simone and Soyez, Gregory and Spannowsky, Michael",
    title = "{Looking inside jets: an introduction to jet substructure and boosted-object phenomenology}",
    eprint = "1901.10342",
    archivePrefix = "arXiv",
    primaryClass = "hep-ph",
    doi = "10.1007/978-3-030-15709-8",
    publisher = "Springer",
    volume = "958",
    year = "2019"
}

@article{Larkoski:2024uoc,
    author = "Larkoski, Andrew J.",
    title = "{QCD masterclass lectures on jet physics and machine learning}",
    eprint = "2407.04897",
    archivePrefix = "arXiv",
    primaryClass = "hep-ph",
    doi = "10.1140/epjc/s10052-024-13341-0",
    journal = "Eur. Phys. J. C",
    volume = "84",
    number = "10",
    pages = "1117",
    year = "2024"
}

@article{Vertesi:2020utz,
    author = "Vertesi, Robert and Gemes, Antal and Barnafoldi, Gergely Gabor",
    title = "{Koba-Nielsen-Olesen-like scaling within a jet in proton-proton collisions at LHC energies}",
    eprint = "2012.01132",
    archivePrefix = "arXiv",
    primaryClass = "hep-ph",
    doi = "10.1103/PhysRevD.103.L051503",
    journal = "Phys. Rev. D",
    volume = "103",
    number = "5",
    pages = "L051503",
    year = "2021"
}

@article{ATLAS:2011eid,
    author = "Aad, Georges and others",
    collaboration = "ATLAS",
    title = "{Properties of jets measured from tracks in proton-proton collisions at center-of-mass energy $\sqrt{s}=7$ TeV with the ATLAS detector}",
    eprint = "1107.3311",
    archivePrefix = "arXiv",
    primaryClass = "hep-ex",
    reportNumber = "CERN-PH-EP-2011-110",
    doi = "10.1103/PhysRevD.84.054001",
    journal = "Phys. Rev. D",
    volume = "84",
    pages = "054001",
    year = "2011"
}

@article{ATLAS:2019rqw,
    author = "Aad, Georges and others",
    collaboration = "ATLAS",
    title = "{Properties of jet fragmentation using charged particles measured with the ATLAS detector in $pp$ collisions at $\sqrt{s}=13$ TeV}",
    eprint = "1906.09254",
    archivePrefix = "arXiv",
    primaryClass = "hep-ex",
    reportNumber = "CERN-EP-2019-090",
    doi = "10.1103/PhysRevD.100.052011",
    journal = "Phys. Rev. D",
    volume = "100",
    number = "5",
    pages = "052011",
    year = "2019"
}

@article{Germano:2024ier,
    author = "Germano, G. R. and Navarra, F. S. and Wilk, G. and Wlodarczyk, Z.",
    title = "{Emergence of Koba-Nielsen-Olsen scaling in multiplicity distributions in jets produced at the LHC}",
    eprint = "2406.04856",
    archivePrefix = "arXiv",
    primaryClass = "hep-ph",
    doi = "10.1103/PhysRevD.110.034026",
    journal = "Phys. Rev. D",
    volume = "110",
    number = "3",
    pages = "034026",
    year = "2024"
}

@article{Dokshitzer:1987nm,
    author = "Dokshitzer, Yuri L. and Khoze, Valery A. and Troian, S. I. and Mueller, Alfred H.",
    title = "{QCD Coherence in High-Energy Reactions}",
    reportNumber = "CU-TP-374",
    doi = "10.1103/RevModPhys.60.373",
    journal = "Rev. Mod. Phys.",
    volume = "60",
    pages = "373",
    year = "1988"
}

@article{Larkoski:2013eya,
    author = "Larkoski, Andrew J. and Salam, Gavin P. and Thaler, Jesse",
    title = "{Energy Correlation Functions for Jet Substructure}",
    eprint = "1305.0007",
    archivePrefix = "arXiv",
    primaryClass = "hep-ph",
    reportNumber = "MIT-CTP-4446, CERN-PH-TH-2013-066, LPN13-026",
    doi = "10.1007/JHEP06(2013)108",
    journal = "JHEP",
    volume = "06",
    pages = "108",
    year = "2013"
}

@article{Dokshitzer:1982xr,
    author = "Dokshitzer, Yuri L. and Fadin, Victor S. and Khoze, Valery A.",
    title = "{Double Logs of Perturbative QCD for Parton Jets and Soft Hadron Spectra}",
    reportNumber = "LENINGRAD-82-745",
    doi = "10.1007/BF01614423",
    journal = "Z. Phys. C",
    volume = "15",
    pages = "325",
    year = "1982"
}

@article{Dokshitzer:1982fh,
    author = "Dokshitzer, Yuri L. and Fadin, Victor S. and Khoze, Valery A.",
    title = "{Coherent Effects in the Perturbative QCD Parton Jets}",
    doi = "10.1016/0370-2693(82)90654-2",
    journal = "Phys. Lett. B",
    volume = "115",
    pages = "242--246",
    year = "1982"
}

@article{Azimov:1984np,
    author = "Azimov, Yakov I. and Dokshitzer, Yuri L. and Khoze, Valery A. and Troyan, S. I.",
    title = "{Similarity of Parton and Hadron Spectra in QCD Jets}",
    reportNumber = "LENINGRAD-84-942",
    doi = "10.1007/BF01642482",
    journal = "Z. Phys. C",
    volume = "27",
    pages = "65--72",
    year = "1985"
}

@article{Dokshitzer:1995ev,
    author = "Dokshitzer, Yuri L. and Khoze, Valery A. and Troian, S. I.",
    title = "{Specific features of heavy quark production. LPHD approach to heavy particle spectra}",
    eprint = "hep-ph/9506425",
    archivePrefix = "arXiv",
    reportNumber = "LU-TP-94-23, LU-TP-94-23---UPDATED",
    doi = "10.1103/PhysRevD.53.89",
    journal = "Phys. Rev. D",
    volume = "53",
    pages = "89--119",
    year = "1996"
}

@article{Capella:1999ms,
    author = "Capella, A. and Dremin, I. M. and Gary, J. W. and Nechitailo, V. A. and Tran Thanh Van, J.",
    title = "{Evolution of average multiplicities of quark and gluon jets}",
    eprint = "hep-ph/9910226",
    archivePrefix = "arXiv",
    reportNumber = "FIAN-TD-22-99, LPT-99-74, UCRHEP-E264",
    doi = "10.1103/PhysRevD.61.074009",
    journal = "Phys. Rev. D",
    volume = "61",
    pages = "074009",
    year = "2000"
}

@article{Buckley:2014ana,
    author = {Buckley, Andy and Ferrando, James and Lloyd, Stephen and Nordstr\"om, Karl and Page, Ben and R\"ufenacht, Martin and Sch\"onherr, Marek and Watt, Graeme},
    title = "{LHAPDF6: parton density access in the LHC precision era}",
    eprint = "1412.7420",
    archivePrefix = "arXiv",
    primaryClass = "hep-ph",
    reportNumber = "GLAS-PPE-2014-05, MCNET-14-29, IPPP-14-111, DCPT-14-222",
    doi = "10.1140/epjc/s10052-015-3318-8",
    journal = "Eur. Phys. J. C",
    volume = "75",
    pages = "132",
    year = "2015"
}

@article{Bierlich:2022pfr,
    author = "Bierlich, Christian and others",
    title = "{A comprehensive guide to the physics and usage of PYTHIA 8.3}",
    eprint = "2203.11601",
    archivePrefix = "arXiv",
    primaryClass = "hep-ph",
    reportNumber = "LU-TP 22-16, MCNET-22-04, FERMILAB-PUB-22-227-SCD",
    doi = "10.21468/SciPostPhysCodeb.8",
    journal = "SciPost Phys. Codeb.",
    volume = "2022",
    pages = "8",
    year = "2022"
}

@article{Cacciari:2008gp,
    author = "Cacciari, Matteo and Salam, Gavin P. and Soyez, Gregory",
    title = "{The anti-$k_t$ jet clustering algorithm}",
    eprint = "0802.1189",
    archivePrefix = "arXiv",
    primaryClass = "hep-ph",
    reportNumber = "LPTHE-07-03",
    doi = "10.1088/1126-6708/2008/04/063",
    journal = "JHEP",
    volume = "04",
    pages = "063",
    year = "2008"
}

@article{Cacciari:2011ma,
    author = "Cacciari, Matteo and Salam, Gavin P. and Soyez, Gregory",
    title = "{FastJet User Manual}",
    eprint = "1111.6097",
    archivePrefix = "arXiv",
    primaryClass = "hep-ph",
    reportNumber = "CERN-PH-TH-2011-297",
    doi = "10.1140/epjc/s10052-012-1896-2",
    journal = "Eur. Phys. J. C",
    volume = "72",
    pages = "1896",
    year = "2012"
}

@article{Duan:2025ngi,
    author = "Duan, Xiang-Pan and Chen, Lin and Ma, Guo-Liang and Salgado, Carlos A. and Wu, Bin",
    title = "{KNO scaling in quark and gluon jets at the LHC}",
    eprint = "2503.24200",
    archivePrefix = "arXiv",
    primaryClass = "hep-ph",
    month = "3",
    year = "2025"
}

@article{Liu:2022bru,
    author = "Liu, Yizhuang and Nowak, Maciej A. and Zahed, Ismail",
    title = "{Mueller\textquoteright{}s dipole wave function in QCD: Emergent Koba-Nielsen-Olesen scaling in the double logarithm limit}",
    eprint = "2211.05169",
    archivePrefix = "arXiv",
    primaryClass = "hep-ph",
    doi = "10.1103/PhysRevD.108.034017",
    journal = "Phys. Rev. D",
    volume = "108",
    number = "3",
    pages = "034017",
    year = "2023"
}

@article{Liu:2023eve,
    author = "Liu, Yizhuang and Nowak, Maciej A. and Zahed, Ismail",
    title = "{Universality of Koba-Nielsen-Olesen scaling in QCD at high energy and entanglement}",
    eprint = "2302.01380",
    archivePrefix = "arXiv",
    primaryClass = "hep-ph",
    month = "2",
    year = "2023"
}

@article{Kang:2023zdx,
    author = "Kang, Zhong-Bo and Kao, Robert and Larkoski, Andrew J.",
    title = "{Multiplicity scaling of fragmentation function}",
    eprint = "2305.13359",
    archivePrefix = "arXiv",
    primaryClass = "hep-ph",
    doi = "10.1103/PhysRevD.109.054039",
    journal = "Phys. Rev. D",
    volume = "109",
    number = "5",
    pages = "054039",
    year = "2024"
}

@article{Medves:2022ccw,
    author = "Medves, Rok and Soto-Ontoso, Alba and Soyez, Gregory",
    title = "{Lund and Cambridge multiplicities for precision physics}",
    eprint = "2205.02861",
    archivePrefix = "arXiv",
    primaryClass = "hep-ph",
    reportNumber = "OUTP-22-07P",
    doi = "10.1007/JHEP10(2022)156",
    journal = "JHEP",
    volume = "10",
    pages = "156",
    year = "2022"
}

@article{Medves:2022uii,
    author = "Medves, Rok and Soto-Ontoso, Alba and Soyez, Gregory",
    title = "{Lund multiplicity in QCD jets}",
    eprint = "2212.05076",
    archivePrefix = "arXiv",
    primaryClass = "hep-ph",
    reportNumber = "CERN-TH-2022-205, OUTP-22-13P",
    doi = "10.1007/JHEP04(2023)104",
    journal = "JHEP",
    volume = "04",
    pages = "104",
    year = "2023"
}

@book{Ellis:1996mzs,
    author = "Ellis, R. Keith and Stirling, W. James and Webber, B. R.",
    title = "{QCD and collider physics}",
    doi = "10.1017/CBO9780511628788",
    isbn = "978-0-511-82328-2, 978-0-521-54589-1",
    publisher = "Cambridge University Press",
    volume = "8",
    month = "2",
    year = "2011"
}

@article{Dremin:1999ji,
    author = "Dremin, I. M. and Gary, J. W.",
    title = "{Energy dependence of mean multiplicities in gluon and quark jets at the next-to-next-to-next-to leading order}",
    eprint = "hep-ph/9905477",
    archivePrefix = "arXiv",
    reportNumber = "FIAN-30-99, UCRHEP-E255",
    doi = "10.1016/S0370-2693(99)00713-3",
    journal = "Phys. Lett. B",
    volume = "459",
    pages = "341--346",
    year = "1999",
    note = "[Erratum: Phys.Lett.B 463, 346--346 (1999)]"
}

@article{Dokshitzer:2025owq,
    author = "Dokshitzer, Yu. L. and Webber, B. R.",
    title = "{Hadron multiplicity fluctuations in perturbative QCD}",
    eprint = "2505.00652",
    archivePrefix = "arXiv",
    primaryClass = "hep-ph",
    doi = "10.1007/JHEP08(2025)168",
    journal = "JHEP",
    volume = "08",
    pages = "168",
    year = "2025"
}

@article{Dokshitzer:2025fky,
    author = "Dokshitzer, Yu. L. and Webber, B. R.",
    title = "{QCD-inspired description of multiplicity distributions in jets}",
    eprint = "2507.07691",
    archivePrefix = "arXiv",
    primaryClass = "hep-ph",
    doi = "10.1007/JHEP10(2025)114",
    journal = "JHEP",
    volume = "10",
    pages = "114",
    year = "2025"
}

@book{Peskin:1995ev,
    author = "Peskin, Michael E. and Schroeder, Daniel V.",
    title = "{An Introduction to quantum field theory}",
    doi = "10.1201/9780429503559",
    isbn = "978-0-201-50397-5, 978-0-429-50355-9, 978-0-429-49417-8",
    publisher = "Addison-Wesley",
    address = "Reading, USA",
    year = "1995"
}

@article{Feynman:1969ej,
    author = "Feynman, Richard P.",
    editor = "Brown, L. M.",
    title = "{Very high-energy collisions of hadrons}",
    reportNumber = "PRINT-69-2817",
    doi = "10.1103/PhysRevLett.23.1415",
    journal = "Phys. Rev. Lett.",
    volume = "23",
    pages = "1415--1417",
    year = "1969"
}

@article{Slattery:1972ni,
    author = "Slattery, P.",
    title = "{Evidence for the Onset of Semiinclusive Scaling in Proton Proton Collisions in the 50-GeV/c - 300-GeV/c Momentum Range}",
    doi = "10.1103/PhysRevLett.29.1624",
    journal = "Phys. Rev. Lett.",
    volume = "29",
    pages = "1624",
    year = "1972"
}

@article{Ames-Bologna-CERN-Dortmund-Heidelberg-Warsaw:1983cqw,
    author = "Breakstone, A. and others",
    collaboration = "Ames-Bologna-CERN-Dortmund-Heidelberg-Warsaw",
    title = "{Charged Multiplicity Distribution in p p Interactions at ISR Energies}",
    reportNumber = "IS-J-1306, CERN-EP-83-165",
    doi = "10.1103/PhysRevD.30.528",
    journal = "Phys. Rev. D",
    volume = "30",
    pages = "528",
    year = "1984"
}

@article{TASSO:1989orr,
    author = "Braunschweig, W. and others",
    collaboration = "TASSO",
    title = "{Charged Multiplicity Distributions and Correlations in e+ e- Annihilation at PETRA Energies}",
    reportNumber = "DESY-89-038",
    doi = "10.1007/BF01674450",
    journal = "Z. Phys. C",
    volume = "45",
    pages = "193",
    year = "1989"
}

@article{ALEPH:1995qic,
    author = "Buskulic, D. and others",
    collaboration = "ALEPH",
    title = "{Measurements of the charged particle multiplicity distribution in restricted rapidity intervals}",
    reportNumber = "CERN-PPE-95-082, CERN-PPE-95-82",
    doi = "10.1007/BF02907382",
    journal = "Z. Phys. C",
    volume = "69",
    pages = "15--26",
    year = "1995"
}

@article{Giovannini:1997ce,
    author = "Giovannini, Alberto and Ugoccioni, Roberto",
    editor = "Capon, G. and Khoze, Valery A. and Pancheri, G. and Sansoni, A.",
    title = "{Soft and semihard components structure in multiparticle production in high-energy collisions}",
    eprint = "hep-ph/9710361",
    archivePrefix = "arXiv",
    reportNumber = "DFTT-65-97",
    doi = "10.1016/S0920-5632(98)00343-0",
    journal = "Nucl. Phys. B Proc. Suppl.",
    volume = "71",
    pages = "201--210",
    year = "1999"
}

@article{Metodiev:2018ftz,
    author = "Metodiev, Eric M. and Thaler, Jesse",
    title = "{Jet Topics: Disentangling Quarks and Gluons at Colliders}",
    eprint = "1802.00008",
    archivePrefix = "arXiv",
    primaryClass = "hep-ph",
    reportNumber = "MIT-CTP-4979",
    doi = "10.1103/PhysRevLett.120.241602",
    journal = "Phys. Rev. Lett.",
    volume = "120",
    number = "24",
    pages = "241602",
    year = "2018"
}

@article{Komiske:2018vkc,
    author = "Komiske, Patrick T. and Metodiev, Eric M. and Thaler, Jesse",
    title = "{An operational definition of quark and gluon jets}",
    eprint = "1809.01140",
    archivePrefix = "arXiv",
    primaryClass = "hep-ph",
    reportNumber = "MIT-CTP 5042",
    doi = "10.1007/JHEP11(2018)059",
    journal = "JHEP",
    volume = "11",
    pages = "059",
    year = "2018"
}

@article{Martins-Fontes:2025iee,
    author = "Martins-Fontes, H. R. and Navarra, F. S.",
    title = "{Soft and semihard components of multiplicity distributions in the $k_T$ factorization approach}",
    eprint = "2506.17127",
    archivePrefix = "arXiv",
    primaryClass = "hep-ph",
    month = "6",
    year = "2025"
}

@article{Gallicchio:2011xq,
    author = "Gallicchio, Jason and Schwartz, Matthew D.",
    title = "{Quark and Gluon Tagging at the LHC}",
    eprint = "1106.3076",
    archivePrefix = "arXiv",
    primaryClass = "hep-ph",
    doi = "10.1103/PhysRevLett.107.172001",
    journal = "Phys. Rev. Lett.",
    volume = "107",
    pages = "172001",
    year = "2011"
}

@article{Gallicchio:2012ez,
    author = "Gallicchio, Jason and Schwartz, Matthew D.",
    title = "{Quark and Gluon Jet Substructure}",
    eprint = "1211.7038",
    archivePrefix = "arXiv",
    primaryClass = "hep-ph",
    doi = "10.1007/JHEP04(2013)090",
    journal = "JHEP",
    volume = "04",
    pages = "090",
    year = "2013"
}

@article{Larkoski:2014pca,
    author = "Larkoski, Andrew J. and Thaler, Jesse and Waalewijn, Wouter J.",
    title = "{Gaining (Mutual) Information about Quark/Gluon Discrimination}",
    eprint = "1408.3122",
    archivePrefix = "arXiv",
    primaryClass = "hep-ph",
    reportNumber = "MIT--CTP-4572, NIKHEF-2014-026",
    doi = "10.1007/JHEP11(2014)129",
    journal = "JHEP",
    volume = "11",
    pages = "129",
    year = "2014"
}

@article{Metodiev:2017vrx,
    author = "Metodiev, Eric M. and Nachman, Benjamin and Thaler, Jesse",
    title = "{Classification without labels: Learning from mixed samples in high energy physics}",
    eprint = "1708.02949",
    archivePrefix = "arXiv",
    primaryClass = "hep-ph",
    reportNumber = "MIT--CTP-4922",
    doi = "10.1007/JHEP10(2017)174",
    journal = "JHEP",
    volume = "10",
    pages = "174",
    year = "2017"
}

@article{Gras:2017jty,
    author = {Gras, Philippe and H{\"o}che, Stefan and Kar, Deepak and Larkoski, Andrew and L{\"o}nnblad, Leif and Pl{\"a}tzer, Simon and Si{\'o}dmok, Andrzej and Skands, Peter and Soyez, Gregory and Thaler, Jesse},
    title = "{Systematics of quark/gluon tagging}",
    eprint = "1704.03878",
    archivePrefix = "arXiv",
    primaryClass = "hep-ph",
    reportNumber = "MIT-CTP-4885, COEPP-MN-17-2, MCNET-17-04",
    doi = "10.1007/JHEP07(2017)091",
    journal = "JHEP",
    volume = "07",
    pages = "091",
    year = "2017"
}

@article{CMS:2013kfa,
    collaboration = "CMS",
    title = "{Performance of quark/gluon discrimination in 8 TeV pp data}",
    reportNumber = "CMS-PAS-JME-13-002",
    year = "2013"
}

@article{ATLAS:2016wzt,
    collaboration = "ATLAS",
    title = "{Discrimination of Light Quark and Gluon Jets in $pp$ collisions at $\sqrt{s} = 8$ TeV with the ATLAS Detector}",
    reportNumber = "ATLAS-CONF-2016-034",
    month = "7",
    year = "2016"
}

@article{ATLAS:2014vax,
    author = "Aad, Georges and others",
    collaboration = "ATLAS",
    title = "{Light-quark and gluon jet discrimination in $pp$ collisions at $\sqrt{s}=7\mathrm {\ TeV}$ with the ATLAS detector}",
    eprint = "1405.6583",
    archivePrefix = "arXiv",
    primaryClass = "hep-ex",
    reportNumber = "CERN-PH-EP-2014-058",
    doi = "10.1140/epjc/s10052-014-3023-z",
    journal = "Eur. Phys. J. C",
    volume = "74",
    number = "8",
    pages = "3023",
    year = "2014"
}

@article{Dreyer:2021hhr,
    author = "Dreyer, Fr{\'e}d{\'e}ric A. and Soyez, Gregory and Takacs, Adam",
    title = "{Quarks and gluons in the Lund plane}",
    eprint = "2112.09140",
    archivePrefix = "arXiv",
    primaryClass = "hep-ph",
    doi = "10.1007/JHEP08(2022)177",
    journal = "JHEP",
    volume = "08",
    pages = "177",
    year = "2022"
}

@article{Currie:2017eqf,
    author = "Currie, James and Gehrmann-De Ridder, Aude and Gehrmann, Thomas and Glover, E. W. N. and Huss, Alexander and Pires, Joao",
    title = "{Precise predictions for dijet production at the LHC}",
    eprint = "1705.10271",
    archivePrefix = "arXiv",
    primaryClass = "hep-ph",
    reportNumber = "IPPP-17-45, ZU-TH-13-17, MPP-2017-107",
    doi = "10.1103/PhysRevLett.119.152001",
    journal = "Phys. Rev. Lett.",
    volume = "119",
    number = "15",
    pages = "152001",
    year = "2017"
}

\end{document}